\documentclass[twocolumn, astrosymb, times,tighten, tikz]{aastex701}

\usepackage{amsmath}
\usepackage{CJK}
\usepackage{graphicx}
\usepackage{hyperref}
\usepackage{enumerate}
\usepackage{subfigure}
\usepackage{makecell}
\usepackage{soul}
\usepackage{xcolor}
\usepackage[T1]{fontenc}

\begin{document}
\begin{CJK*}{UTF8}{gbsn}
\shortauthors{B.\ Peng, D.\ Valencia }

\title{Modeling carbon outgassing from chondritic planetesimals}

\author{Bo Peng (彭博)\email{bpengeps@stanford.edu}}
\email{bpengeps@stanford.edu}
\affiliation{Department of Earth \& Planetary Sciences, Stanford University, 450 Jane Stanford Way, Stanford CA, 94305, USA}
\affiliation{Department of Astronomy \& Astrophysics, University of Toronto, 50 St. George St., Toronto, ON M5S 3H4, Canada}
\author{Diana Valencia\email{diana.valencia@utoronto.ca} }
\affiliation{Department of Astronomy \& Astrophysics, University of Toronto, 50 St. George St., Toronto, ON M5S 3H4, Canada}
\affiliation{Department of Physical and Earth Sciences, University of Toronto Scarborough, 1065 Military Trail, Toronto, ON, M1C 1A4, Canada}
\vspace{3cm}

\begin{abstract}

The thermochemical evolution of planetesimals is an underprobed stage of volatile delivery to terrestrial planets during their formation, and may contribute to the volatile depletion of the Earth relative to primitive chondrites. We have developed a model of C outgassing from porous, chondritic planetesimals. Our model tracks the thermal evolution and the production of CO/CO$_2$ gas using the redox states of ordinary and enstatite chondrites (OC and EC, respectively, collectively the ``NCs"), and CI and CV carbonaceous chondrites (``CCs"). We posit the formation of global fractures when local gas pressure exceeds confinement levels, which vent the excess directly to space, leading to efficient C depletion. We also account for sintering and the enthalpy of dehydration from wet carbonaceous chondrite bodies. We find that C depletion is more efficient on CC planetesimals than NCs due to the former's oxidized environment: for 10-100 km planetesimals formed at 2 Myr after CAI formation, $>50\%$ of C is depleted in almost all CC bodies while $<50\%$ is depleted in almost all NC bodies. Both the largest and the smallest bodies tend to preserve more C, the former due to sintering locking condensed C in against escape, while the latter due to efficient conductive cooling. Earlier accreted planetesimals deplete more C: bodies formed before $\sim$2 My deplete most of their C. Our results favor NC planetesimals as the C carriers during terrestrial planets' accretion. Terrestrial planets likely accreted from a mix of C-depleted and C-rich bodies from both CC and NC reservoirs.
\end{abstract}

\section{Introduction}\label{Sec:Intro} 

The carbon budget of terrestrial planets impacts their long-term evolution and habitability. How carbon was delivered to the solar system's terrestrial planets during their formation remains unclear (e.g., \citealt{Hirschmann_2021PNAS}). Compared to the chondritic record, the modern Earth mantle is 2-3 magnitude depleted in carbon \citep{bergin15, Hirschmann_2018_deep_Earth_HCN}. If our meteoritic record is a representative sample of the terrestrial building blocks, then understanding carbon delivery to the terrestrial planets requires elucidating the pathways of carbon (and broadly, volatile) depletion at each stage of planet assembly. 

Previous works probed diverse carbon loss mechanisms during planet formation. These include (a) the pyrolysis, photolysis and oxidation of carbon-rich grains in the protoplanetary disk \citep{LeeDisk,bergin15, Anderson_2017ApJ_disk_C...845...13A, Li_2021SciA_C_deficit_early_subli,Binkert_23_C_in_disc, Okamoto_2024A&A...692A..11O}, (b) carbon preferentially dissolving into the metallic melt and thus buried during protoplanet core formation (e.g. \citealt{HIRSCHMANN2012, Fischer20}), and (c) the erosion of carbon in the atmospheres of protoplanets, either via atmosphere-protoplanetary disk recycling \citep{Johansen21}, or via atmosphere loss after disk dispersal (e.g. \citealt{Tucker_14_multipleOutgassing}).

An under-examined but significant process of carbon depletion is outgassing from  planetesimals. These $\sim$100-km-sized bodies were large enough and accreted early enough that short-lived radioactive isotopes $^{26}$Al and $^{60}$Fe heated their interiors to enable silicate melting (e.g., \citealt{Hevey_2006M&PS_differentiation...41...95H, Moskovitz_2011M&PS...46..903M}). This high-temperature yet low-gravity environment can efficiently deplete volatiles. Recent theoretical and meteoritic studies indicate that planetesimal outgassing may have been prevalent. \cite{Li_2021SciA_C_deficit_early_subli} showed that the protoplanetary disk likely delivered more carbon to the terrestrial planet formation zone than the reasonable maximum of bulk Earth abundances. While \cite{Gu_2024_Earth_Volatiles} combined magma ocean protoplanet models with realistic accretion histories and demonstrated that core burial and atmospheric escape during the giant impact phase are insufficient to explain the bulk silicate Earth's carbon depletion without an already depleted planetesimal reservoir. Further, chondrites that experienced higher metamorphic temperatures tend to have lower carbon content, while iron meteorites, remnants of planetesimal cores, point to open-system carbon loss during core-mantle differentiation \citep{Hirschmann_2021PNAS}.  

Meanwhile, detailed modeling of planetesimal outgassing that accounts for the diversity of chondritic compositions has yet to be done. Pioneering works by \cite{Sugiura_1984LPSC...14..641S, Sugiura_1986E&PSL..78..148S} and \cite{Hashizume_1998M&PS...33.1181H} modeled carbon migration and loss in ordinary chondrite (OC) planetesimals that assumed constant gas permeabilities and porosities, relevant to intact OC matrices. Later works examined low temperature, aqueous fluid migration on carbonaceous chondrite (CC) bodies \citep{Grimm_1989Icar_McSween...82..244G, Young_1999Sci...286.1331Y, Young_2001RSPTA.359.2095Y, Young_2003E&PSL.213..249Y, Palguta_2010E&PSL.296..235P, Fu_2017pedc.book..115F}. In particular, \cite{Grimm_1989Icar_McSween...82..244G} first highlighted the role of fluid venting due to gas overpressure fracturing the matrix. \cite{Fu_2017pedc.book..115F} compiled chondrite permeabilities and showed that efficient fluid migration is not guaranteed without fracturing, and the chondritic record likely corresponds to diverse outgassing histories. The latter is expected, as planetesimals may have accreted in distinct batches during the first few million years of the solar system's lifetime and sampled varied disk locations \citep{Kruijer_2017PNAS..114.6712K, Lichtenberg_2021Sci...371..365L}. 

Here, we present a thermo-mechanical-chemical model of chondritic planetesimals that trace their carbon depletion via overpressure-induced fracturing and venting. We utilize the redox states of typical CC, OC, and enstatite chondrites (EH) to probe the C depletion efficiencies of diverse planetary building blocks. We also examine the ability of sintering, the viscous compaction of an initially porous matrix, in quenching C outgassing. 

In wet CC bodies, aqueous evolution may deplete some soluble organics (e.g., \citealt{Young_1999Sci...286.1331Y}), while early-formed planetesimals' melting metal-silicate differentiation may also drive C loss \citep{Lichtenberg_2021Sci...371..365L}. As a first step we do not explicitly model aqueous evolution, besides the thermal effect of phyllosilicates' dehydration in CCs, because aqueous fluid migration tends to happen at lower temperatures ($\lesssim$200 $^{\circ}$C, \citealt{Fu_2017pedc.book..115F}) than C outgassing ($\sim$700 K - 1100 K), as we will show. Our model is thus strictly apropos for most NCs and some CCs that accreted close to, or within the ice line, and likely did not form with significant water \citep{Bermingham20, Doyle_2015NatCo...6.7444D, Hirakawa_2021EP&S...73...16H}. We also do not consider melting and melt migration because both sintering and C outgassing occur below the silicate solidus $\sim$1400 K \citep{Taylor_1992JGR....9714717T}. Chondrites have diverse carbon carriers and complex volatilization behaviors \citep{Christ_2024CmChe...7..118C,Alexander_2017ChEG...77..227A}. To make the problem tractable, we use graphite in chemical equilibrium calculations while assuming fast kinetics appropriate for simulating the pyrolysis behaviors of chondritic organics \citep{Kebukawa_2010M&PS...45...99K}. Thus, this work seeks to understand the C depletion behaviors of both individual planetesimals and population-level trends, which are sculpted by the interplay of self-gravity-induced sintering and overpressure-driven outgassing. Under this framework, we explore how these population-level trends are influenced by composition, radius, and time of formation.

\section{Planetesimal Outgassing model}\label{Sec:Method}

Our model simulates the 1-D thermal-mechanical-chemical evolution of primitive planetesimals to trace the fate of carbon in their interiors. We assume the planetesimals start as loose, homogeneous aggregates of chondrite, graphite, and pore space. We trace their thermal evolution, controlled by radiogenic heating, conductive cooling, the heat consumed by phyllosilicates' dehydration, and finally the heat of graphite oxidation. The temperature profile informs the chemical equilibrium between C, CO, and CO$_2$. We test different chondritic compositions—enstatite, ordinary, and carbonaceous chondrites—by adopting their corresponding oxygen fugacities and densities.

The key, competing mechanisms in our model are pressure-valve outgassing and sintering. As temperature rises in planetesimal interiors, pore gas accumulates following its chemical equilibrium, until it exceeds the lithostatic confinement pressure in certain layers of the planetesimal. We posit that, this situation leads to the formation of extensive fractures in the matrix which efficiently vents the excess gas to space. We adopt the most straightforward representation of this process in our model, a local pressure valve effect, where the overpressure is directly removed to space, resulting in local C loss. 
We also model sintering, in which lithostatic pressure drives the viscous deformation of matrix grains, collapsing the pore space. Sintering preserves the C reservoir against outgassing, because it effectively lithifies the initially loosely-packed planetesimals (e.g., \citealt{henke12, neumann14}), preventing fracturing and efficient gas venting. Finally, the pore gas pressure also supports the local matrix against sintering by countering the lithostatic pressure. Thus, gas venting and sintering are mutually limiting; their interactions shape the eventual C depletion in chondritic planetesimals. We ignore the contribution of background outflow through diffusion, making our C depletion predictions conservative. Caveats for our approach are discussed in Section \ref{sec_2:Disc_caveats}. 
  
\subsection{Chemical equilibrium}\label{sec_2:chem_eq}

\begin{figure}
    \centering
    \includegraphics[width =\linewidth]{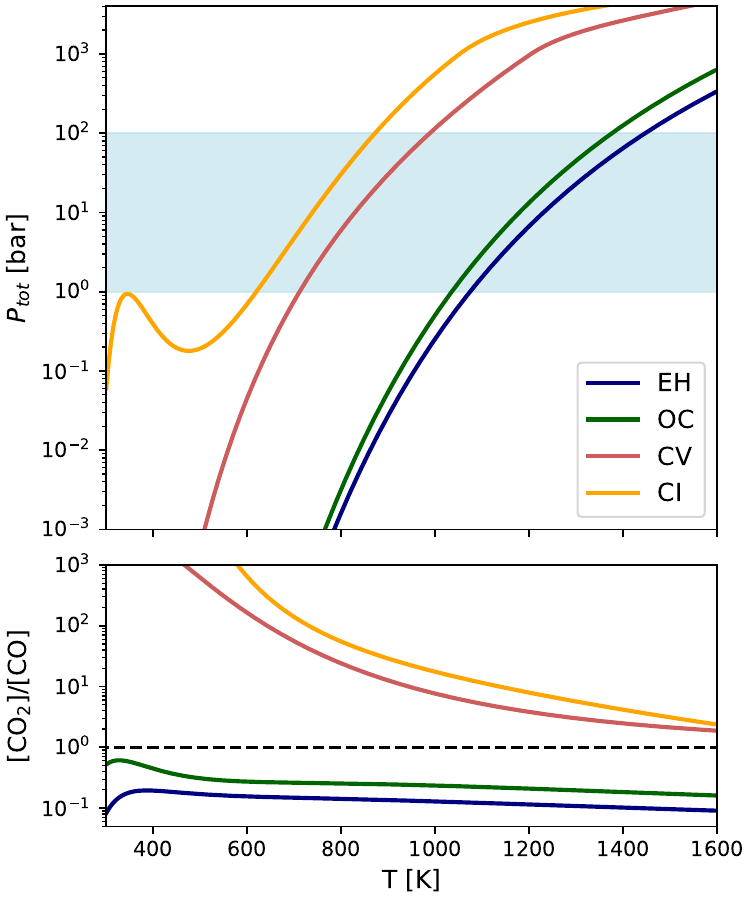}
    \caption{Chemical equilibrium with four chondritic redox states. Top: total equilibrium pressures vs. temperature. The highlighted region represents the lithostatic pressures of $\sim$10 - 100 km planetesimals. Bottom: molar CO/CO$_2$ ratio in equilibrium with the four chondritic compositions. }
    \label{fig:Chem_Eq}
\end{figure}

Our model calculates the equilibrium partial pressures of CO and CO$_2$, $P_{\mathrm{CO, eq}}$ and $P_{\mathrm{CO_2, eq}}$ in contact with graphite (Figure \ref{fig:Chem_Eq}).   Conceptually, the timescale for chemical kinetics is faster than that of the radiogenic heating, which sets the timescale of the thermal evolution of a planetesimal ($\lesssim 10^3$ years vs. $\gtrsim10^6$ years, see Sections \ref{sec_2:Rad_heat} and \ref{sec_2:reac_rate}). Therefore, in the layers of a planetesimal that are not actively venting carbon, the pore gas pressures are practically governed by local chemical equilibrium (meanwhile, in the locations that are actively outgassing, the valve mechanism sets the gas pressure). 
Functionally, we use $P_{\mathrm{CO, eq}}$ and $P_{\mathrm{CO_2, eq}}$ to calculate the kinetic reaction rate (Section \ref{sec_2:reac_rate}), which helps regulate the actual pore gas pressures, thus impacting the compositional and thermal evolution (Sections \ref{sec_2:mass_vol_cons} and \ref{sec_2:Eng_cons}). The equilibrium is controlled by the following reactions
\begin{equation}
    \mathrm{C(graphite) +O_2(g) \rightleftharpoons CO_2(g), \quad(R1)} 
\end{equation}
\begin{equation}
    \mathrm{C(graphite) + \frac{1}{2}O_2(g). \rightleftharpoons CO(g) \quad(R2)} 
\end{equation}

These reactions are also known as the CCO buffer \citep{keppler19}. The equilibrium constants of the above reactions are  
\begin{equation}\label{eq_2:K_1_CCO}
    K_1(T) \equiv \frac{f_{\mathrm{CO_2}}}{f_{\mathrm{O_2}}},
\end{equation}
\begin{equation}\label{eq_2:K_2_CCO}
    K_2(T) \equiv \frac{f_{\mathrm{CO}}}
    {f_{\mathrm{O_2}}^{0.5}},
\end{equation}
where $f_{\mathrm{CO}}$ and $f_{\mathrm{CO_2}}$ are the fugacities of CO and CO$_2$, describing their chemical potency in a reaction. $f_{\mathrm{O_2}}$ is the oxygen fugacity, a measurement of the environment's oxidation state. In our system, $f_{\mathrm{O_2}}$ is controlled by the redox state of the chondrites. For this, we use the chondritic $f_{\mathrm{O_2}}$'s as functions of temperature from \cite{Schaefer17}, where they calculated the equilibrium $f_{\mathrm{O_2}}$ of atomic mixtures corresponding to the typical composition of four chondrite classes: carbonaceous chondrites CI and CV, H-type ordinary chondrite, and EH-type enstatite chondrites. We denote the oxygen fugacity of composition $c$ as $f_{\mathrm{O_2}}^\mathrm{c}(T)$.

In the low-pressure regime, where the total gas pressure $P_{tot}< 1$ kbar, we calculate $K_1$ and $K_2$ following the recipe of \cite{Robie_1995_book_therm}, using the thermochemical data for CO and CO$_2$ from \cite{NIST}, while we use \cite{Robie_1995_book_therm} for graphite. Since the actual gas pressure in the pore space is limited by the lithostatic pressure of these minor bodies and scarcely exceeds 100 bars, we use the ideal gas equations of state (EoS), where a species' fugacity is equivalent to its partial pressure
\begin{equation}\label{eq_2:EoS_CO2}
    f_{\mathrm{CO}_2} \equiv P_{\mathrm{CO}_2}, 
\end{equation}
\begin{equation}\label{eq_2:EoS_CO}
    f_{\mathrm{CO}} \equiv P_{\mathrm{CO}}, 
\end{equation}
where $P_{\mathrm{CO}}$ and $P_{\mathrm{CO}_2}$ are the partial pressures of CO and CO$_2$.

Rewriting Equations \ref{eq_2:K_1_CCO} and \ref{eq_2:K_2_CCO}, we can calculate the equilibrium CO and CO$_2$ partial pressures with known equilibrium constants and $f^\mathrm{c}_{\mathrm{O_2}}(T)$ 

\begin{equation}
    \label{eq_2:PCO2_low}
      P_{\mathrm{CO_2, eq}} = K_1(T)f^\mathrm{c}_{\mathrm{O_2}}(T),
\end{equation}
\begin{equation}\label{eq_2:PCO_low}
    P_{\mathrm{CO, eq}} = K_2(T)[f^\mathrm{c}_{\mathrm{O_2}}(T)]^{\frac{1}{2}}.
\end{equation}

To simplify the mass conservation calculation, we assume that gas-phase reactions are fast, and CO and CO$_2$ are always in equilibrium with O$_2$ (e.g., \citealt{Mentser_1967Carbo...5..331M, Li_2001Carbo_kinetics..39..725L}). Their relative molar fraction is thus set by the reaction 
\begin{equation}
    \mathrm{CO +\frac{1}{2}O_2 \rightleftharpoons CO_2,\quad(R3)},
\end{equation}
with its equilibrium constant defined as
\begin{equation}\label{eq_2:K3}
    K_{3}(T) = \frac{f_{\mathrm{CO_2}}}{f_{\mathrm{O_2}}^{0.5} f_{\mathrm{CO}}}.
\end{equation}
Again, we use the thermochemistry data from \cite{NIST} to calculate $K_3(T)$. 
With $f_{\mathrm{O_2}}$ known at a particular temperature, we utilize the ideal EoS  (Equations \ref{eq_2:EoS_CO2} and \ref{eq_2:EoS_CO}) and Equation \ref{eq_2:K3} to find the gas ratio 
\begin{equation}\label{eq_2:r_Cg}
    r^\mathrm{c}_{\mathrm{Cg}}(T) \equiv\frac{P_{\mathrm{CO_2}}}{P_{\mathrm{CO}}} = K_3(T) [f^\mathrm{c}_{\mathrm{O_2}}(T)]^{\frac{1}{2}}.
\end{equation}

With $r\mathrm{^c_{Cg}}(T)$ known (Figure \ref{fig:Chem_Eq}, bottom), we treat the two gases as one fluid, the ``carbon gas", with the subscript $\mathrm{Cg}$. The carbon gas pressure is
\begin{equation}\label{eq_2:P_tot_NIST}
    P_{\mathrm{Cg}} \equiv P_{\mathrm{CO}} + P_{\mathrm{CO_2}}.
\end{equation}
We treat $P_{\mathrm{Cg}}$ as the total gas pressure in the pore space and ignore the partial pressure of O$_2$ because $P_{\mathrm{O}_2} \sim f_{\mathrm{O}_2}\ll P_\mathrm{{tot}}$ for all compositions and temperatures of interest.

As the chondritic $f_{\mathrm{O_2}}$ rises with temperature, generally so does $P_\mathrm{Cg,\,eq}$ (Figure \ref{fig:Chem_Eq}, top). At high temperatures, for CI and CV chondrites, the equilibrium $P_\mathrm{Cg,\,eq}$ rises beyond one kbar at $\sim$1100 K. In these higher pressure ranges, the \cite{NIST} thermochemical data, calibrated at 0.1 MPa, become insufficient. Instead, we adopt the P-T-$f_{\mathrm{O_2}}$ relation of the C-O-graphite system from \cite{Jakobsson_1994GeCoA..58....9J}, which is derived from experimental results at 7.5 to 20 kbars and 1100 to 1500 $^\circ$C
\begin{equation}\label{eq_2:JO_CCO}
   \log_{10} f_{\mathrm{O_2}} = 4.325 - \frac{21803}{T} + 0.171 \frac{P - 1}{T} ,
\end{equation}
where $P$ is the total pressure of the environment, which in our context is $P_\mathrm{Cg,\,eq}$.

Combining Eq. \ref{eq_2:JO_CCO} and the chondritic $f\mathrm{^c}_{\mathrm{O_2}}(T)$, we arrive at the high-pressure solution for $P_\mathrm{Cg,\,eq}$ - 
\begin{equation}\label{eq_2:P_tot_JO}
\begin{split}
        P_\mathrm{Cg,\,eq}(T_{high}) = &5.85\log_{10}f\mathrm{^c}_{\mathrm{O_2}}(T)T \\&- 25.29 T + 1.275\times10^5.
\end{split}
\end{equation}

In practice, we calculate $P_\mathrm{Cg,\,eq}$ for planetesimals of CI and CV compositions piece-wise: we use Equations \ref{eq_2:PCO2_low}, \ref{eq_2:PCO_low} and \ref{eq_2:P_tot_NIST} for $P_\mathrm{Cg,\,eq}<1$ kbar and Eq. \ref{eq_2:P_tot_JO} for $P_\mathrm{Cg,\,eq}>7.5$ kbar. We fit a third-degree polynomial to interpolate between the two solutions smoothly
\begin{equation}\label{eq_2:P_trans}
    P_\mathrm{{Cg,\, eq,\, trans}}= 
      a_{\mathrm{ch}}T^3 + b_{\mathrm{ch}}T^2 + c_{\mathrm{ch}}T + d_{\mathrm{ch}},
\end{equation}
where the fitting constants $a_{\mathrm{ch}}$ through $d_{\mathrm{ch}}$, as well as the transition temperatures $T_0$ and $T_1$ are given in Table \ref{Tab:transition}.

\begin{deluxetable*}{lCCCCCC}
    \tablecaption{Fitting parameters for Equation \ref{eq_2:P_trans}.\label{Tab:transition}}
    \tablehead{
    \colhead{\textbf{Comp.}} & \colhead{$a_{\mathrm{ch}}$} & \colhead{$b_{\mathrm{ch}}$} & \colhead{$c_{\mathrm{ch}}$} & \colhead{$d_{\mathrm{ch}}$} & \colhead{$T_0$} & \colhead{$T_1$}
    }
    \startdata
    \textbf{CI} & -8.55\cdot10^{-7}& 1.328\cdot 10^{-3}& 10.48& -1.049\cdot 10^{4}& 1051.8 & 1774.3\\
    \textbf{CV} & 2.407\cdot 10^{-5}& 9.569\cdot 10^{-2} & 134.9& -6.477\cdot 10^{4}& 1203.8 & 1757.8\\
    \enddata
\end{deluxetable*}

For EH and OC chondrite compositions, we use the low-temperature $P_\mathrm{Cg,\,eq}$ because it does not rise above one kbar at $T \lesssim 1500$ K. Above $\sim$1500 K, matrix melting would commence, and our model would not be strictly applicable.

\subsection{Mass and Volume Conservation}\label{sec_2:mass_vol_cons}
Our model tracks mass and volume conservation for graphite and carbon gas on a 1-D radial grid of the planetesimal. We assume that the pore spaces at each layer initially act as closed systems. When the gas pressure exceeds the local confinement pressure, we postulate that an extensive crack network forms within the local matrix, allowing gas to vent directly to the surface. This supposition is motivated by the fact that extensive brecciation is common in chondritic samples \citep{Bischoff_2006_brecciation_mess.book..679B} and that the propagation of cracks drastically improves the permeability of the chondritic matrix \citep{Fu_2017pedc.book..115F}. We simulate this scenario by introducing a pressure valve mechanism, whereby a sink term removes excess gas, effectively capping the gas pressure at the confinement level. This open-system scenario modifies the mass conservation equation. 

Below, we establish the mass and volume conservation equations for both open and closed systems and derive the time rates of change for the variables the model numerically integrates. These include the rate of change of the volume fractions of graphite and pore space ($\dot{\phi}_{\mathrm{Cs}}$ and $\dot{\phi}_\mathrm{por}$, respectively), as well as that of the density of carbon gas ($\dot{\rho}_\mathrm{Cg}$). Since carbon is a minor constituent in chondrites, while oxygen is abundant, we only track the C atoms in graphite, CO and CO$_2$ when considering mass conservation, effectively assume unlimited O supply. The volumetric masses of graphite (``$\mathrm{Cs}$") and carbon gas   (``$\mathrm{Cg}$") are 
\begin{equation}\label{eq_2:mass_vol_den_Cs}
    m_\mathrm{Cs} = \phi_\mathrm{Cs} \rho_\mathrm{Cs},
\end{equation}
\begin{equation}\label{eq_2:mass_vol_den_Cg}
    m_\mathrm{Cg} = \phi_\mathrm{por} \rho_\mathrm{Cg}.
\end{equation}

\subsubsection{Closed system}\label{sec_2:mass_vol_cons_CL}
We assume the system is closed before the pressure valve is turned on. In this case, as temperature rises, graphite is converted to carbon gas, both raising $\phi_\mathrm{por}$ and liberating pore space previously occupied by graphite. Below we derive the corresponding $\dot{\phi}_\mathrm{Cs}$, $\dot{\phi}_\mathrm{por}$ and $\dot{\rho}_\mathrm{Cg}$.

Volume conservation dictates 
\begin{equation}\label{eq_2:vol_cons}
    1 = \phi_\mathrm{por} + \phi_\mathrm{Cs} + \phi_\mathrm{chond}.
\end{equation}
where $\phi_\mathrm{por}$ is the porosity or the volume fraction occupied by the carbon gas. $\phi_\mathrm{Cs}$ is the volume fraction of graphite, and $\phi_\mathrm{chond}$ is that of the chondritic matrix. Oxidizing graphite liberates pore space. Thus, we take the time derivative of Equation \ref{eq_2:vol_cons}, assuming the chondritic matrix does not deform, or $\dot{\phi}_\mathrm{chond} = 0$, and obtain 
\begin{equation}\label{eq_2:volume_cons_used}
    \dot{\phi}_\mathrm{por} = -\dot{\phi}_\mathrm{Cs}.
\end{equation}
Here $\dot{\phi}_i$ denotes the time rate of change of phase $i$'s volume fraction. In reality, the chondritic matrix does undergo sintering, the hydrostatic compression that closes the pore space. But this effect is active at a larger timescale than chemical equilibration. Thus, we simulate this effect independently of the system of conservation equations here. We describe our sintering prescription in Section \ref{Sec_2:sintering}. 

Mass conservation for C in a closed system is simply the C lost in graphite being gained in the gas phase
\begin{equation}\label{eq_2:m_CG_closed_simple}
    \dot{m}_\mathrm{Cg,\,closed} = -\dot{m}_\mathrm{Cs,\, closed},
\end{equation}
where $\dot{m}_i$ is the rate of change of the mass of phase $i$, in $\mathrm{kg/(m^3\cdot s)}$. This relation is valid as we only track the C atoms in carbon gas and not the oxygen for simplicity. For simplicity We write the kinetic rate as $\dot{m}\mathrm{_{Cg}|_{burn}}$, the volumetric rate of carbon burning in the units of $\mathrm{kg/(m^3\cdot s)}$, or the mass of gas generated per m$^3$ of the material mixture (chondrite, graphite, and pore space), per second.
As detailed in Section \ref{sec_2:reac_rate}, $\dot{m}\mathrm{_{Cg}|_{burn}}(T, P_\mathrm{Cg})$ is a function of temperature, pore gas pressure, and chondrite composition. In a closed system, C burning is the only source term for the gas:
\begin{equation}\label{eq_2:m_CG_closed}
    \dot{m}_\mathrm{Cg,\,closed} = \dot{m}\mathrm{_{Cg}|_{burn}},
\end{equation}

Applying the chain rule to $\dot{m}_\mathrm{Cg,\,closed}$ (using Equation \ref{eq_2:mass_vol_den_Cg}), we have
\begin{equation}\label{eq_2:m_CG_chain_closed}
    \dot{m}_\mathrm{Cg,\,closed} = \dot{\phi}_\mathrm{por}\rho_\mathrm{Cg} + 
    \phi_\mathrm{por}\dot{\rho}_\mathrm{Cg}
\end{equation}

Meanwhile the mass conservation for graphite is
\begin{equation}\label{eq_2:graphite_burn}
    \dot{m}_\mathrm{Cs} = - \dot{m}_\mathrm{Cg}\mathrm{|_{burn}}.
\end{equation}
Note that we dropped the ``$\mathrm{closed}$" label because burning is the only graphite sink, making this expression true for both open and closed systems. We assume that graphite's density is constant (i.e. ${\rho}_\mathrm{Cs} = 2$g/cm$^3$; $\dot{\rho}_\mathrm{Cs} = 0$), thus, the chain-rule expansion of $\dot{m}_\mathrm{Cs}$ 
  (Equation \ref{eq_2:mass_vol_den_Cs}) is simply:
\begin{equation}\label{eq_2:graphite_mass_conserve}
    \dot{m}_\mathrm{Cs} = \dot{\phi}_\mathrm{Cs} \rho_\mathrm{Cs}.
\end{equation}

We track the time evolution of graphite and gas by integrating $\dot{\phi}_\mathrm{por}$, $\dot{\phi}_\mathrm{Cs}$, and $\dot{\rho}_\mathrm{Cg}$. The first two of which are found by first calculating the burning rate $\dot{m}_\mathrm{Cg}|_\mathrm{burn}$. Then using Equations \ref{eq_2:volume_cons_used}, \ref{eq_2:graphite_burn} and \ref{eq_2:graphite_mass_conserve}, we obtain 
\begin{equation}\label{eqn:dphi_C}
    \boxed{
    \dot{\phi}_\mathrm{por} = -\dot{\phi}_\mathrm{Cs} = \frac{\dot{m}_\mathrm{Cg}|_\mathrm{burn}}{\rho_\mathrm{Cs}}.}
\end{equation}

We then express the rate of gas density change in the closed system, $\dot{\rho}_\mathrm{Cg, \,closed}$, by manipulating Equation \ref{eq_2:m_CG_chain_closed}, and obtain
\begin{equation}\label{eq_2:rho_CG_closed}
    \boxed{\dot{\rho}_\mathrm{Cg, \,closed} = \frac{\dot{m}_\mathrm{Cg}|_\mathrm{burn}}{\phi_\mathrm{por}}\Big(1 - \frac{\rho_\mathrm{Cg}}{\rho_\mathrm{Cs}}\Big).}
\end{equation}

\subsubsection{Open system}\label{sec_2:mass_vol_cons_OP}

When the local gas pressure reaches the confinement level, a pressure valve effect is triggered. That is, an open-system outflow commences, removing the excess and capping the gas pressure at the confinement level. In this regime, the model calculates a different $\dot{\rho}_\mathrm{Cg}$. Meanwhile, since the gas outflow also decreases the local thermal energy (see Section \ref{Sec_2:Q_valve}), the model also tracks the rate of mass outflow, $\dot{m}_\mathrm{Cg}|_\mathrm{valve}$. Below, we derive these two quantities from the constraints and conservation equations of this regime. 

As $P_\mathrm{Cg}$ is fixed at the confinement level, $\rho_\mathrm{Cg}$ only responds to the change in local temperature and becomes decoupled from the chemical reaction. In other words, we require
\begin{equation}
    P_\mathrm{Cg} = P_\mathrm{valve},
\end{equation}
The ideal EoS links $P_\mathrm{Cg}$ to $\rho_\mathrm{Cg}$,
\begin{equation}
    P_\mathrm{Cg} = \rho_\mathrm{Cg}R_\mathrm{gas} T/\mu.
\end{equation}
where $\mu = 0.012$ kg/mol is the nominal ``molecular" weight of carbon gas, and $R_\mathrm{gas}$ is the gas constant. Taking the time derivative of the above, noting that $\dot{P}_\mathrm{Cg}=0$, we have
\begin{equation}
    0 = \frac{\dot{\rho}_\mathrm{Cg, \,open}}{\mu} R_\mathrm{gas} T + \frac{\rho_\mathrm{Cg}}{\mu}\dot{T} R_\mathrm{gas}.
\end{equation}
where $\dot{T} \equiv \partial T/\partial t$ is the rate of temperature change. Rearranging the above for the rate of gas density change, $\dot{\rho}_\mathrm{Cg, \,open}$, we have
\begin{equation}\label{eqn:valve_rho_CG}
    \boxed{\dot{\rho}_\mathrm{Cg,\,open} = -\frac{\rho_\mathrm{Cg}}{T}\dot{T}.}
\end{equation}

To derive the mass removal rate due to outflow, $\dot{m}_\mathrm{Cg}|_\mathrm{valve}$, we first note that, this term is essentially the difference between the kinetically-controlled gas production rate,  $\dot{m}_\mathrm{Cg}|_\mathrm{burn}$, and the valve-controlled, total rate of change in pore gas mass, $\dot{m}_\mathrm{Cg, \,open}$. This is simply reflecting gas mass conservation: 
\begin{equation}\label{eq_2:m_CG_open}
    \dot{m}_\mathrm{Cg, \,open} = \dot{m}_\mathrm{Cg}|_\mathrm{burn} + \dot{m}_\mathrm{Cg}|_\mathrm{valve}.
\end{equation}
 In other words, the $\dot{m}_\mathrm{Cg}|_\mathrm{burn}$ in excess to $\dot{m}_\mathrm{Cg, \,open}$ is removed by $\dot{m}_\mathrm{Cg}|_\mathrm{valve}$ ($\dot{m}_\mathrm{Cg}|_\mathrm{valve}\leq0$ by convention). Again, $\dot{m}_\mathrm{Cg}|_\mathrm{burn}$ is directly calculated from kinetics (Section \ref{sec_2:reac_rate}), so to solve for $\dot{m}_\mathrm{Cg}|_\mathrm{valve}$, one only needs to find $\dot{m}_\mathrm{Cg, \,open}$. We do this by again utilizing the chain rule expansion of Equation \ref{eq_2:mass_vol_den_Cg}, and write 
\begin{equation}\label{eq_2:chain_m_cg_open}
    \dot{m}_\mathrm{Cg, open} = \dot{\phi}_\mathrm{por}\rho_\mathrm{Cg} + 
    \phi_\mathrm{por}\dot{\rho}_\mathrm{Cg,\,open}.
\end{equation}
Here, because $\dot{\phi}_\mathrm{por}$ and $\dot{\rho}_\mathrm{Cg,\,open}$ are given by Equations \ref{eqn:dphi_C} and \ref{eqn:valve_rho_CG} respectively, $\dot{m}_\mathrm{Cg, open}$ can be found. Thus, we combine Equations \ref{eqn:dphi_C}, \ref{eqn:valve_rho_CG}, \ref{eq_2:m_CG_open} and \ref{eq_2:chain_m_cg_open}, and obtain the rate of mass outflow 
\begin{equation}\label{eq_2:m_rate_valve}
     \boxed{\begin{split}
         \dot{m}_\mathrm{Cg}|_\mathrm{valve} = &\, \dot{m}_\mathrm{Cg}|_\mathrm{burn}\Big(\frac{\rho_\mathrm{Cg}}{\rho_\mathrm{Cs}} - 1\Big) \\ &\,- \phi_\mathrm{por}\rho_\mathrm{Cg}\frac{\dot{T}}{T}.
     \end{split}}
\end{equation}

One can gain some physical intuition for the above expression by comparing it with Equations \ref{eq_2:rho_CG_closed} and \ref{eqn:valve_rho_CG}, and arrive at 
\begin{equation}
    \dot{m}_\mathrm{Cg}|_\mathrm{valve} = \phi_\mathrm{por} \big( \dot{\rho}_\mathrm{Cg, \,open} - \dot{\rho}_\mathrm{Cg, \,closed}\big).
\end{equation}
In other words, the mass extracted by the pressure valve is the difference between what the valve dictates (first term on the right) and what the chemical equilibrium would have produced (second term). The absence of $\dot{\phi}$ terms demonstrates that the valve term does not modify the pore volume, and only decreases gas density. 

\subsection{Energy conservation}\label{sec_2:Eng_cons}
Besides mass and volume composition, the model also integrates for a planetesimal's thermal evolution $T(r, t)$. Here we derive the temperature change rate $\frac{\partial T}{\partial t}(r, t)$ using the energy conservation equation
\begin{equation}\label{eq_2:eng_conservation}
\begin{split}
    \frac{\partial \overline{\rho c_\mathrm{p}}T}{\partial t} = &\nabla\cdot (\overline{k}\nabla T ) + Q_\mathrm{rad}  
    \\ & + Q_\mathrm{burn} + Q_\mathrm{valve} + Q_\mathrm{dehy},
\end{split}
\end{equation}
where $\overline{\rho c_\mathrm{p}}$ is the bulk specific heat per unit volume (in $\mathrm{J/(K \cdot m^3)}$, and $\overline{k}$ is the bulk thermal conductivity (in W/(m $\cdot$ K)  , see Section \ref{Sec_2:material_p_k}).  
$Q_\mathrm{rad}$ is the heat released by $^{26}$Al and $^{60}$Fe decay  (Section \ref{sec_2:Rad_heat}), $Q_\mathrm{burn}$ is the heat released by carbon burning  (Sections \ref{Sec_2:carbon_burning} and \ref{sec_2:reac_rate}), $Q_\mathrm{valve}$ is the heat taken away with the excess gas when the pressure valve mechanism is active  (Equations \ref{eq_2:m_rate_valve} and \ref{eq_2:Q_valve}), and $Q_\mathrm{dehy}$ is the heat consumed by the endothermic dehydration reaction  (Equation \ref{eq_2:Q_dehy_simp}). Before describing each of these terms, we first isolate $\frac{\partial T}{\partial t}$. Here we note that $Q_\mathrm{dehy}$ contains an implicit $\frac{\partial T}{\partial t}$ dependence because we simulate dehydration as a temperature-controlled reaction (see Section \ref{Sec_2:Q_dehy}). That is, the dehydration term can be expressed as
\begin{equation}\label{eq_2:Q_dehy_simp}
    Q_{dehy} = \frac{\mathrm{d}Q}{\mathrm{d}T}\Big|_\mathrm{dehy}\frac{\partial T}{\partial t},
\end{equation}
where $\frac{\mathrm{d}Q}{\mathrm{d}T}\Big|_\mathrm{dehy}\leq0$ is the heat consumed by dehydration per degree of temperature rise.

To isolate $\frac{\partial T}{\partial t}$ from Equation \ref{eq_2:eng_conservation}, we recognize that $\overline{\rho c_\mathrm{p}}$ is a function of the local chondrite-graphite-gas mixture, which evolves over time 
\begin{equation}
    \overline{\rho c_\mathrm{p}} = \sum_i \phi_i(r, t) \rho_i(r, t) c_{\mathrm{p},i}(T(r, t)),
\end{equation}
where $c_{\mathrm{p},i}$ is the specific heat capacity of phase $i$, and in general is a function of temperature, while $T(r, t)$ is a function of radial location and time. We thus expand the left-hand side of Equation \ref{eq_2:eng_conservation}
\begin{equation}\label{eq_2:Eng_LHS}
\begin{split}
    \frac{\partial }{\partial t}\big[\sum_i(\phi_i \rho_i c_{\mathrm{p},i}) T\big] = & \sum_i(\dot{\phi_i} \rho_i c_{\mathrm{p},i}) T
    \\
    & + \sum_i(\phi_i \dot{\rho_i} c_{\mathrm{p},i}) T 
    \\
    & + \sum_i(\phi_i \rho_i \frac{\partial c_{\mathrm{p},i}}{\partial T}\frac{\partial T}{\partial t}) T
    \\& + \sum_i(\phi_i \rho_i c_{\mathrm{p},i}) \frac{\partial T}{\partial t}.
\end{split}
\end{equation}

On the right-hand side of this equation, the first two terms can be rewritten using $m_i = \phi_i\rho_i$, as
\begin{equation}\label{eq_2:Eng_LHS_simple}
    \sum_i(\dot{\phi_i} \rho_i c_{\mathrm{p},i}) T + \sum_i(\phi_i \dot{\rho_i} c_{\mathrm{p},i}) T = 
    \sum_i(\dot{m_i} c_{\mathrm{p},i}) T.
\end{equation}
While the time derivative of temperature can be separated from the last two terms.

We define an effective heat capacity by collecting the $\frac{\partial T}{\partial t}$ terms' prefactors from Equations \ref{eq_2:Q_dehy_simp} and \ref{eq_2:Eng_LHS}
\begin{equation}\label{eq_2:C_eff}
    C_{\mathrm{eff}} \equiv \sum_i\phi_i \rho_i(c_{\mathrm{p},i} + T\frac{\partial c_{\mathrm{p},i}}{\partial T}) + \frac{\mathrm{d}Q}{\mathrm{d}T}\Big|_\mathrm{dehy},
\end{equation}
where $C_{\mathrm{eff}}$ is in $\mathrm{J/(K \cdot m^3)}$. This enables us to write the time evolution of temperature as 
\begin{equation}\label{eq_2:dTdt}
    \boxed{\begin{split}
    \frac{\partial T}{\partial t} = \frac{1}{C_{\mathrm{eff}}} \Big[ & \nabla\cdot (\overline{k}\nabla T ) + Q_\mathrm{rad} 
    + Q_\mathrm{burn} \\ & + Q_\mathrm{valve}  -  \sum_i(\dot{m_i} c_{\mathrm{p},i}) T\Big]
    \end{split}.}
\end{equation}

\subsubsection{Radiogenic heating}\label{sec_2:Rad_heat}
The decay of $^{26}$Al and $^{60}$Fe drove the metamorphism and differentiation of the chondritic parent bodies (e.g. \citealt{Hevey_2006M&PS_differentiation...41...95H, Moskovitz_2011M&PS...46..903M, Neumann_2012A&A...543A.141N}. The radiogenic heating rate we use  (Equation \ref{eq_2:dTdt}) is 

\begin{equation}\label{eq_2:Q_rad_sum}
    Q_{rad} = Q_{^{26}Al} + Q_{^{60}Fe}.
\end{equation}

For each isotope $^j\mathrm{A}$, its heating rate $Q_\mathrm{A}$ is
\begin{equation}\label{eq_2:Q_A(t)}
    Q_\mathrm{A}(t) = \rho_\mathrm{ch}\phi_\mathrm{ch}w_\mathrm{A}\mathrm{\Big[\frac{^{j}\mathrm{A}}{^s\mathrm{A}}}\Big]\frac{N_\mathrm{A}}{\mu_{^{s}\mathrm{A}}}\frac{E_\mathrm{A}}{\tau_j}\,\exp\big({-\frac{t}{\tau_\mathrm{A}}}\big),
\end{equation}
where $w_\mathrm{A}$ is the concentration of element A by mass, $\mathrm{\big[\frac{^j\mathrm{A}}{^s\mathrm{A}}}\big]$, is the abundance of the radioactive isotope, $N_\mathrm{A}$ is the Avogadro constant, $E_\mathrm{A}$ is the decay energy per atom, $\tau_\mathrm{A}$ is the decay's e-folding time, $\mu_{^s\mathrm{A}}$ is the atomic mass of the stable isotope of element A, $^s\mathrm{A}$. 
Following the convention of Section \ref{sec_2:Eng_cons}, $Q_\mathrm{A}$ has units of $\mathrm{J/(s\cdot m^3)}$.
We assume constant abundances of $^{26}$Al and $^{60}$Fe, yet each chondrite type has its own $w_{\mathrm{Al}}$ and $w_{\mathrm{Fe}}$, which results in different total radiogenic heat budgets (see Tables \ref{tab_2:rad_heating}  and  \ref{tab_2:material_props} for values). At the time of calcium aluminum-rich inclusion condensation, or the ``time zero" of chondritic materials, CV chondrites have the most radiogenic heat per mass, followed by ordinary (OC), enstatite (EH), and CI chondrites, respectively. While at 10 Myr, due to its high iron abundance, EH overtakes the other chondritic compositions so that $\mathrm{EH>CV>OC>CI}$. 
  
\begin{table}
\begin{center}
    \centering
    \begin{tabular}{cccc}
         & $E_\mathrm{A}$ [J] & $[\frac{^{j}\mathrm{A}}{^{s}\mathrm{A}}]$  & $\tau$ [Myr] \\
        \hline
        $^{26}$Al & $6.415\times10^{-13}$ & $5\times 10^{-5}$ & 1.034\\
        $^{60}$Fe & $4.87\times 10^{-13}$ & $1.6\times10^{-6}$ & 3.75\\
        \hline
    \end{tabular}
    \caption{Radiogenic heating parameters from \cite{Neumann_2012A&A...543A.141N} and sources cited within. }\label{tab_2:rad_heating}
\end{center}
\end{table}

\subsubsection{Carbon burning}\label{Sec_2:carbon_burning}
In tracking a planetesimal's thermal evolution (Equation \ref{eq_2:dTdt}), we calculate the heat of carbon burning $Q_\mathrm{burn}$, by
\begin{equation}
    Q_\mathrm{burn} = \frac{\dot{m}_\mathrm{Cg}|_\mathrm{burn}}{\mu_\mathrm{Cg}} (x_{\mathrm{CO}} \Delta_\mathrm{f} H_{\mathrm{CO}} + x_{\mathrm{CO_2}}\Delta_\mathrm{f} H_{\mathrm{CO_2}}),
\end{equation}
where $x_{\mathrm{CO}}$ and $x_{\mathrm{CO_2}}$ are molar fractions of CO and CO$_2$, calculated from Equation \ref{eq_2:r_Cg}, and $\Delta_\mathrm{f} H_{\mathrm{CO_2}}$ and  $\Delta_\mathrm{f} H_{\mathrm{CO}}$ are the enthalpies of reactions 1 and 2, respectively, in J/mol. These are found using the thermodynamic data from \cite{NIST}. Specifically, we use the enthalpy change fitted to the Shomate equation 
\begin{equation}
\begin{split}
    \Delta H_i^{\circ}(T) \equiv& H_i^{\circ} - {H_i^{\circ}}(298.15) 
    \\=& A_i T' + \frac{1}{2}B_iT'^2 + \frac{1}{3}C_iT'^3 
    \\& + \frac{1}{4}D_i T'^4 - \frac{E_i}{t} + F_i - H_i,
\end{split}
\end{equation}
where $H_i^{\circ}$ is the standard enthalpy for phase $i$ as a function of temperature, and $A_i$ to $H_i$ are fitting parameters. The standard enthalpy of formation for a species is then
\begin{equation}
    \Delta_\mathrm{f} H_i = \Delta H_i^{\circ}(T) + \Delta {H_{\mathrm{f},i}^{\circ}}(298.15),
\end{equation}
where $\Delta {H_{\mathrm{f},i}^{\circ}}(298.15)$ is the standard enthalpy of formation for phase $i$. We then find the enthalpy of reaction, $\Delta_\mathrm{f} H^{\circ}$, via

\begin{equation}\label{eq_2:reaction heat}
    \Delta_\mathrm{f} H^{\circ} = \Sigma_\mathrm{p} v_\mathrm{p} \Delta_\mathrm{f} H_\mathrm{prod} - \Sigma_\mathrm{r} v_\mathrm{r} \Delta_\mathrm{f} H_\mathrm{reac},
\end{equation}
where the subscripts $\mathrm{p}$ and $\mathrm{r}$ denote the reaction's products and reactants, and $v_i$ are their stoichiometric coefficients.

The standard enthalpies of formation for CO and CO$_2$ at 298.15 K are $\Delta_\mathrm{f} H^{\circ}_{\mathrm{CO_2}}(298.15) = -393.52$ kJ/mol and $\Delta_\mathrm{f} H^{\circ}_{\mathrm{CO}}(298.15) = -110.53$ kJ/mol. For CI and CV chondrites that mainly outgas CO$_2$, 1 wt\% of carbon approximately corresponds to 320 kJ/kg of chondrites. For EH and OC chondrites, which mostly produce CO, this comes to 92 kJ/kg. 

\subsubsection{Heat of gas advection}\label{Sec_2:Q_valve}

When the pressure valve is turned on, and open-system flow is permitted, the gas vented to space removes some heat with it. This energy sink is
\begin{equation}\label{eq_2:Q_valve}
\begin{split}
Q_\mathrm{valve} = \dot{m}_\mathrm{Cg}|_\mathrm{valve} \Big(&\frac{44}{12}x_{\mathrm{CO_2}} c_{\mathrm{p}, \mathrm{CO_2}} \\
&+ \frac{28}{12}x_{\mathrm{CO}}c_{\mathrm{p}, \mathrm{CO}}\Big)T.
\end{split}
\end{equation}
where $c_{\mathrm{p,CO_2}}$ and $c_{\mathrm{p,CO}}$ are the specific heat capacities in J/(kg$\cdot$K).  This is used in Equation \ref{eq_2:dTdt}.

\subsubsection{Heat of Dehydration}\label{Sec_2:Q_dehy}
Due to low-temperature aqueous alteration, some carbonaceous chondrite (CC) matrices are rich in hydrated phyllosilicates \citep{Weisberg_2006mess.book...19W}. 
In the context of high-temperature carbon outgassing, the dehydration of phyllosilicates is expected to alter the gas composition, matrix texture, and thermal histories of the parent bodies in complex ways. Here, we focus on accounting for the standard enthalpy of reaction to capture the effect of dehydration on a planetesimal's thermal evolution (Equations \ref{eq_2:eng_conservation}, \ref{eq_2:C_eff}, and \ref{eq_2:dTdt}) as a first step. We discuss other aspects in Section \ref{sec_2:discussion}. We use the dehydration of serpentine as a representative reaction, as serpentine has been reported as the dominant hydrated mineral in some CC samples. This reaction is

\begin{equation}\label{eq_2:dehy_reac}
    \begin{split}
        \mathrm{Mg_3Si_2O_5(OH)_4 \,(Srp) =}&  \mathrm{MgSiO_3 \,(En)} \\&+  
        \mathrm{Mg_2SiO_4 \,(Fo)} \\&+ \mathrm{2H_2O \,(g)},
    \end{split}
\end{equation}
where serpentine (Srp) breaks down to enstatite (En) and forsterite (Fo) while releasing water vapor. We use the serpentine thermochemistry data from \cite{Robie_1995_book_therm} and use \cite{NIST} for the rest. We calculate the heat per mole of reaction $\Delta H_{\mathrm{dehy}}(T)$ using \ref{eq_2:reaction heat}. 

The kinetics of serpentine dehydration has been the topic of long-established literature \citep{King_2021GeCoA.298..167K}. Significant nuance exists, such as the different kinetic behaviors of serpentine's three different mineral structures or the role of composition (e.g., bulk Fe/Mg ratio, \citealt{Lindgren_2020GeCoA.289...69L}). Experimental works characterize serpentine dehydration as a temperature-activated reaction, with characteristic temperatures of 600 $\sim$ 800 $^{\circ}$C \citep{Viti_2010AmMin..95..631V, Gualtieri_2012AmMin..97..666G}. Meanwhile, works on naturally or artificially heated carbonaceous chondrites typically find a lower dehydration temperature at $<$400 $^{\circ}$C to 600 $^{\circ}$C, followed by the crystallization of the anhydrous olivine at 600 $^{\circ}$C to $\sim$900 $^{\circ}$C \citep{Nakamura_2005JMPeS.100..260N, Lindgren_2020GeCoA.289...69L, Matsuoka_2022GeCoA.316..150M}. Yet a more detailed, quantitative picture is hindered by the diversity in the chondritic samples and experimental techniques  (see a review in \citealt{King_2021GeCoA.298..167K}). 

The varied experimental results prompt us to adopt a simpler parameterized treatment: the degree of dehydration $X_\mathrm{dehy}(T)$ increases smoothly with temperature. We use a smoothed step function for $X_\mathrm{dehy}(T)$ that rises from 0\% at $T_\mathrm{0,dehy}$ to 100\% at $T_\mathrm{1,dehy}$
\begin{equation}\label{eq_2:X}
    X= 6\theta^5 - 15\theta^4 + 10\theta^3 ,
\end{equation}
where $\theta$ is the normalized temperature 
\begin{equation}
    \theta\equiv \frac{T - T_\mathrm{0, dehy}}{T_\mathrm{1, dehy} - T_\mathrm{0, dehy}}.
\end{equation}
We take $T_\mathrm{0,dehy} = 400\, ^{\circ}$C and $T_\mathrm{1,dehy} = 600\, ^{\circ}$C. The volumetric heat of dehydration is
\begin{equation}\label{eq_2:Q_dehy}
\begin{aligned}
    Q_\mathrm{dehy} &= \frac{\mathrm{d}Q}{\mathrm{d}T}\Big|_\mathrm{dehy} \frac{\partial T}{\partial t}\\
            & = -\Delta H_{\mathrm{dehy}}(T)\frac{w_{\mathrm{H_2O},0}m_{\mathrm{tot},0}}{\mu_{\mathrm{H_2O}}}\frac{\mathrm{d}X}{\mathrm{d}T}\frac{\partial T}{\partial t}.
\end{aligned}
\end{equation}
where $w_{\mathrm{H_2O},0}$ is the initial water abundance by mass, $m_{\mathrm{tot},0}$ is the initial volumetric mass in kg/m$^3$, $\mu_{\mathrm{H_2O}} = 0.018\mathrm{kg/mol}$ is water's molar mass, and $\frac{\mathrm{d}X}{\mathrm{d}T}$ is the differential of Equation \ref{eq_2:X}.   We assume a CI-like wet composition with $w_{\mathrm{H_2O},0} = 15 \mathrm{wt}\%$. Further, we assume that the water released by the reaction leaves the system and dehydration is irreversible. This means $X$ does not decrease when the local temperature drops. In other words, $\frac{\mathrm{d}Q}{\mathrm{d}T}\Big|_\mathrm{dehy} = 0$ when $\frac{\partial\mathrm T}{\partial t}\leq 0$.

\subsection{Kinetic Reaction Rate}\label{sec_2:reac_rate}

Our model calculates the kinetic reaction rate, the rate at which graphite is converted to CO and CO$_2$. This is then used to calculate $\dot{\phi}_\mathrm{por}$ and $\dot{\rho}_\mathrm{Cg}$ by mass and volume conservation  (Section \ref{sec_2:mass_vol_cons}), as well as the thermal evolution (Section \ref{sec_2:Eng_cons}). Realistically, one can envision C volatilization in chondritic planetesimals in two steps: C gasification followed by rock-gas equilibration. First, C-rich gases are released from the condensed C carriers, primarily insoluble organic materials (IOMs), but also include amorphous carbon, carbonates and diamonds. IOMs' gasification is likely a complicated, multi-step process  but accelerates at high temperatures \citep{Alexander_2017ChEG...77..227A}. Existing measurements suggest these reactions are geologically fast (a few centuries) at elevated temperatures $\geq$500 K \citep{Kebukawa_2010M&PS...45...99K}. Next, the gases liberated from IOMs proceed to equilibrate with the chondritic matrix. This step is also likely fast at high temperatures \citep{Thompson_2021NatAs...5..575T}, although experimental data remain sparse. We thus assume both processes are fast, and employ a parameterized kinetic rate normalized to the equilibrium gas density, $\rho_\mathrm{Cg,\,eq} (T, f\mathrm{^c}_{\mathrm{O_2}})$ (see Section \ref{sec_2:chem_eq}):
\begin{equation}\label{eq_2:kinetic_burning_rate}
    \dot{m_\mathrm{Cg}}|_\mathrm{burn} = R\cdot \rho_\mathrm{Cg,\,eq}\cdot\Big(1 - \frac{P_\mathrm{Cg}}{P_\mathrm{Cg,\,eq}}\Big),
\end{equation}
where $R = 1$/year is our choice of rate constant. This parameterization is analogous to the treatment of \cite{Sugiura_1986E&PSL..78..148S}, where the kinetics responds to the deviation of pore gas pressure from chemical equilibrium ($1 - \frac{P_\mathrm{Cg}}{P_\mathrm{Cg,\,eq}}$), such that equilibrium pressure is quickly reached in closed systems. At locations with open-system flow, the high $R$ ensures efficient local C depletion. Meanwhile the $\rho_\mathrm{Cg,\,eq} (T, f\mathrm{^c}_{\mathrm{O_2}})$ dependence enables the reaction to speed up as the planetesimal heats up, because the equilibrium $\rho_\mathrm{Cg,\,eq}$ generally rises with temperature, see Figure \ref{fig:Chem_Eq}. Our treatment also ensures faster rates in more oxidizing chondrites. We discuss the caveats of our approach in Section \ref{Sec_2:Disc_ChemEq}.  

\subsection{Sintering}\label{Sec_2:sintering}

Sintering describes the process of compacting the porous matrix by the viscous deformation of individual grains under lithostatic pressure. This process transforms a planetesimal from loosely packed pebbles and dust particles into one integral solid. We invoke sintering primarily as a mechanism for stopping carbon outgassing. We envision sintering closing the networks of cracks that enable outgassing, therefore preserving carbon. We mimic this effect by stopping the pressure valve mechanism at a layer where the local porosity falls below the numerical floor of $\phi_\mathrm{crit} = 10^{-5}$.

Previous works (e.g. \citealt{henke12, neumann14}) on planetesimal compaction used experimentally measured olivine rheology from \cite{Schwenn_1978Tectp..48...41S}. Although the experimental setup of \cite{Schwenn_1978Tectp..48...41S} is directly relevant to sintering, the rheology they determine contains large experimental uncertainties (e.g. $>30\%$ for the activation energy). Therefore, we opt for an updated upper mantle rheology from \cite{Jain_2020JGRB..12519896J}. They combined experimental data and geophysical observations within a rigorous statistical framework, thereby tightening the constraints on upper mantle rheology. Assuming dry conditions, the constitutive equations for diffusion and dislocation creep of individual olivine grains are 
\begin{equation}\label{eq_2:sint_diff}
    \dot{\epsilon}_\mathrm{diff} = A_1 d^{-p_1}\sigma\exp\Big(-\frac{E_1 + PV_1}{R_\mathrm{gas}T}\Big),
\end{equation}
\begin{equation}\label{Eq_2:Sint_disl}
    \dot{\epsilon}_\mathrm{disl} = A_3 \sigma^{n_3}\exp\Big(-\frac{E_3 + PV_3}{R_\mathrm{gas}T}\Big).
\end{equation}
Here, $\dot{\epsilon}_i$(in s$^{-1}$) is the strain rate for each creep mechanism. It is a function of grain size, $d$, the deviatoric stress, $\sigma$, the pressure, $P$, and the material properties that \citep{Jain_2020JGRB..12519896J} constrained. The latter includes the activation energies $E_i$, the activation volumes $V_i$, the scaling coefficients $A_i$, and the exponents $p_1$ and $n_3$. The total strain rate is 

\begin{equation}
    \dot{\epsilon} = \dot{\epsilon}_\mathrm{diff} + \dot{\epsilon}_\mathrm{disl}.
\end{equation}

We recognize that using experimental olivine rheologies calibrated for upper mantle conditions inevitably involves extrapolation, which our model shares in its approach with earlier works (e.g., \citealt{henke12, neumann14}). The caveats in this methodology are discussed in Section \ref{Sec_2:Disc_sint}.

The strain rate is then related to the evolution of porosity by \cite{neumann14}

\begin{equation}\label{eq_2:sint_rate}
    \frac{\partial}{\partial t}\log (1- \phi_\mathrm{por}) = \dot{\epsilon}(T, P, \sigma, d).
\end{equation}

We interpret $P$ as the lithostatic pressure $P_\mathrm{lith}$ for calculating the strain rate. Grain size is an important parameter. \cite{neumann14} showed that assuming different grain sizes significantly changes a planetesimal's final porosity. Larger grains are more resistant to sintering, which leads to higher sintering temperatures and less compacted planetesimals.   Chondrites, however, are aggregates of particles with a bimodal size distribution: mm-sized chondrules and micron-sized matrix grains \citep{Weisberg_2006mess.book...19W, Brearley_1989GeCoA..53.2081B, Vaccaro_2023M&PS...58..688V}.
We note that mm-sized chondrules constitute up to 80\% by volume in inner solar system chondrites (ordinary and enstatite), while carbonaceous chondrites contain $\gtrsim$50 vol.\%  micron-sized matrix grains \citep{Weisberg_2006mess.book...19W}. Thus, the sintering behavior for both groups can be better captured by utilizing the grain sizes of their major components. We, therefore, adopt a 1 mm grain size for OC and EH planetesimals and 1 micron for the CI and CV bodies, but discuss the impact of alternative choices in Section \ref{sec_2:discussion}.

The deviatoric stress $\sigma$ felt by individual grains is a function of both the overbearing pressure and the geometry of grain packing. \cite{henke12, neumann14} envisioned spherical grains packed in polyhedral unit cells. The unit cell contracts symmetrically (i.e., maintaining its shape), while the grain particle gradually deforms into the intersection of a larger sphere and the shrunken unit cell. They form growing circular interfaces with neighboring grains, eventually occupying the entire unit cell. While \cite{neumann14} considered multiple packing geometries, for simplicity, we assume simple cubic packing (SCP), where the unit cell is a cube. The lithostatic, compressive force $A \times P_\mathrm{lith}$ acts upon the upper interface of the unit cell, where $A$ is the surface area of the interface. This force is separated into a component acting on the pore space $A_\mathrm{por} \times P_\mathrm{Cg}$, and another deforming the grain $(A - A_\mathrm{por}) \sigma$, where $A_\mathrm{por}$ is the interface surface area occupied by the pore space. Force balance, therefore, leads to
\begin{equation}\label{eq_2:P_lith_P_Cg}
    P_\mathrm{lith} = x_\mathrm{por}(\phi_{por}) P_\mathrm{Cg} + (1- x_\mathrm{por}(\phi_\mathrm{por})) \sigma.
\end{equation}
where $x_\mathrm{por}(\phi_\mathrm{por}) \equiv A_\mathrm{por}/A$ is the fraction of the unit cell's surface area that the pore space occupies, and $(1 - x_\mathrm{por})$ is the fraction of the surface area occupied by the circular interface between grains. We integrate the lithostatic pressure profile using 
\begin{equation}\label{eq_2:P_lith}
    P(r) = \int_{R}^{r}\frac{G M(r)}{r^2}\sum_{i} \phi_i(r)\rho_i(r).
\end{equation}
where $r$ is the radial location, $R$ is the radius of the planetesimal, and the summation term is the mean density. $M(r)$ is the mass enclosed at $r$ - 
\begin{equation}
    M(r) = \int_{0}^{r} 4 \pi r^2 \sum_{i} \phi_i(r)\rho_i(r)\mathrm{dr}.
\end{equation}
We find $x_\mathrm{por}$ for the SCP packing numerically through the cell geometry following \cite{neumann14}, and use a polynomial fit for it
\begin{equation}\label{eq_2:x_por_phi_por}
\begin{split}
    x_\mathrm{por} = &3.132\phi_\mathrm{por}^3 - 2.648\phi_\mathrm{por}^{2} 
    \\ &- 2.786\phi_\mathrm{por} + 4.689\phi_\mathrm{por}^{0.8}.
\end{split}
\end{equation}
Equations \ref{eq_2:P_lith_P_Cg} and \ref{eq_2:x_por_phi_por} connect several different model components. Sintering rate is controlled by the deviatoric stress, $\sigma$, which we find by first calculating $P_\mathrm{lith}$ via Equation \ref{eq_2:P_lith} and $P_\mathrm{Cg}$ via mass and energy conservation, then using Equation \ref{eq_2:P_lith_P_Cg}, we have

\begin{equation}\label{eq_2:sigma_P_lith}
    \sigma = \frac{P_\mathrm{lith} - x_\mathrm{por} P_\mathrm{Cg}}{1 - x_\mathrm{por}}.
\end{equation}
Equation \ref{eq_2:P_lith_P_Cg} also defines the minimum pressure for open-pore outflow. That is, when the pore gas pressure becomes high enough to support the matrix on its own and halt sintering, then we posit that $P_\mathrm{Cg}$ has reached the confinement pressure and has generated the brecciation to enable open-pore flow. Thus,
\begin{equation}\label{eq_2:P_conf}
    P_\mathrm{conf} = \frac{P_\mathrm{lith}}{x_\mathrm{por}}.
\end{equation}.
We adopt an initial porosity of $\phi_\mathrm{por,0}=0.4$ following \citep{henke12, neumann14}, and set a numerical porosity floor of $10^{-5}$. Once local porosity falls below it, the layer is considered fully sintered, and gas venting is disabled. 

\begin{table}
\begin{center}
    \centering
    \begin{tabular}{ccccc}
       \hline
         & EH & OC  & CV  & CI\\
        \hline
        $\rho_\mathrm{ch}$ [kg/m$^3$] & 3670$^{1}$ & 3640$^{1}$ & 3540$\,^2$ & 2420$\,^2$\\
        $w_\mathrm{Al}$ [wt\%]&  0.77$\,^3$&  1.24$\,^3$&  1.73$\,^3$& 0.858$\,^3$\\
        $w_\mathrm{Fe}$ [wt\%]&  33.15$\,^3$&  22.41$\,^3$&  23.54$\,^3$& 18.35$\,^3$\\
        $w_\mathrm{Cs}$ [ppmw]&  4100$\,^3$& 1200$\,^3$ & 2700$\,^3$ & 2700$\,^3$$^*$ \\
        \hline
    \end{tabular}
    \caption{Material properties of chondrites. Sources: $^1$\cite{Britt_2003M&PS...38.1161B}, $^2$\cite{Macke_2011M&PS...46.1842M}, $^3$\cite{Schaefer17}.\\
    $^*$ We used CV bulk carbon content for CI bodies; see text for details.}\label{tab_2:material_props}
\end{center}
\end{table}

\subsection{Material Properties}\label{Sec_2:material_properties} 

In tracking the thermal evolution (Section \ref{sec_2:Eng_cons}), our model utilizes literature results for the properties of each material: chondrites, graphite, CO, and CO$_2$ gases. We then use the corresponding mixing laws for these properties, reported below. 

\subsubsection{Specific heat capacities and densities}\label{Sec_2:material_p_Cp&rho}
We use temperature-dependent specific heat capacities in J/(kg$\cdot$K). Chondritic heat capacity follows \citep{Neumann_2012A&A...543A.141N}
\begin{equation}
    c_\mathrm{p, ch} = 800. + 0.25T - 1.5\times10^7 T^{-2}.
\end{equation} 
For CO and CO$_2$, we use the Shomate equation fit from NIST
\begin{equation}
    c_\mathrm{p, Sh} = \frac{1}{\mu_i}(A_i + B_i T' + C_i T'^2 + D_i T'^3 + E_i T'^{-2}),
\end{equation}
where $T'\equiv T/1000\,K$ , $\mu_i$ is the molecular weight for gas $i$, and $A_i$ through $E_i$ are fitting parameters. We use a constant specific heat for solid carbon $c_\mathrm{p, Cs} = 8.58\,\, \mathrm{J/(mol\cdot K)}$.

The specific heat capacity of the mixture is found by mass-weighted averaging $ c_\mathrm{p} = \sum_i \phi_i \rho_i c_{\mathrm{p},i}$.

We assume constant densities for chondrites, summarized in Table \ref{tab_2:material_props}. We find the densities of CO and CO$_2$ using their molar ratio $r_\mathrm{Cg}$ and $\rho_\mathrm{Cg}$, the ``carbon gas" density, which we integrate over time.

\subsubsection{Thermal conductivity}\label{Sec_2:material_p_k}
The thermal conductivity of porous materials is a function of their porosity. While previous works have used experimentally-calibrated thermal conductivities of porous chondrites in vacuum (e.g., \citealt{henke12}), we account for gas conductivity by using a geometric mixing law \cite{Beck_1976Geop...41..133B}

\begin{equation}
    \bar{k} = k_\mathrm{ch}^{\phi_\mathrm{ch}}k_\mathrm{C}^{\phi_\mathrm{Cs}}k_\mathrm{Cg}^{\phi_\mathrm{por}},
\end{equation}
where $k_\mathrm{ch} = 4.56$ W/(m$\cdot$K) is the thermal conductivity of fully-compacted chondrites \citep{Neumann_2012A&A...543A.141N},  $k_\mathrm{C} = 2$ W/(m$\cdot$K) is that of amorphous carbon \citep{Giri_2022npjCM...8...55G}, and $k_\mathrm{Cg}$ is the mean gas conductivity, which we find via \cite{Mathur_1967MolPh..12..569M}

\begin{equation}
    k_\mathrm{Cg} = \frac{1}{2}\Big[\sum_i x_i k_i  + \big(\sum_{i}\frac{x_i}{k_i}\big)^{-1} \Big],
\end{equation}
where $x_i$ and $k_i$ are the molar mixing ratio and thermal conductivity of gas $i$. We use the low-density limit thermal conductivities for CO$_2$ from \citep{Huber_2016JPCRD..45a3102H} 
\begin{equation}
    k = 0.001\cdot \sqrt{T_\mathrm{r}} \Big( k_0 + \frac{k_1}{T_\mathrm{r}} + \frac{k_2}{T_\mathrm{r}^2}+ \frac{k_3}{T_\mathrm{r}^3}\big)^{-1},
\end{equation}
where $T_\mathrm{r}\equiv T/T_\mathrm{c}$ is a reduced temperature, and $T_\mathrm{c}$ is the critical temperature, and $k_0$ through $k_{3}$ are fitting constants \citep{Huber_2016JPCRD..45a3102H}. For the thermal conductivity of CO, we used tabulated data from \cite{Millat_1989JPCRD..18..565M}. 

\subsubsection{Bulk C abundance}

We used bulk C abundances of each chondritic group from those tabulated by \cite{Schaefer17} (see Table \ref{tab_2:material_props}). Initially, graphite is uniformly distributed. Because the confinement pressure is 0 at the surface, in our framework, the devolatilization near the surface should be controlled solely by reaction kinetics. In reality, asteroid sample return missions have recovered volatile-rich material from small bodies, indicating volatiles can survive on the surface of chondritic planetesimals against space weathering on at least million-year timescales (e.g. \citealt{Okazaki_2023Sci...379.0431O_Ryugu}). Thus, we disable graphite burning at the surface layer.

CI chondrites have $\sim$3 wt\% C \citep{Lodders_03_SolarAbun}, yet adopting such a high C content results in $\Delta T\sim$500 K of extra heating due to graphite burning, leading to numerically unreliable behaviors. Such a large $\Delta T$ would also result in early melting, which we do not simulate. As a result, we elected to use CV-level C concentration for CI bodies. We note that using 3 wt\% graphite does not qualitatively change the outcome of near-complete outgassing from early-formed CI bodies. We discuss the more complex reaction kinetics and energetics in real chondritic planetesimals in Section \ref{sec_2:discussion}.

\subsection{Numerical procedure}

\begin{figure*}
    \centering
    \includegraphics[width=\linewidth]{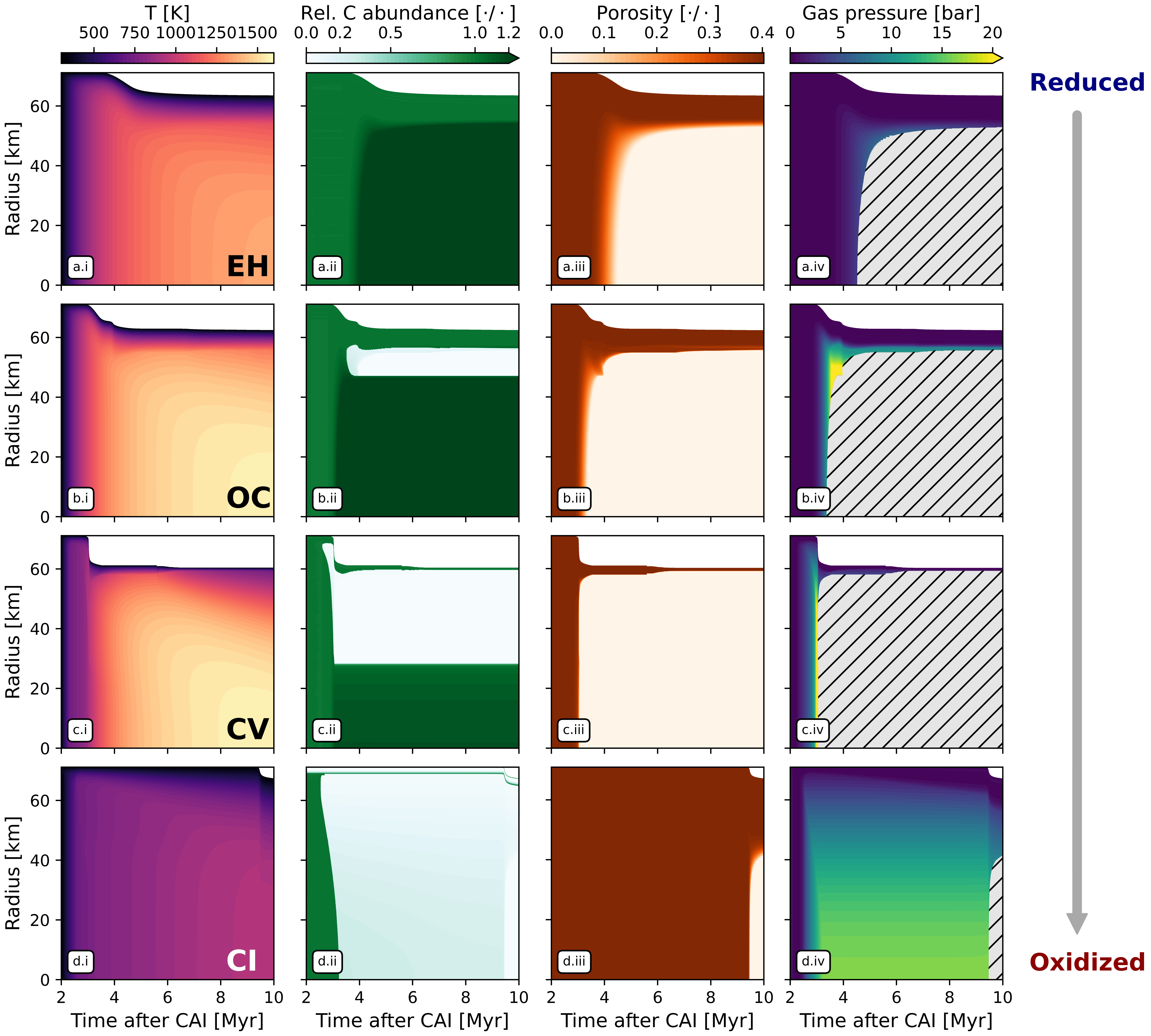}
    \caption{Evolutions of fiducial planetesimals with different chondritic compositions. Each plot is the contour of one quantity across time and radial location. Note that the total radius decreases due to sintering. From top row to bottom: EH, OC, CV, and CI bodies. Physical quantities from left to right: (i) temperature, (ii) condensed + gaseous C abundance normalized to the initial reservoir; $>1$ relative abundance occurs due to sintering concentrating C, (iii) porosity and (iv) gas pressure; hatched areas are completely sintered with no meaningful gas pressure.}
    \label{fig:sample_profiles}
\end{figure*}

We solve the mass, volume and energy conservation equations (Equations \ref{eqn:dphi_C}, \ref{eq_2:rho_CG_closed}, \ref{eqn:valve_rho_CG}, and \ref{eq_2:dTdt})  to track the evolution of $\phi_\mathrm{Cs}$, $\phi_\mathrm{por}$, $\rho_\mathrm{Cg}$ and $T$. Our model utilizes the Radau integrator from the SciPy software package \citep{SciPy_2020-NMeth}, which is an implicit 5th-order Runge-Kutta algorithm with adaptive time steps. 

The main variables we test are a planetesimal's chondritic composition, radius, and time of accretion, as detailed in Section \ref{Sec_2:Results}. The fiducial case for each composition has a compacted radius $R_\mathrm{c} = 60$ km and accreted at 2 Myrs after CAI (Section \ref{Sec_2:results_composition}). $R_\mathrm{c}$ is the radius of a body if it had no porosity. Both choices are typical values for planetesimals within and beyond the snow line \citep{Lichtenberg_2023ASPC_review..534..907L, Neumann_2024NatSR..1414017N}.

We assume an initially isothermal interior and fix the surface temperature \citep{Neumann_2012A&A...543A.141N}
\begin{equation}\label{eq_2:surface_init_conds}
    T(t = 0) = T_\mathrm{surf} = 300\mathrm{\,K}.
\end{equation}
Continuity is enforced at the center \citep{Neumann_2012A&A...543A.141N}, such that
\begin{equation}
    \frac{\partial T}{\partial r}(r = 0) = 0.
\end{equation}
We assume an initially homogeneous composition, which includes 40 vol\% pore space (Section \ref{Sec_2:sintering}), a fraction of graphite depending on the chondritic type (Table \ref{tab_2:material_props}), and chondritic matrix for the rest.

Our governing equations are solved on a radial grid, which has a resolution of 500 m (after complete compaction). The outer 20 km of larger bodies ($R_\mathrm{c} \geq 30$ km), and the entirety of smaller bodies, have a higher resolution of $\leq$250 m. We tested doubling the spatial resolution for our CV fiducial runs, and the difference in the pattern and extent of C depletion is small.

We implement sintering using an explicit scheme that periodically restarts the numerical simulation with updated spacings of each grid point, as well as $\phi_\mathrm{Cs}$ and $\phi_\mathrm{por}$. Specifically, we track the cumulative change of porosity  $\Delta \phi_\mathrm{por}|_\mathrm{sint} = \Sigma\dot{\phi}_\mathrm{por}|_\mathrm{sint}\delta t$, where $\dot{\phi}_\mathrm{por}|_\mathrm{sint}\leq0$ is the instantaneous sintering rate calculated from Equation \ref{eq_2:sint_rate}, and $\delta$ t is the duration of each time step. Once $\Delta \phi_\mathrm{por}|_\mathrm{sint}$ of the body at any location exceeds 0.002, we stop the integration and update the grid, $\mathrm{\phi_\mathrm{Cs}}$ and $\phi_\mathrm{por}$. Since the entire layer shrinks by a fraction of $\Delta \phi_\mathrm{por}|_\mathrm{sint}$, the updated porosity is
\begin{equation}
    \phi_\mathrm{por} = \frac{\phi_\mathrm{por,0} + \Delta \phi_\mathrm{por}|_\mathrm{sint}}{1 + \Delta \phi_\mathrm{por}|_\mathrm{sint}},
\end{equation}
where $\phi_\mathrm{por,0}$ is the pre-update porosity. Since graphite is incompressible, as the local layer of material decreases in volume, the volume fraction of graphite increases (note $\Delta \phi_\mathrm{por}|_\mathrm{sint}\leq0$):
\begin{equation}
    \phi_\mathrm{Cs} = \frac{\phi_\mathrm{Cs}}{1 + \Delta \phi_\mathrm{por}|_\mathrm{sint}}.
\end{equation}

Our choice for sintering resolution ensures that a planetesimal with an initial porosity of 0.4 can be resolved by at least 200 steps. We tested 10$\times$ higher porosity resolution for the fiducial OC body, but saw no appreciable difference. Our treatment assumes that sintering produces no thermal feedback, which has been demonstrated to be minor relative to radiogenic heating for these low-mass bodies \citep{henke12}.

\section{Results}\label{Sec_2:Results}

Our model tracks the evolution of a planetesimal in terms of temperature, porosity, and carbon reservoir in both graphite and gas phases. 

In a typical simulation (see, for instance, Figures \ref{fig:sample_profiles}b.i to \ref{fig:sample_profiles}b.iv), a planetesimal heats up globally due to radiogenic heating, while developing a deepening surface temperature gradient due to conductive cooling, congruous to previous literature (e.g. \citealt{henke12, Moskovitz_2011M&PS...46..903M, neumann14}). Layers at depth (i.e., those close to the center and with high $P_\mathrm{lith}$) tend to reach high temperatures before conductive cooling takes over. Their relatively high $P_\mathrm{conf}$-T environment leads to earlier sintering, preserving their C reservoir against outgassing (e.g., Figures \ref{fig:sample_profiles}b.ii and \ref{fig:sample_profiles}b.iii). While those close to the surface (low $P_\mathrm{lith}$) but still experience significant heating (high $P_\mathrm{Cg,\, eq}$) tend to trigger the pressure valve mechanism and see severe C lost to outgassing (e.g. Figures \ref{fig:sample_profiles}b.ii and \ref{fig:sample_profiles}b.iv). The competition of sintering and outgassing leads to diverse outcomes in C abundance and porosity profiles. 

Here, we explore how these outcomes emerge from local thermochemical evolution, which, in turn, is shaped by the planetesimal's composition, radius, and time of formation.

\subsection{Effects of composition}\label{Sec_2:results_composition}

A planetesimal's composition dictates its redox state, bulk density, radiogenic heat budget, and carbon content. To probe the effects of different compositions on the thermochemical histories of planetesimals, we simulate four planetesimals composed of enstatite (EH), ordinary (OC), and carbonaceous chondrites (CV and CI), using our fiducial initial conditions: accreted at 2 Myr after CAI formation, with a compressed radius $R_\mathrm{c}$ of 60 km. $R_\mathrm{c}$ is the size of the planetesimal if it were sintered entirely or had zero porosity. Each run terminates at 10 Myrs. The evolution of these bodies' temperature profiles, carbon inventories, porosities, and pore gas pressures is shown in Figure \ref{fig:sample_profiles}. 

The competition between sintering and pressure-driven venting generally sculpts the degree of C depletion. Overall, carbon loss prevails in oxidized CC bodies (CV and CI, Figures \ref{fig:sample_profiles}c.ii and \ref{fig:sample_profiles}d.ii) while sintering halts most C loss in more reduced NC bodies (EH and OC, Figures \ref{fig:sample_profiles}a.ii and \ref{fig:sample_profiles}b.ii). This is foremost a result of the bodies' redox states, while also influenced by their matrix densities and heating budget.  

To understand this outcome, we first examine the thermal evolution of the four bodies since gas venting and sintering are both temperature-activated processes. The thermal histories (Figures \ref{fig:sample_profiles}a.i to \ref{fig:sample_profiles}d.i) are qualitatively similar across compositions. The bodies heat up mostly isothermally while a conductive gradient develops at their surface. As radiogenic heat sources deplete over time, the interior temperature's rise slows down, and surface cooling prevails in increasingly deeper layers.

\begin{figure}
    \centering
    \includegraphics[width=\linewidth]{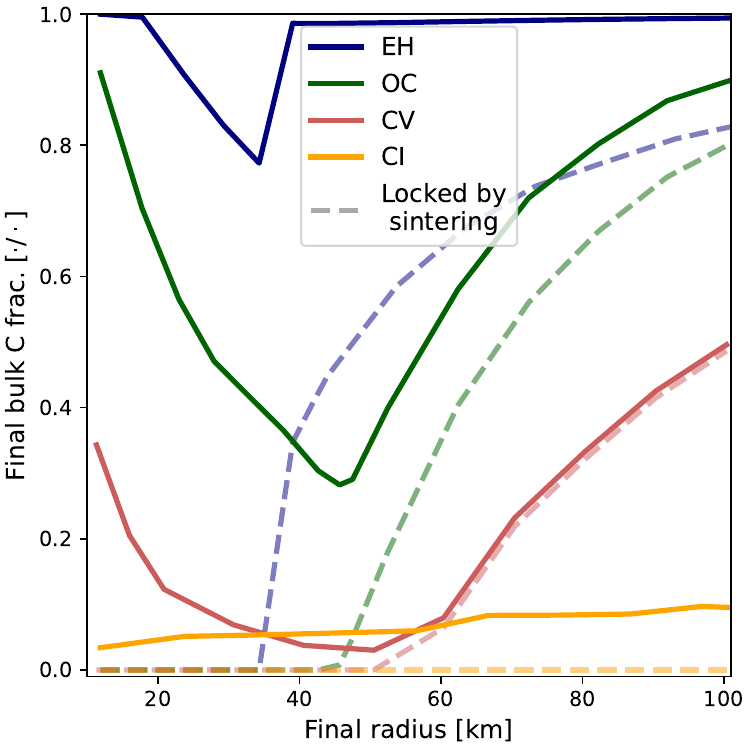}
    \caption{The fraction of total carbon reservoir left at 10 Myr, relative to a planetesimal's initial C abundance, for planetesimals with varying radii. Blue, green, red, and orange lines are bodies of EH, OC, CV, and CI compositions. Dashed lines are the fraction of bulk carbon in completely sintered layers.}
    \label{fig:comp_rad_C_frac}
\end{figure}

The pore gas pressure, $P_\mathrm{Cg}$, rises with temperature until it exceeds the local confinement pressure $P_\mathrm{conf}$, and open-system venting starts. Locally, this results in quick and complete carbon removal. This is because venting keeps the local gas pressure below equilibrium, promoting graphite burning. The heat released by the reaction increases the local temperature and the equilibrium gas pressure, which in turn leads to higher burning rates and ultimately results in runaway graphite depletion. 

Spatially, gas venting begins near the boundary of the surface $T$ gradient and the deep isothermal layer, where low $P_\mathrm{conf}$ meets high $T$. We term this event ``ignition": the earliest time and its corresponding layer where $\dot{m}\mathrm{_{Cg}|_{valve}}<0$, and the local mass conservation equations switch from the closed-system solution to the open-system one (see Section \ref{sec_2:mass_vol_cons}). As the body heats up further, neighboring layers also experience quick carbon removal. This leads to two burning fronts moving upwards and downwards, respectively (see, e.g., Figure \ref{fig:sample_profiles}c.ii). The upward front is slowed and stopped by conductive cooling, while the downward front progresses quickly ($\lesssim$1 Myr) to reach the center (CI, Figure \ref{fig:sample_profiles}d.ii), or stopped by sintering (OC and CV, Figures \ref{fig:sample_profiles}b.ii and \ref{fig:sample_profiles}c.ii). 

On the other hand, sintering, in general, occurs in the isothermal interior and progresses faster near the center (see e.g., Figure \ref{fig:sample_profiles}a.iii) because higher $P_\mathrm{lith}$ and $T$ leads to faster matrix deformation (Equations \ref{eq_2:sint_diff} and \ref{Eq_2:Sint_disl}). Combined with the C venting behavior, we see that, relative to the sintering progress, the earlier a planetesimal reaches ignition, the deeper its C depletion front can reach and the greater its extent of bulk C loss. Meanwhile, the more oxidizing the material, the higher local $P_\mathrm{Cg}$ reaches at the same temperature, and thus the lower T for $P_\mathrm{Cg}$ to exceed $P_\mathrm{conf}$ and drastic C loss to occur. Further, more oxidized bodies also tend to reach ignition closer to the surface, which is congruous with their earlier ignition. This is because ignition tends to occur at the base of the conductive layer, which moves deeper with time (See, e.g., Figure \ref{fig:sample_profiles}a.i). Having shallower ignition points lowers $P_\mathrm{conf}$, promoting faster outgassing. These combined effects explain the trend that more oxidizing planetesimals (CC) tend to deplete more of their C inventory. 

Secondary to the redox effect is the effect of different heat source budgets and bulk densities. In the case of the EH body, weaker internal heating means more time for conduction to push the ignition point deeper, given the same ignition temperature. This likely contributes to the EH planetesimal's compaction, completely preventing carbon loss. The CI chondrite body also significantly lacks internal heating, but its bulk density is also $\sim$30\% lower. This leads to lower $P_\mathrm{conf}$, which slows sintering and promotes quicker ignition. This, along with its highly oxidizing environment, results in the venting front reaching its center.

The CCs have the additional heat sink from the endothermic dehydration reaction compared to the NC bodies. They also have a smaller matrix grain size by a factor of $10^{3}$. Both factors discourage carbon loss. Dehydration slows interior heating, while smaller grains promote faster sintering. Yet neither factor reverses the trend of more complete carbon removal in CCs than NCs. If these were not considered, then the bifurcation of volatile loss efficiency between these two groups would be wider. 

Finally, the feedback of carbon loss on the sintering histories of these planetesimals is complex and varied, which we discuss in Section \ref{sec_2:discussion}. Here, we note that while the OC and CV bodies' carbon-depleted layers are sintered quickly afterward, the CI body's interior stays gas-saturated (Figure \ref{fig:sample_profiles}d.iv), which delays sintering for $\sim$6 Myrs after C loss.
\begin{figure}
    \centering
    \includegraphics[width=\linewidth]{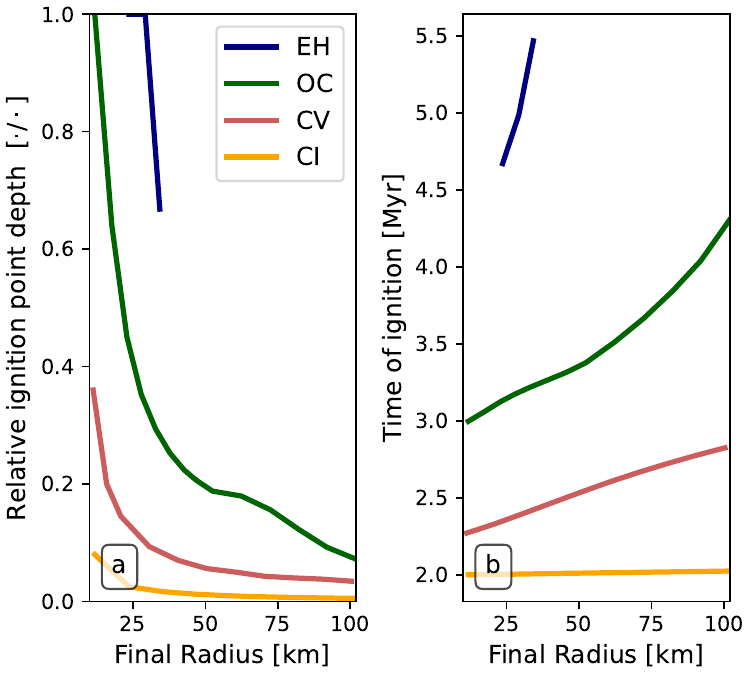}
    \caption{a: depth of the ignition point relative to the total radius at the time of ignition vs. the radius at 10 Myr after CAI. b: timing of ignition after CAI vs. the final radius. Same color codes as Figure \ref{fig:comp_rad_C_frac}.}
    \label{fig:comp_rad_ignition_st}
\end{figure}

\subsection{Effects of Radius}
\begin{figure}
    \centering
    \includegraphics[width=\linewidth]{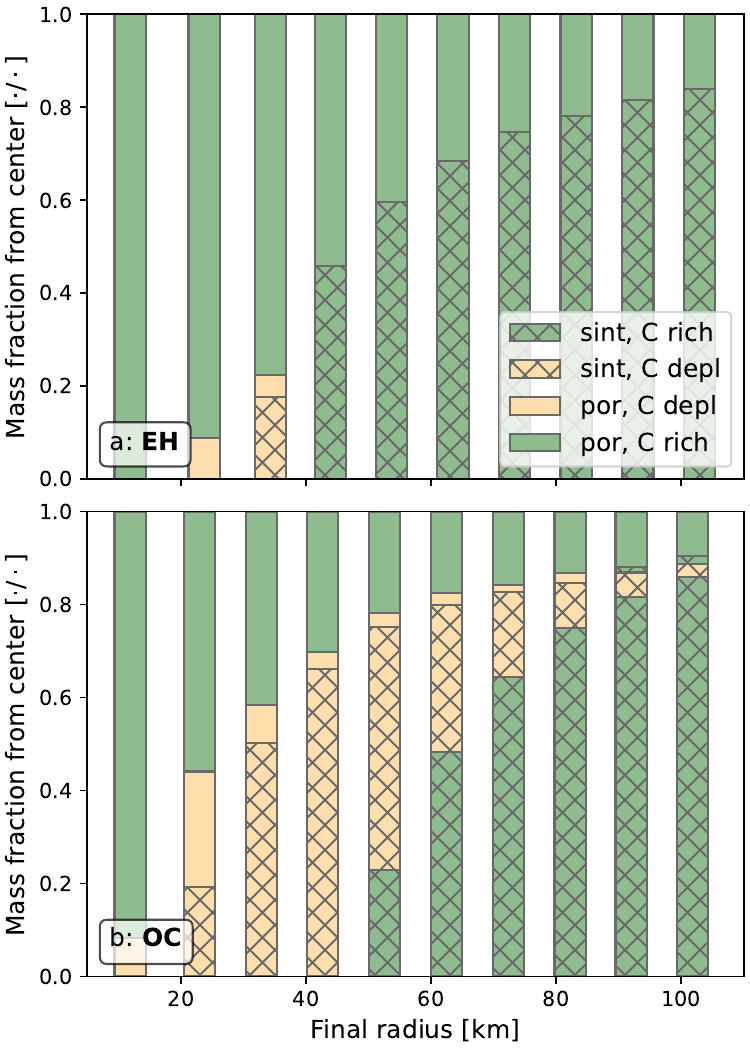}
    \caption{C depletion and sintering outcomes of NC planetesimals of varying sizes. The bodies have $R_\mathrm{c} = $10, 20,...100 km, but their x locations are their final radii. Each bar represents the layering of a body in mass fraction, from its center (bottom) to its surface (top). The green portions of each bar have more than 10\% of solid carbon retained, the ivory parts are the opposite. The cross-pattern shaded region is completely sintered with a porosity of $<0.03$, below the pore closure porosity predicted by SCP packing geometry.}
    \label{fig:outcome_prfl_R_NC}
\end{figure}

We probe the effect of planetesimal radius on its C depletion by simulating planetesimals of all four compositions with $R_\mathrm{c}\in[10, 100]$km, all accreted at 2 Myrs after CAI. Figure \ref{fig:comp_rad_C_frac} shows their bulk C content at 10 Myr after CAI, relative to their initial inventories (solid lines). The dashed lines mark the fraction of C locked in sintered layers. 

The final C budgets of EH, OC, and CV bodies show a similar trend with radius. Both the largest and smallest planetesimals can preserve more carbon.  This is because the smallest bodies have the strongest conductive cooling, pushing ignition to deeper locations relative to their size, as shown in Figure \ref{fig:comp_rad_ignition_st}a. Larger, more massive bodies have higher $P_\mathrm{conf}$, which boosts sintering and locks in more C (Figure \ref{fig:comp_rad_C_frac}, dashed lines). It also raises the ignition temperature and thus delays the timing of ignition, as shown in Figure \ref{fig:comp_rad_ignition_st}b. The sweet-spot radii where carbon loss is maximized are $\sim$30, 40, and 50 km for EH, OC, and CV planetesimals, respectively. As seen before, carbon loss is generally more efficient across all radii for more oxidizing bodies. For EH chondrites, only $\sim20-40$ km bodies experience C loss of more than a couple of percent, while all but the largest CV planetesimals lose more than half of their C inventory. The most C-depleted NC planetesimal (a 40-km OC body) retains $\sim$30\% of its original inventory, more than half of the most C-rich CC object (a 100-km CV body). Thus, a planetesimal's composition exerts greater influence on its fate of carbon than the body's radius. 

For CI bodies formed at 2 Myrs, ignition is always reached before sintering, resulting in almost constant, $>90\%$ C depletion throughout the range of $R_\mathrm{c}$ we consider (Figure \ref{fig:comp_rad_C_frac}, solid orange line). Ignition in CI bodies tends to occur in two separate locations: one is almost at the start of the simulation and near the surface, while the other occurs deeper within the body. We discuss the cause and reliability of this behavior in Section \ref{sec_2:discussion}. 

\begin{figure}
    \centering
    \includegraphics[width=\linewidth]{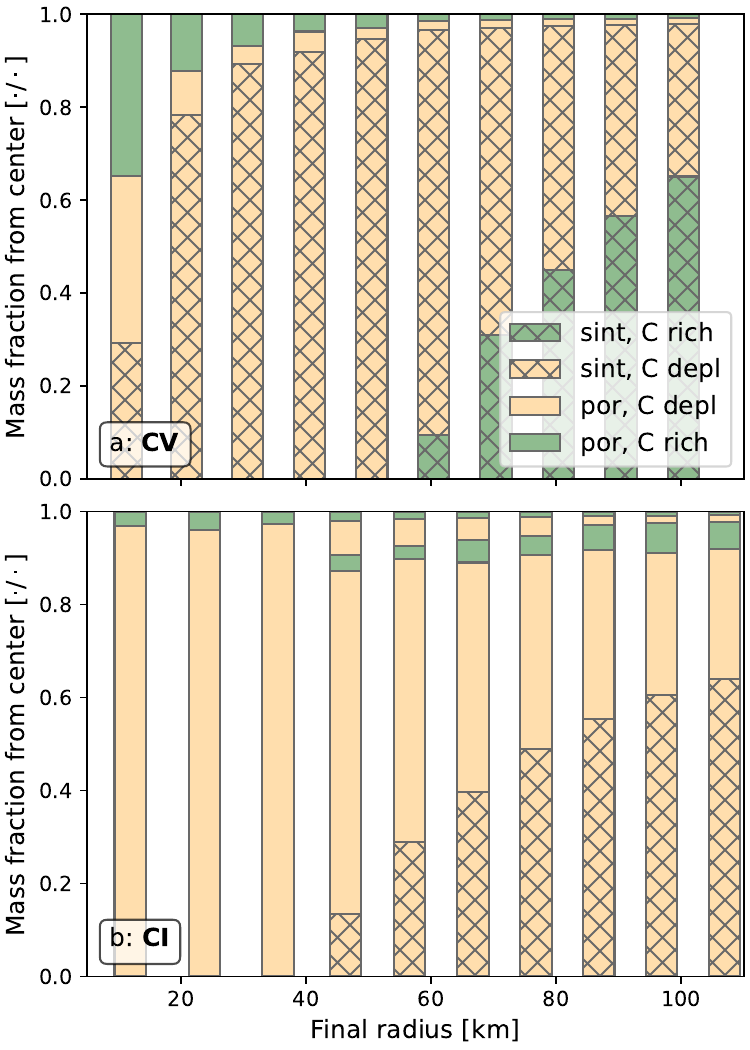}
    \caption{Same as Figure \ref{fig:outcome_prfl_R_NC}, but with CC bodies.}
    \label{fig:outcome_prfl_R_CC}
\end{figure}
To show the interplay of sintering and C depletion with more nuance, we plot the final radial profiles of planetesimals with the same range of $R_\mathrm{c}$'s and compositions. Figure \ref{fig:outcome_prfl_R_NC} shows the final profiles of the NC planetesimals, while Figure \ref{fig:outcome_prfl_R_CC} shows those of the CC bodies. For all compositions, larger planetesimals have thicker deep, sintered (shaded) layers. Smaller bodies have proportionally more extended primordial surface layers that experienced neither sintering nor C loss (green, not shaded). For OC and CV bodies (Figures \ref{fig:outcome_prfl_R_NC}b and \ref{fig:outcome_prfl_R_CC}b), the C-depleted layer (yellow) occupies a progressively thin sliver close to the surface in progressively larger bodies, sandwiched by the primordial surface and the sintered, C-rich layer (green, shaded). 

In all compositions, smaller planetesimals tend to have C-depleted but porous layers (yellow, not shaded). These are especially prominent in CI bodies smaller than $\sim$70 km (Figure \ref{fig:outcome_prfl_R_CC}b). These layers are pressurized by CO/CO$_2$ gas against compaction when the interior temperature peaks, and sintering would have been the most effective. Thus, gas pressure may help the planetesimal retain its bulk porosity long term, i.e., after conductive cooling takes over. On the other hand, CV bodies are the most compacted among all compositions, with $\gtrsim$60 km bodies sintered almost to the surface. This contrast is due to CV bodies having 2$\times$ the ${^{26}}$Al abundance than CI and $\sim$30\% higher density. 

Large CV bodies also deplete carbon almost to the surface (Figure \ref{fig:outcome_prfl_R_CC}a). The vast majority of carbon is stored at depth. This contrasts with the NC bodies, where the primordial surface accounts for a significant portion of the C reservoir left. Again, despite the shared trends across different compositions, the final profiles are qualitatively different between NC and CC bodies. Large NC bodies do not have significant C-depleted layers, unlike all CC bodies. In contrast, extensive primordial surface layers are common among medium-to-small NC bodies but are absent in CC planetesimals other than $<$20 km CV bodies. 

\subsection{Effects of accretion time}

\begin{figure}
    \centering
    \includegraphics[width=\linewidth]{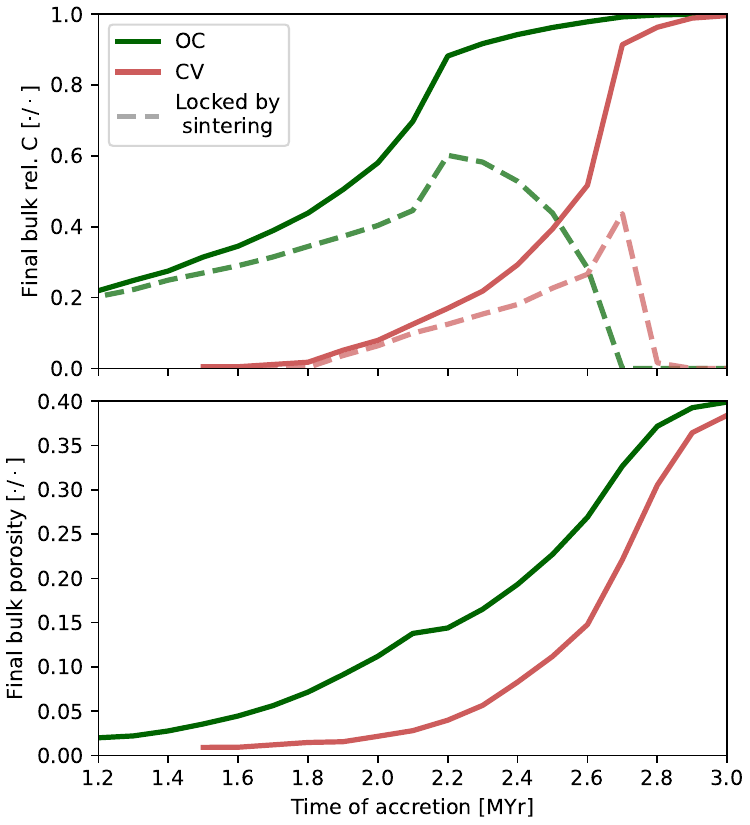}
    \caption{Top: final bulk C abundances relative to the initial reservoir of CV (red lines) and OC (green lines) bodies formed at varied times. The dashed lines are the fraction of C locked by sintering. Bottom: the final bulk porosities of the same bodies. }
    \label{fig:bulk_quantities_t0}
\end{figure}

We examine the effects of varying the timing of accretion on the outcome of C-depletion by simulating the evolutions of our fiducial OC and CV bodies with a range of $t_0\in[1.5,3]$ Myr. The bulk C inventories and porosities at 10 Myrs are shown in Figure \ref{fig:bulk_quantities_t0}. As expected, early-formed planetesimals deplete more carbon and sinter more completely for both compositions (Figure \ref{fig:bulk_quantities_t0}, top). This shows that although more vigorous heating accelerates both sintering and C venting, the net effect favors outgassing. The earlier a planetesimal accretes, the greater fraction of its preserved carbon is in its deep sintered layer (Figure \ref{fig:bulk_quantities_t0}, top, dashed lines vs solid lines). Again, we note that devolatilization is more efficient for CC bodies. OC bodies accreted earlier than 1.9 Myr after CAI depleted more than half of their C inventory, while this transition happens at 2.6 Myr for CV bodies. However, this transition for both compositions is fairly sharp: bodies with both compositions would preserve all their C inventories if they formed after 3 Myr. In other words, C outgassing only applies to planetesimals that accreted quickly. In terms of the final bulk porosity (Figure \ref{fig:bulk_quantities_t0}, bottom), sintering is also more complete in earlier-accreted bodies. OC bodies preserve more porosity than CV ones. 

\begin{figure}
    \centering
    \includegraphics[width=\linewidth]{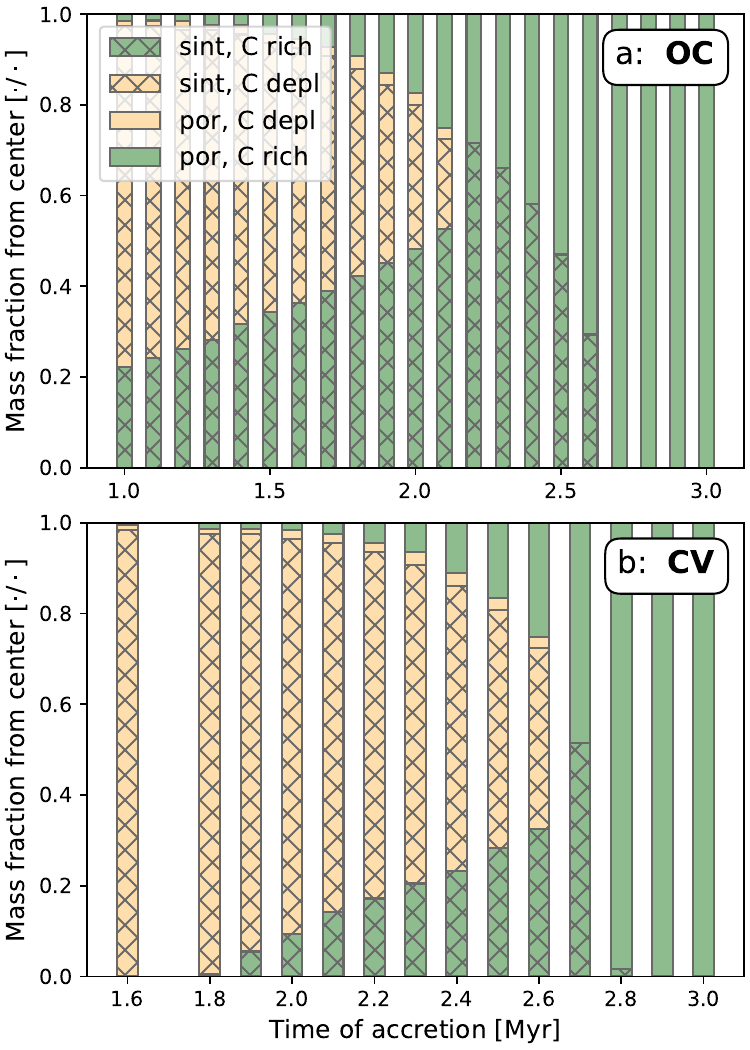}
    \caption{C depletion and sintering outcomes of (a) OC and (b) CV bodies with varying timings of accretion. Same color codes as Figure \ref{fig:outcome_prfl_R_NC}.}
    \label{fig:outcome_prfl_t0_HCV_122024}
\end{figure}

We show the final sintering and C-depletion structures for these OC and CV bodies of varied $t_0$ in Figure \ref{fig:outcome_prfl_t0_HCV_122024}. The shared trend is that earlier-accreted bodies have a smaller primordial surface layer and a smaller deep sintered C-rich layer. C depletion can almost reach the surface if a planetesimal accretes early enough ($\lesssim1.5$ Myrs and $\lesssim2$ Myrs for OC and CV bodies, respectively; the surface layer is kept primordial by the boundary conditions).

\subsection{Semi-analytical predictor of C outcome in planetesimals}

The clear trends in C depletion as a function of radius, composition and time of accretion are due to the interplay among thermal evolution, sintering and pressure valve venting. We can approximate the simulation results of C-depletion by estimating the characteristic temperatures of these three processes given a planetesimal's size and composition. This approach offers an efficient, semi-analytical parameterization to our model and could be integrated to larger-scale planet formation models.

For any layer of a planetesimal to deplete C, it needs to reach an ``ignition temperature" $T_\mathrm{ig}$ such that its pore gas pressure reaches hydrostatic confinement level, that is
\begin{equation}\label{eq_2:T_ig}
    P_\mathrm{Cg,\,eq}(T_\mathrm{ig}) = P_\mathrm{conf}.
\end{equation}
Necessarily, the local heating should be strong enough, and local cooling weak enough, that the peak temperature $T_\mathrm{peak}$ this layer can reach is higher than its ignition temperature,
\begin{equation}
    T_\mathrm{ig} \leq T_\mathrm{peak}.
\end{equation}

The layer also needs to be not completely compacted when its ignition temperature is reached, otherwise the fully lithified matrix prevents the formation of outgassing pathways and C depletion. Since the sintering rate is exponential in temperature, one can define a sintering temperature $T_\mathrm{sint}$, above which sintering becomes faster than the typical thermal evolution timescale (which we detail later). So C depletion also requires 
\begin{equation}
    T_\mathrm{ig} \leq T_\mathrm{sint}.
\end{equation}

\begin{figure}
    \centering
    \includegraphics[width=\linewidth]{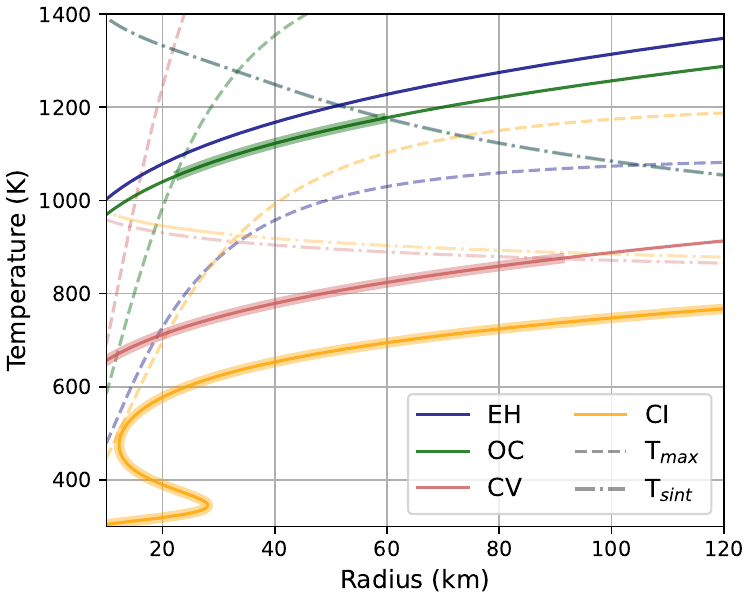}
    \caption{Estimated characteristic temperatures for outgassing (solid lines), sintering (dash-dot lines) and peak heating (dashed lines) in bodies accreted at 2 Myrs after CAI, with varying total radii and composition. The highlighted parts of the solid lines indicate the composition and radii for which significant C depletion is expected in the planetesimal. }
    \label{fig:T_characteristic}
\end{figure}

While $T_\mathrm{ig}$ is clearly defined from the equilibrium relation $P_\mathrm{Cg,\, eq}(T)$ (see Section \ref{sec_2:chem_eq}), we need to adopt some simplifying assumptions to estimate $T_\mathrm{peak}$. When $T_\mathrm{peak}$ is reached at time $t_\mathrm{peak}$, $\partial T/\partial t (t_\mathrm{peak})= 0$ by definition. Assuming thermal evolution is only controlled by $^{26}$Al heating and conductive cooling, and taking the first-order approximation for the spatial derivatives, we can simplify Equation \ref{eq_2:dTdt} to 
\begin{equation}\label{eq_2:t_peak_est}
     C_\mathrm{eff}\frac{\partial T}{\partial t} \approx -\frac{k\Delta T}{\Delta r^2} + Q_\mathrm{0, Al}\exp(-\frac{t_\mathrm{peak}}{\tau_\mathrm{Al}}) =0
\end{equation}
where $\Delta T = T_\mathrm{peak} - T_0$ is the net temperature rise from the initial state. It is also the temperature differential between the planetesimal surface and the location $r$ (Equation \ref{eq_2:surface_init_conds}). $\Delta r = R - r$ is the distance of location $r$ to the surface, and $Q_\mathrm{0, Al}$ is the collection of prefactors for $^{26}$Al decay heating in Equation \ref{eq_2:Q_A(t)}. To first order, $\Delta T$ can be estimated by the sum of the decay heating alone, since much of a body's interior is isothermal during the heating phase of its thermal history. Therefore,
\begin{equation}\label{eq_2:Delta_T_est}
    \begin{split}
        \Delta T &= \int_{t_0}^{t_\mathrm{peak}} \frac{Q_0}{C_\mathrm{eff}}\exp(-\frac{t}{\tau_\mathrm{Al}}) \mathrm{d}t\\
        & = \frac{Q_0 \tau_\mathrm{Al}}{C_\mathrm{eff}}\big[\exp(-\frac{t_0}{\tau_\mathrm{Al}}) - \exp(-\frac{t_\mathrm{peak}}{\tau_\mathrm{Al}})\big]
    \end{split}
\end{equation}
where $t_0$ is the time of formation, for which we use 2 Myrs for our test case.
Substituting Equation \ref{eq_2:Delta_T_est} into Equation \ref{eq_2:t_peak_est}, we solve for $t_\mathrm{peak}$:
\begin{equation}
    t_\mathrm{peak} = t_0 + \tau_\mathrm{Al} \ln\Big(1 + \frac{\Delta r^2 C_\mathrm{eff}}{\tau_\mathrm{Al}k}\Big).
\end{equation}
Thus, $T_\mathrm{peak}$ of a layer of a body can be estimated by first finding $t_\mathrm{peak}$ given $\Delta r$, as well as constant estimates of $C_\mathrm{eff}$ and $k$. For our test case, we use the layer of 50\% mass fraction, resulting in $\Delta r = 0.206 R$. For the heat capacity, we ignore the contribution of pore gas and graphite, assume 40\% porosity and use $C_\mathrm{eff} = 1000 \mathrm{J/(kg\cdot K)} \cdot 0.6\cdot \rho_\mathrm{ch}$ (see Table \ref{tab_2:material_props} for $\rho_\mathrm{ch}$'s). For the thermal conductivity we use $k = 1 \mathrm{W/(m\cdot K)}$ to account for the effect of porosity. The resultant $T_\mathrm{peak}$ for bodies with a range of $R\in[10, 120]$km and 4 chondritic compositions are shown in Figure \ref{fig:T_characteristic} as the dashed lines. 

In terms of sintering, we define $T_\mathrm{sint}$ as the temperature at which the sintering timescale $t_\mathrm{sint}$ becomes lower than $10^5$ years, which is the lower end of typical thermal evolution timescales: for instance, a planetesimal accreted close to $t = 0$ years after CAI would likely melt and differentiate in $\sim 10^5$ years \citep{neumann17}. The sintering timescale is 
\begin{equation}
    t_\mathrm{sint} = \frac{\phi_\mathrm{por,0}}{\dot{\phi}_\mathrm{por}(\phi_\mathrm{por,0}, r, R, \rho, T_\mathrm{sint})}
\end{equation}
where $\phi_\mathrm{por,0} = 0.4$ is the initial fiducial porosity and $\dot{\phi}_\mathrm{por}(\phi_\mathrm{por,0}, r, R, \rho, T_\mathrm{sint})$ is the sintering rate calculated from Equations \ref{eq_2:sint_diff} to \ref{eq_2:sint_rate}, assuming bulk density $\rho = 0.6\rho_\mathrm{ch}$ (i.e. ignoring gas and graphite). For the radial location, we again use the layer of 50\% mass such that $\Delta r = 0.206R$ (i.e. $r \equiv  R - \Delta r = 0.796 R$). Fixing $t_\mathrm{sint} = 10^5$ years, we numerically find $T_\mathrm{sint}$ for the same $R$ and compositions as those of $T_\mathrm{peak}$ (dash-dot lines in Figure \ref{fig:T_characteristic}). Finally, the $T_\mathrm{ig}$ of the same bodies are found by numerically inverting Equation \ref{eq_2:T_ig}, and plotted as solid lines in Figure \ref{fig:T_characteristic}. 

We highlight the range of planetesimals with low enough ignition temperatures - $T_\mathrm{ig}\leq T_\mathrm{sint}$; $T_\mathrm{ig}\leq T_\mathrm{peak}$ - to experience outgassing before compaction or cool-down (Figure \ref{fig:T_characteristic}, thick solid lines). We estimate these highlighted bodies to be C-depleted. These include all of CI bodies, CV bodies smaller than $\sim$90 km, OC bodies with $R\in [25,60]$ km, and none of EH bodies. These results match well with the planetesimal simulations in Figure \ref{fig:comp_rad_C_frac}, where the simulated bodies that depletes more than half of their initial carbon lie in roughly the same radius ranges as our semi-analytic argument predicts. This argument can thus be an efficient predictor to the outcomes of planetesimal outgassing. 

\section{Discussion}\label{sec_2:discussion}

\subsection{Model Caveats}\label{sec_2:Disc_caveats}
Here, we discuss in depth the main design choices of our model and their corresponding caveats. These include using equilibrium chemistry, kinetically fast reaction rates, and the pressure valve mechanism as the mode of outgassing. We also discuss our assumptions around sintering, specifically the effects of grain sizes, as well as a low initial C content for CI chondrites. These discussions highlight that while future experimental and modeling work can relax our assumptions, our predictions for the extent of C depletion in these planetary building blocks are qualitatively sound and should be seen as conservative.

\subsubsection{Chemical equilibrium and fast reactions}\label{Sec_2:Disc_ChemEq}
We use the fast equilibrium burning of graphite as a simplified representation of the diverse C volatilization processes in actual planetesimals. 
We can infer the latter from the diverse phases of carbon found in chondrites, which include organics, carbonates, carbides, amorphous C/graphite, and diamonds \citep{Christ_2024CmChe...7..118C}. One can conceptualize the volatilization process with two timescales: one for liberating carbon from the condensed phase to gas phase - gasification - and another for the C-bearing gas to equilibrate with the matrix's oxidation state. Our simplified reaction rate (Equation \ref{eq_2:kinetic_burning_rate}) effectively assumes both timescales are short since it is modulated by $P_\mathrm{Cg,\,eq}$ and a fast rate constant. The validity of these assumptions can be checked with chondrite heating experiments. 

In terms of quick chemical equilibration (CE), recent CM chondrite heating experiment \cite{Thompson_2021NatAs...5..575T} (hereafter T21) demonstrated that outgassed species are generally similar to the CE predictions, and that CO and CO$_2$ are the main C-bearing species, supporting our treatment. However, they also found that the outgassed C is more reducing than CE predicts at T $<$ 1000 K (Figure 4 of T21). In other words, the equilibrium timescale is slower than that of gas removal at low temperatures. Since T21's experiments are conducted in low pressure and with continuous gas removal, interpreting this is somewhat ambiguous: it can either be that the production and removal of a reduced C-bearing gas mixture are faster than its equilibration with the matrix, or that C-rich species equilibrate faster with some reducing minerals and actually reflect a more reducing environment in CE. The latter possibility is more problematic than the former, if this result is generalizable to all CCs. This is because, for the former option of more efficient outgassing and removal, the experimental gas removal rate is likely not comparable to gas venting from planetesimals. While for the latter option, a reducing environment (but in CE) lowers $P_\mathrm{tot}$ which discourages venting. However, this issue is probably minor since our model predicts that C venting in CCs tends to commence at 700 - 800 K, where the experimentally measured $f_{\mathrm{O_2}}$ is lower by $<10\times$ (T21). Meanwhile CCs' equilibrium $P_\mathrm{tot}$ is $>100\times$ higher than NCs (Figure \ref{fig:Chem_Eq}). So correcting for a $<10\times$ overestimation in the $f_{\mathrm{O_2}}$ of CCs does not close the gap in $P_\mathrm{tot}$ between CCs and NCs. Thus, our conclusion that CC planetesimals are more prone to C depletion should not be qualitatively impacted. It is also probable that chemical equilibrium can be reached at lower temperatures in a closed, pressurized system, a closer analogy to planetesimal interiors. 

Besides the issue of gas-matrix CE efficiency, the redox state of the matrix may also evolve as oxygen leaves the system with carbon. For most chondrites other than CI (and CM), this is likely a minor effect since oxygen is a major rock-forming species while C is a minor component. [C/O] is 0.10, 0.0097, 0.0047, 0.019 for CI, CV, OC and EH chondrites, respectively \citep{Schaefer17}. Nevertheless, for especially C-rich chondritic bodies such as those of CI and CM (e.g., Figure 3 of T21), C outgassing may reduce the matrix, lowering $P_\mathrm{tot}$, and become more self-limiting than our results suggest. Future works accounting for the redox evolution for these chondritic bodies can improve their realism.

Distinct from the gas-rock CE timescale, the timescale of C gasification is directly controlled by the nature of the condensed C phase. Our results assuming quick gasification, are not sensitive to the exact reaction rate constant we choose. As a test, we ran the fiducial OC simulation at a 10$\times$ faster rate and found the results unchanged. Our approach is appropriate when the primary carbon carrier is organics that volatilize at low temperatures \citep{Kebukawa_2010M&PS...45...99K, Alexander_2017ChEG...77..227A}. This is the case for CCs and OC, but less so for EH, where the primary C carrier is amorphous C and graphite \citep{Grady_2003SSRv..106..231G}, although a fraction of the graphite is possibly the residue of organics after thermal modification (\citealt{Piani_2012M&PS...47....8P}, but also see \citealt{Remusat_2012GeCoA..96..319R}).

\begin{figure}
    \centering
    \includegraphics[width=\linewidth]{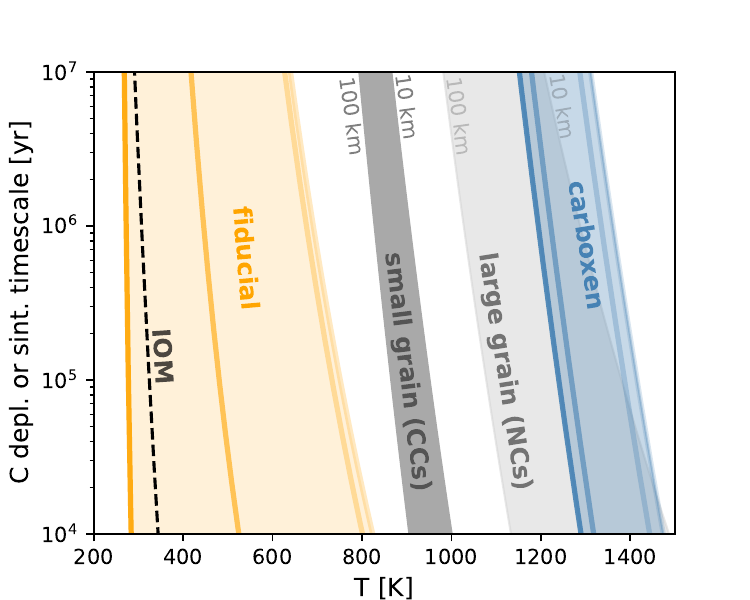}
    \caption{Comparing timescales of reaction kinetics with those of sintering across temperatures. Orange and blue bands: timescales of depletion using the rates that assume open-system flow with $P_\mathrm{conf} = 0$, and an initial C content of 0.3 vol\%. The orange band utilizes our fiducial reaction rates, while the blue band employs oxidation rates of carboxen from \cite{Jaramillo_2014CoFl..161.2951J}, a standard carbon black analogous to chondritic amorphous C and soot. The four lines in each band are, from left to right, timescales for CI, CV, OC, and EH redox states. The dashed black line is the depletion timescale using the destruction of aliphatic C-H in chondritic IOM from \cite{Kebukawa_2010M&PS...45...99K}. The two gray bands are sintering timescales at the center of a $R_\mathrm{c} = $10 - 100 km body with 40\% porosity. The lighter gray band uses a grain size $d = 1\mathrm{mm}$, while the darker band uses $d = 1\mathrm{\mu m}$.}
    \label{fig:kinetics}
\end{figure}

To see the effects of amorphous C vs. organics as the carbon carrier, we calculate the C depletion timescales using the experimentally measured soot oxidation rate (\citealt{Jaramillo_2014CoFl..161.2951J}, see Figure \ref{fig:kinetics}, blue shaded band), and compare them to the sintering timescales (Figure \ref{fig:kinetics}, gray bands), as well as the measured kinetic timescale for the thermal degradation of insoluble organic matter (IOM) from CM chondrites (\citealt{Kebukawa_2010M&PS...45...99K}; Figure \ref{fig:kinetics}, dashed line). We can see that, overall, reactions speed up with temperature, so reactions that reach the same C depletion timescale at higher temperatures are slower. Our fiducial, parameterized C depletion timescale encapsulates the kinetic IOM rate. In the range of timescales relevant to the thermal evolution of planetesimals, $10^5$ to $10^7$ years, our fiducial reaction rates are faster than the sintering rates, no matter the latter's grain size choices. Meanwhile, the more sluggish soot oxidation kinetics have timescales longer than their corresponding sintering kinetics. Specifically, the soot oxidation timescales in CC bodies (Figure \ref{fig:kinetics}, the two leftmost blue lines) are longer than the sintering rates of small-grain, CC bodies but comparable to the sintering rates of large-grain bodies. The soot oxidation timescales in NC bodies (Figure \ref{fig:kinetics}, the two rightmost blue lines) are longer than the sintering timescales of bodies with both grain sizes, except the smallest large-grain bodies. Therefore, if a planetesimal's carbon is in amorphous C or graphite and comparable to soot, then even without the confinement pressure discouraging C loss and venting is always active, the oxidation kinetics are so slow that sintering would almost always prevail, and most of the carbon would be preserved.

Therefore, our results may represent an upper limit of C loss from organics-poor EH bodies. Yet, since we already predict minimal outgassing from EH, this does not qualitatively change our results. C volatilization kinetics of EH bodies may also behave differently from our estimates because of kinetic effects similar to those observed by T21, or outgassing of more reduced species such as atomic carbon \citep{Anzures_2024LPICo3040.1250A}.  

The fact that amorphous C and graphite in reduced bodies may survive planetesimal metamorphism is also evidenced by the presence of ureilites. These are graphite-rich (up to 8 wt\%) remanent from the mantle(s) of differentiated planetesimals \citep{Storz_2021GeCoA.307...86S}. Since their carbon content can survive silicate melting ($\geq$1450 K), they must have survived the C outgassing phase at lower temperatures. However, their high C content may result from the concentration of graphite via its flotation in the silicate melt. Thus, they might not be representative of the entire ureilite parent body \citep{keppler19, Goodrich_2022M&PS...57.1589G}.  

A phenomenon resulting from our choices in chemistry that warrants more clarification is the near-surface C-depletion in CI chondrites. This is a result of the equilibrium $P_\mathrm{tot}$ curve for CI having a peak at 350 K. Since the surface temperature of our planetesimals is set at 300 K, the $\sim$ 1 bar gas pressure that a CI body can generate quickly exceed the confinement pressure near the surface. In reality, it is possible that C outgassing is kinetically inhibited at these low temperatures, which our parameterized kinetics do not capture. Our model, therefore, may underestimate the fraction of C preserved near a CI body's surface.  

More generally, the underlying mechanism of C gasification in CCs and OCs, the thermal processing of chondritic organics, is a complex, multi-step process \citep{Alexander_2017ChEG...77..227A}, as evidenced by the divergent kinetic parameters measured by \cite{Kebukawa_2010M&PS...45...99K} and \cite{Kiryu_2020M&PS...55.1848K} (see Figure 8 of \citealt{Kiryu_2020M&PS...55.1848K}). Beyond that, the microphysics of gasification likely also matters: the diffusion efficiency of the product is a function of IOM porosity and pore geometry, which feedback to the overall volatilization rate. Yet kinetic experiments have not converged on this aspect, either: some works such as \cite{Kebukawa_2010M&PS...45...99K} (and amorphous C oxidation works such as \citealt{Jaramillo_2014CoFl..161.2951J}) considered this aspect, while others did not \citep{COdy_2008E&PSL.272..446C_Organic_reaction, Kiryu_2020M&PS...55.1848K}. A more nuanced planetesimal outgassing model calls for a more systematic and quantitative understanding of the kinetics of this process, as well as how the product of gasification interacts with the chondritic matrix. Further, previous open-system chondrite outgassing experiments, in which pulverized CC samples are heated in a low-pressure environment, while important, may be less apropos to simulating the fluid-matrix interaction in planetesimal interiors, where gas may have sufficient time to equilibrate with the matrix before being extracted. The applicability of CC heating experiments in NC bodies is also uncertain. Our work, therefore, motivates chondritic outgassing experiments in a closed, pressurized environment as another end-member case to calibrate future models of planetesimal outgassing. We also advocate for more experimental efforts in NC thermal processing and outgassing.

\subsubsection{Metal dissolution buffering}\label{Sec_2:Disc_Metal_disl}

As a first step, in our chemical equilibrium treatment, we did not consider the effects of metallic phases. Here we discuss its implications. \cite{Hashizume_1998M&PS...33.1181H} (hereafter HS98) considered C and H migration and outgassing in OC planetesimals by accounting for the dissolution of C into the Fe-Ni metal. Since C is siderophile, dissolving into the metallic phases during the peak heating episode reduces pore gas pressure, possibly buffering the majority of C against escape. While we recognize that incorporating this effect might abate the extent of C depletion, its impact is likely secondary, for two reasons. Firstly, primitive chondrites that we consider typically do not contain enough Fe-Ni metals to dissolve their total carbon inventory. In fact, the limited C dissolution likely strengthens our argument for an NC-CC dichotomy in C retention/loss. Secondly, significant C buffering by metallic phases against outgassing is not supported by observation.

C dissolution in metal is likely minor in chondrites due to both low chondritic metal abundance and carbon's limited solubility in Fe-Ni alloy at low temperatures. CI chondrites do not contain metal phases, while CV chondrites have 0-5 vol\% metals \citep{Weisberg_2006mess.book...19W}. At the CV bodies' characteristic outgassing temperature of 800 K (Figure \ref{fig:T_characteristic}, thick pink line), the Fe-Ni alloy contains low-Ni $\mathrm{\alpha-(Fe,Ni)}$, or kamacite, and high-Ni $\mathrm{\gamma-(Fe,Ni)}$, or taenite (see e.g. \citealt{Wasson_1974_Meteorites_book}, pp 56). Both phases have negligible C solubility $<$0.15 wt\% of the metal, or $<$0.02 wt\% of the bulk CV \citep{Okamoto_1992_CFe, Wada_1971_FeNiC}. 
NC metals could preserve more C due to both higher outgassing temperature ($\sim$1100 K, Figure 9) – and thus higher C solubility, and higher metallic fraction. A typical Fe-Ni alloy with 5 wt\% Ni \citep{Schaefer17, kimura_2008M&PS...43.1161K_FeNi_in_chondrites} can dissolve 0.8 wt\% C at 1100 K, at which the alloy is in its gamma (fcc) phase \cite{Wada_1971_FeNiC}. The metal fraction in OC samples range widely: 2 - 8 vol\% \citep{Weisberg_2006mess.book...19W}, and so are their C abundances: 0.08 - 0.6 wt\% among the most primitive petrologic type 3 samples \citep{Moore_1967JGR....72.6289M_C_in_OC}. Choosing a median metal abundance of 5 vol\% and a conservative 0.2 wt\% C abundance (HS98 for the latter), we estimate that at most 44\% of the C reservoir in OC bodies can be dissolved in metals. While at 10 vol\% metal and 0.4 wt\% C \citep{Weisberg_2006mess.book...19W, Schaefer17}, the metal in EH bodies can accommodate 43\% of its C budget at most. 

The fact that likely only a fraction of C can be dissolved in metals supports our overpressure-driven outgassing: the undissolved C can generate the overpressure and be lost. Further, not all of the soluble C can be saved against outgassing. Dissolved C still equilibrates with gas phases: assuming chemical equilibration among the C phases and the matrix, CO/CO$_2$ gas simply cannot be eliminated by dissolution. Only when dissolved C becomes significantly undersaturated would the gas overpressure become untenable. Such a decrease in activity at a given temperature effectively increases the outgassing temperature (but only for this fraction of dissolved C). Yet, since the equilibrium gas pressure for OC and EH grow near exponentially with temperature (Figure \ref{fig:Chem_Eq}a), this effect should only marginally change the outgassing temperature and our results in general. Nevertheless, we can estimate the fraction of C buffered against loss by dissolution by assuming a critical chemical activity, $a_{min}$: when dissolved C’s activity drops below $a_{min}$, outgassing stops. At $a_{min}$ = 0.5, OC and EH can preserve 26\% and 25\% of their initial C, while at $a_{min}$ = 0.1, the preserved fractions are 6\% for both compositions (as calculated from \citealt{Wada_1971_FeNiC}). Thus in NCs, a non-negligible yet still minor fraction of their C abundance could be buffered against total loss.

The above arguments are not necessarily inconsistent with HS98, whose models show complete C dissolution in the metal phases of OC bodies. This is possible in some chondritic parent bodies due to the wide range of C and metal abundance and alloy composition (especially Ni abundance, strongly impacting C solubility \citealt{Wada_1971_FeNiC}) among the chondritic samples. Nevertheless, HS98 demonstrates gas pressure buildup and global C loss comparable to our OC results(see their Figures 8 and 9). 

More importantly, observation is consistent with C outgassing rather than burial in metallic phases. \cite{Hirschmann_2021PNAS} demonstrated that C/S and C abundance decrease with metamorphic level. Significantly, this trend extends to remnants of planetesimal cores - meteoritic iron samples: both C/S and bulk C wt\% decrease further in these samples relative to chondrites. If C were to be removed by the metallic phases before planetesimal core - mantle differentiation occurs, one would expect the opposite, where the metallic phases would become C-enriched relative to S. In fact, \cite{Hirschmann_2021PNAS} suggests that degassing before and during planetesimal differentiation sculpted their observations. 

Taken together, metal buffering against C loss is likely insignificant in CCs, and non-negligible but still minor in NCs. This effect, therefore, likely widens the NC-CC dichotomy in C preservation/loss that this work demonstrates. Nevertheless, moving beyond carbon CE with bulk chondritic content and accounting for more nuanced chondritic mineralogy is a substantive direction of future work.

\subsubsection{Pressure-valve gas venting}\label{Sec_2:Disc_Pres_valve_venting}

Our model assumes that gas venting is binary and only responds to local pressure exceeding the confinement level. Either the local layer is a closed system or is venting directly to space. This is the most straightforward approach that is consistent with a theoretical picture of extensive fracturing that dramatically increases the local gas permeability. An alternative outgassing scenario would be the slow diffusion of gas throughout the planetesimal matrix. Outgassing in real primordial planetesimals is likely a combination of both. Some pore fluid would slowly seep upwards, equilibrating with the local matrix, while the rest might outgas rapidly to space with minimal re-equilibration. We argue the latter plays a dominating role for two reasons. Firstly, pre-sintering, planetesimals are likely poorly lithified due to their low self-gravity and thus prone to fracturing. In fact, sintering, lithification, and metamorphism should be seen as describing the same underlying response of the planetesimal matrix to elevated temperatures. Secondly, although positive permeabilities have been measured for diverse chondritic samples (e.g. \citealt{Matsui_1986Metic..21..109M_permeability, Sugiura_1986PolRe..41..358S_permeability, Corrigan_1997M&PS...32..509C_permeability, MacPherson_2014M&PS...49.1250M_permeability}), these permissibilities span a wide range between $<10^{-21}$ and $10^{-14}$ $\mathrm{m}^2$, as compiled by \cite{fu_2017}. Much of the low-permeability chondrites likely inhibits efficient fluid migration (e.g. \citealt{Bland}), making global fractures, which can improve permeability by up to $\sim$4 orders of magnitude, necessary for C gas venting. To demonstrate this point, we calculate the diffusion timescales for the interior of our fiducial $R_\mathrm{c}=60$ km bodies with 40\% porosity for all 4 compositions.

For this, we follow the treatment of \cite{Marboeuf_2012A&A...542A..82M_comet_flow} (hereafter M12), who modeled comet outgassing considering both Knudsen flow and viscous flow. For conciseness we briefly describe their framework while detailing the components where our calculations differ from theirs. Following M12, we define a gas diffusion flux, $\Phi$, in response to its pressure gradient, in $\mathrm{mol \,\, m^{-2}\,\, s^{-1}}$ 

\begin{equation}
    \Phi = G_\mathrm{d} \frac{\partial P}{\partial r}
\end{equation}
where $P$ is the total gas pressure and $G_\mathrm{d}$ is the diffusion coefficient in $\mathrm{mol \,\, m^{-1}\,\, s^{-1} Pa^{-1}}$. We assume the gas pressure throughout the body is at the local confinement level, $P(r) = P_\mathrm{conf}(r)$, the latter calculated via Equations \ref{eq_2:P_lith} to \ref{eq_2:P_conf}. We further assume that thermochemical equilibrium is reached throughout the body, therefore the temperature profile is such that $P_\mathrm{Cg,\,eq}(T(r)) = P_\mathrm{conf}(r)$. 

Depending on the diffusion regime, $G_\mathrm{d}$ can represent either a Knudsen or viscous flow, or a combination of both. Knudsen flow occurs when the mean free path of gas molecules, $\lambda$, becomes comparable or larger than the average width of the pore space, $r_\mathrm{p}$. When the reverse is true, then the flow is in the viscous regime. The criterion delineating these regimes is the Knudsen number $K^\mathrm{n}$
\begin{equation}\label{eq_2: kn_number}
    K^\mathrm{n} = \frac{\lambda}{2 r_\mathrm{p}},
\end{equation}
where we take the representative pore size, $r_\mathrm{p}$, to be the grain sizes of each composition: 1 mm for NCs and 1 $\mathrm{\mu m}$ for CCs (see Section \ref{Sec_2:sintering}). Radial locations where $K^n>1$ is in the Knudsen regime, $G_\mathrm{d} = G\mathrm{_d^k}$, where the Knudsen diffusion coefficent, $G\mathrm{_{d}^k}$, is obtained following Equation 28 of M12. Locations where $K^\mathrm{n}<0.01$ is in the viscous regime, $G\mathrm{_d} = G\mathrm{_d^v}$. We adopt a Darcy's law form for $G\mathrm{_d^v}$ that explicitly includes matrix permeability, $K$:
\begin{equation}
    G\mathrm{_d^v} = K\frac{n_\mathrm{eq}}{\eta},
\end{equation}
where $n_\mathrm{eq} = \rho_\mathrm{Cg,\,eq}(T)/\mu$ is the molar density of the gas, and $\eta (P, T)$ is the dynamic viscosity, calculated from Equations 30 and 31 of M12. In calculating these, we have continued to assume that CO and CO$_2$ are in equilibrium described by Equation \ref{eq_2:r_Cg}. In the intermediate regime where $K^\mathrm{n}\in[0.01, 1]$, we uses Equation 32 of M12.

For simplicity we test the effects of matrix fracturing by using an explicit permeability (rather than following M12 and utilizing a permeability parameterized by $r_\mathrm{p}$). We choose $K = 10^{-17}\,\mathrm{m^2}$ for an intact matrix, while $K = 10^{-13}\,\mathrm{m^2}$ for a fractured one. These choices are roughly the midpoint in the wide possible $K$ range in \cite{Fu_2017pedc.book..115F}. We find the diffusion timescale as 
\begin{equation}
    \tau_\mathrm{d} = \frac{(R - r)n_\mathrm{eq}}{\Phi},
\end{equation}
which corresponds to the time required to transport the gas with a local molar density $n_\mathrm{eq}$ from location $r$ to the surface at $R$. Since the flow law is in terms of the pressure gradient and not the traditional density, the diffusion timescale takes a different form than the traditional $L^2/D$ one.

\begin{figure}
    \centering
    \includegraphics[width=\linewidth]{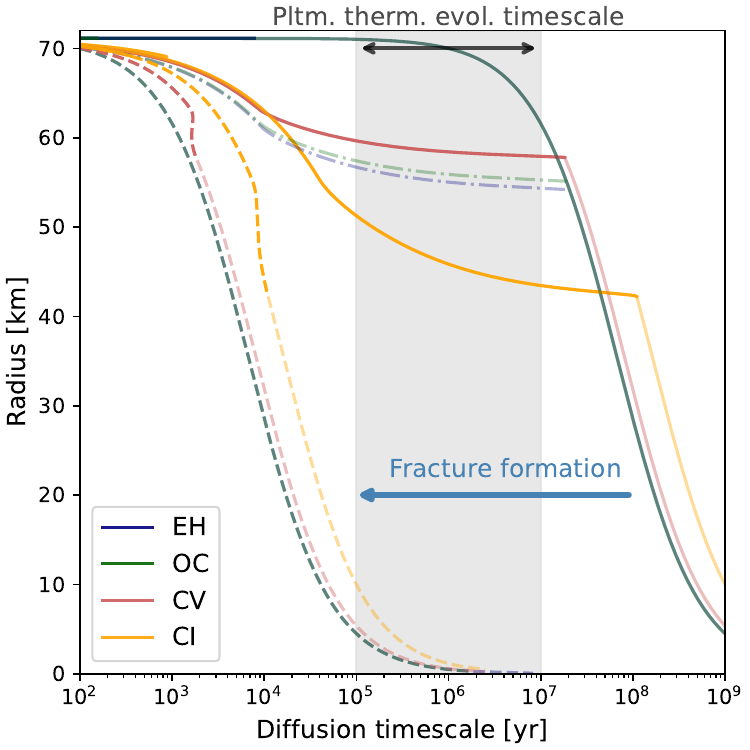}
    \caption{Diffusion timescales in fiducial $R_\mathrm{c} = 60$ km bodies' interiors, for different chondritic compositions. EH and OC lines are mostly overlapping. Solid lines are bodies with intact matrices ($K = 10^{-17} \mathrm{m^2}$), while dashed lines are fractured bodies ($K = 10^{-13} \mathrm{m^2}$). Dash-dotted lines are EH and OC bodies if they had the same grain size as CCs, $d = 1\mathrm{\mu}$m. The shaded region denotes timescales relevant for planetesimal thermal evolution.}
    \label{fig:diff_timescale}
\end{figure}

The resultant $\tau_\mathrm{d}$ in fiducial bodies of different compositions are shown in Figure \ref{fig:diff_timescale}. Compared to the timescales of planetesimal thermal evolution, $10^5$ - $10^7$ years, diffusion is slower in most of the interiors for all intact bodies (solid lines). In colder layers closer to the surface (highlighted portions of the solid lines), Knudsen flow enables more efficient diffusion. the relative depth of this outgassing surface layer depends on composition: $\sim$2\% for NC bodies, $\sim$18\% for CV and $\sim$30 - 40\% for CI. While in fractured bodies, diffusion is generally faster than thermal evolution except near the center, where our lithostatic pressure profile dictates $\Phi \propto \partial P/\partial t\rightarrow0$. This contrast of sluggish versus efficient outgassing pre- and post-fracturing corroborates our closed-until-forced-open approach to outgassing. Nevertheless, given the uncertainty and likely heterogeneity in permissibilities of primitive planetesimals, there can still be bodies where diffusion alone suffices in depleting their volatile inventories (the reverse scenario is less likely, see \citealt{Hirschmann_2021PNAS}). Our results in C depletion are therefore conservative.

A consequence of our binary outgassing mechanism is the binary outcome of local C loss. At a given location, the carbon reservoir is either almost wholly preserved by sintering or almost completely removed. There is no clear evidence for this behavior in the chondritic record, where chondrites show a relatively weak trend of progressively less carbon with higher metamorphic temperatures \citep{Weisberg_2006mess.book...19W, VACHER_2024_Geochimica_et_Cosmochimica_Acta}. As discussed in Section \ref{Sec_2:results_composition}, the binary outcome we predict is a consequence of open-system loss and the exothermic nature of the reaction. Accounting for slower diffusion could create smoother final C profiles (e.g., \citealt{Sugiura86}). Since the heated chondritic organics likely carbonize into more refractory soot and graphite, it is possible that the progressively sluggish kinetics can avoid complete C removal in open systems. Further, the thermal processing of organics is not necessarily exothermic. If the reaction is strongly endothermic, such that the radiogenic heating can be effectively stalled, C outgassing can become self-limiting. Given these unknowns, our model is better characterized as an estimate of the bulk C content evolution for planetesimals rather than an explanation for the diverse outcomes of chondrite metamorphism. 

\subsubsection{Sintering}\label{Sec_2:Disc_sint}

Our model assumes complete sintering can prevent outgassing, preserving the C reservoir. Our sintering model follows the treatment of well-established works \citep{henke12,neumann14} and contains similar assumptions. One of these is allowing for complete sintering. From first glance, this seems incongruent with the fact that most chondrites have measurable porosity. No apparent trend is present between porosity and the degree of metamorphism for OC samples \citep{Corrigan_1997M&PS...32..509C_permeability, Consolmagno_2008ChEG...68....1C_porosity, Soini_2020M&PS...55..402S}, but such trend seem to exist in CCs \citep{Mache_2011M&PS...46.1842M_porosity_CC}. We note that, however, sintering on a particular chondritic sample is not only a function of its peak metamorphic temperature but also its original depth in the planetesimal, as well as the planetesimal's mass. Therefore, thermally processed chondrites in small planetesimals or near the surface of large ones can preserve their porosities. Severely compacted portions of planetesimals could experience further evolution such as melting and differentiation, and become achondrites, thus no longer represented in the chondritic record. Nevertheless, if sintering - and its countering effects to de-volatilization - are overestimated in our models, then our results represent a conservative estimate of C depletion. 

Following \cite{henke12, neumann14}, we used an olivine-dominated upper-mantle rheology for chondrites, albeit an updated one (\citealt{Jain, Jain_2020JGRB..12519896J} vs. \citealt{Schwenn_1978Tectp..48...41S}). We believe this is sufficient for our purposes, while recognizing that we are extrapolating a model calibrated at higher P-T to a low-gravity context. Future experimental work can also improve our understanding of chondritic rheology by utilizing better analogues to chondritic mineralogy beyond olivine. 

There can also be other sintering mechanisms beyond pressure-driven viscous deformation, such as grain surface diffusion driven by crystalline surface energy (e.g. \citealt{Mackenzie_1949PPSB_sintering_theory, KANCHIKA_20184283_sintering_pressure}). Yet this effect is likely minor compared to pressure-driven sintering. For instance, the surface energy of olivine is on the order of 1 J/m$^2$ (e.g. \citealt{Swain_Atkinson_1978PApGe_Olivine_SE}). For grains with sizes between 1 mm and 1 $\mu$m, this corresponds to an energy density, or effective pressure of $10^3$ to $10^6$ Pa. This is lower than the typical lithostatic pressure at the center of a $\sim100$ km body, $P_{conf}\sim 10^7$ Pa. Incorporating these effects may marginally speed up sintering, but likely would not qualitatively impact our conclusions.

A sintering behavior that requires further discussion is that of the C-depleted layers (see, for instance, Figures \ref{fig:sample_profiles}c.iii and \ref{fig:sample_profiles}d.iii). Instead of sintering gradually like the EH body shown in Figure \ref{fig:sample_profiles}a.iii, pores collapse fast and entirely in C-depleted regions of OC, CV, and CI chondrites. This is due to positive feedback in (nearly) gas-saturated but C-depleted matrix. 

When C venting is active, sintering is paused because the matrix is supported by pore gas. After C depletion, the matrix may remain gas-saturated as long as its temperature continues to rise and the gas pressure holds. This is especially present in the fiducial CI simulation (Figure \ref{fig:sample_profiles}d.iii), where much of the body stays gas-saturated and porous for 6 Myrs after C outgassing. Thus, the local porosity of a C-depleted layer can remain high when the matrix is hot and plastic enough that, if there were no gas support, sintering would have progressed rapidly. Once the layer cools by conduction, gas pressure support diminishes, and the layer sinters rapidly. Further, as the porosity decreases, the surface area that the pore gas exerts pressure on, $x_\mathrm{por}$, also decreases, and the deviatoric stress $\sigma$ increases (Equation \ref{eq_2:sigma_P_lith}), boosting the sintering rate. This positive feedback thus causes the runaway sintering in this layer. As one layer starts to sinter, the mass above it is brought closer to the body's center, increasing $P_\mathrm{lith}$ throughout the body and triggering sintering in all layers that are close to gas saturation. This explains the drastic, global sintering behavior of the fiducial CV body and, to a lesser extent, the analogous behaviors of the fiducial OC and CI bodies. 

Moving beyond our simple, binary approach, future work could build on a better quantitative understanding of the interplay between fracturing, pore fluid migration, and sintering. A more nuanced understanding of the chondritic organics' thermodynamics and kinetics would also improve the realism of future models. Our work thus underscores the multidisciplinary experimental effort required to advance our understanding of the planetesimals' geophysical and geochemical evolution.

\begin{figure}
    \centering
    \includegraphics[width=\linewidth]{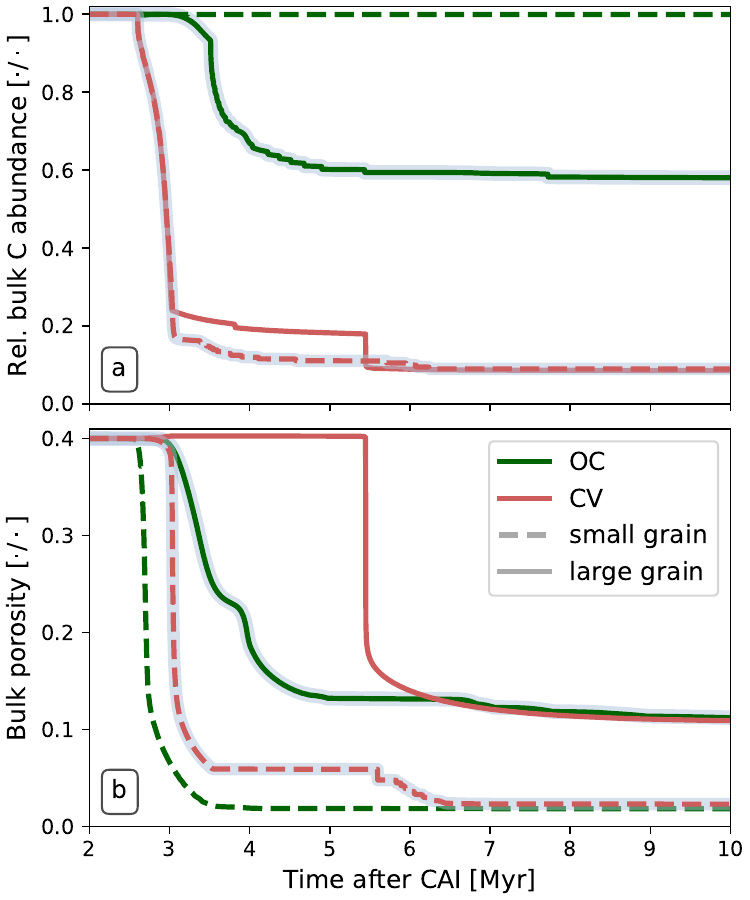}
    \caption{Evolution of (a) bulk C reservoirs and (b) porosities of OC and CV bodies with different grain sizes. The red lines are CV bodies with $R_\mathrm{c} = 60$ km, while the green lines are OC bodies of the same size. Solid lines are bodies with grain sizes $d = 1\mathrm{mm}$, while dashed lines are bodies with $d = 1\mathrm{\mu m}$. The lines highlighted with gray shadows are our fiducial runs. }
    \label{fig:grain_size}
\end{figure}

\subsubsection{Effects of grain size}

The grain sizes we adopt have a limited impact on our results. We used a $10^3$ larger grain size for NCs bodies than CCs, reflecting the higher chondrule content of the former. This results in more efficient sintering in CC bodies, which favors narrowing the gap in C depletion between the two groups.

To probe the impact of this choice, we rerun our fiducial OC body with CC-like grain size (Figure \ref{fig:grain_size}, dashed green lines) and our fiducial CV body with NC-like grain size (Figure \ref{fig:grain_size}, solid red lines). Contrasting the fiducial, large grain OC run (Figure \ref{fig:grain_size}, solid green lines) with the small grain run, we found the latter to preserve almost all C, while the fiducial run depletes 42\% of its reservoir (Figure \ref{fig:grain_size}a, solid green vs. dashed green lines). The OC run with the larger grain size experiences protracted sintering that starts 0.4 Myr later and finishes 1.5 Myr later than the run with smaller grains (Figure \ref{fig:grain_size}b, solid green vs. dashed green lines). Meanwhile, the CV runs of both grain sizes yield a similar final C reservoir of 10\%. However, we note that the larger-grain CV body preserves most of its C in the primordial surface layer, while the (fiducial) smaller-grain CV body preserves C  near the center due to sintering. This difference is caused by the earlier and stronger sintering in the smaller-grain body. This both locks the C at depth and improves the thermal conductivity closer to the surface, increasing near-surface temperature to enable C venting (See the near-surface contrast between Figures \ref{fig:sample_profiles}c.ii and \ref{fig:sample_profiles}b.ii for example). In terms of the porosity evolution, we note that the combined effect of larger grains and post-C-venting gas support delays the main sintering episode in the CV body by 2.5 Myrs (Figure \ref{fig:grain_size}b solid red vs. dashed red lines). For both compositions, larger grains result in 10\% more bulk porosity at 10 Myrs. Thus, the gap in C depletion at 10 Myrs between the CV and OC bodies is robust against different grain size choices (Figure \ref{fig:grain_size}a, green vs. red lines). Our fiducial choices (Figure \ref{fig:grain_size}a, highlighted lines) narrow the gap, although it remains qualitatively significant.   Real chondritic planetesimals have a distribution of grain sizes \citep{Weisberg_2006mess.book...19W, Brearley_1989GeCoA..53.2081B, Vaccaro_2023M&PS...58..688V}, whose compacting behavior would be more complex, but the two cases tested here likely encapsulate the range of possibilities.

\subsubsection{C content in CI bodies and the enthalpy of devolatilization}
\begin{figure}
    \centering
    \includegraphics[width=\linewidth]{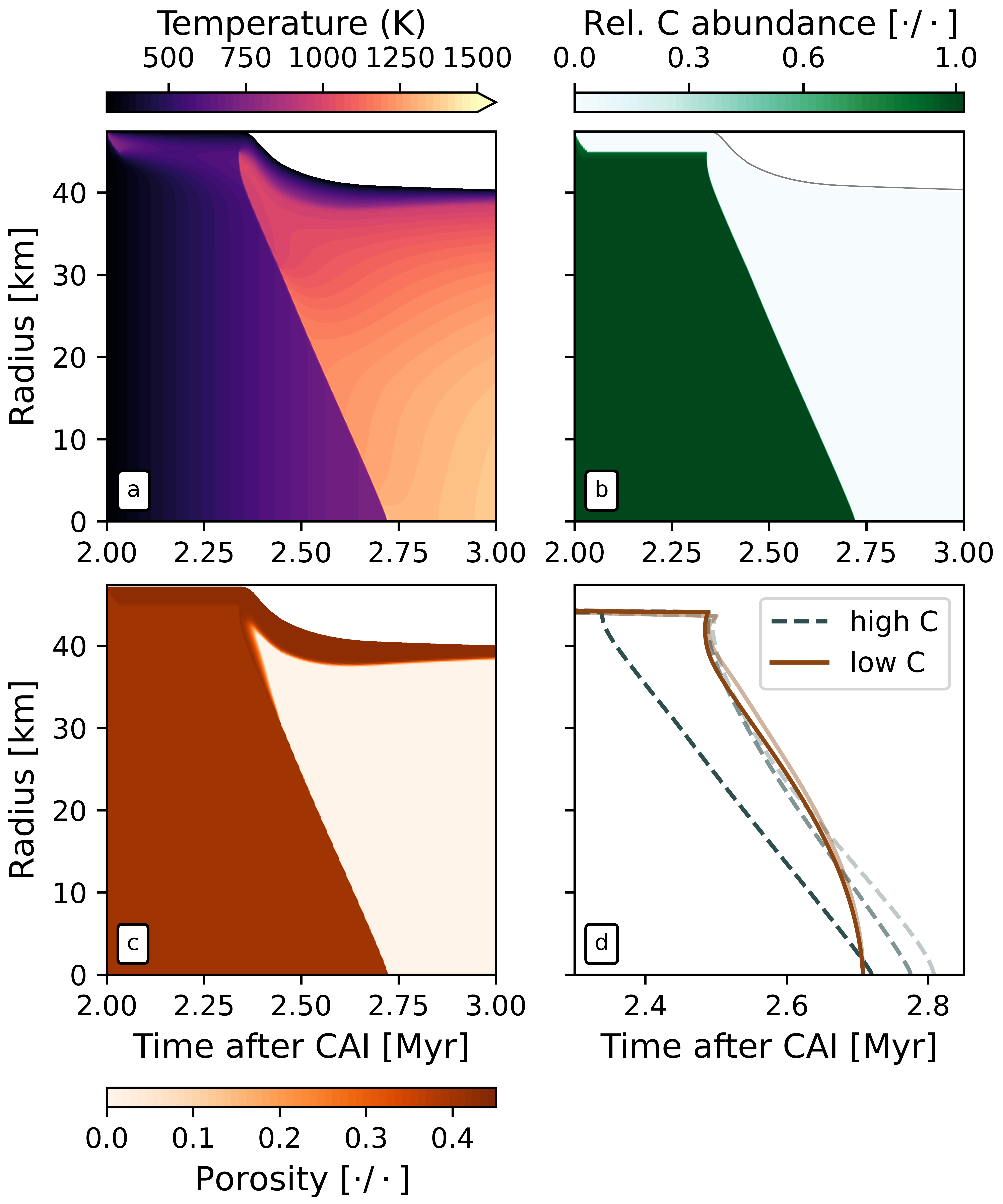}
    \caption{Sample evolution of a $R_c = 40$ km CI body, with 3.5 wt\% initial C abundance and comparison with low spatial resolution runs, as well as fiducial low C abundance ones. (a) Temperature evolution of the first million year since the planetesimal's accretion. (b) Evolution of its C reservoir normalized to the initial abundance. (c) Evolution of its porosity. (d) Comparing runs with high vs. low C abundances and different spatial resolutions. Each line is a run's C depletion front, the location and time where valve venting is first active. The dark green, dashed lines have 3.5 wt\% C, while the brown solid lines have fiducial 0.27 wt\% C. The three high-C runs, from left to right, have spatial resolutions of 200, 400 and 800 m, respectively, While from left to right, the low-C runs have 200 and 800 m resolutions, respectively.}
    \label{fig:resolution}
\end{figure}

For our fiducial CI bodies, we adopt a low initial C content of 0.27 wt\%, roughly an order of magnitude lower than that of real CI chondrites, at 3.5 wt\% \citep{Schaefer17}. We choose this to balance the self-consistency and realism of the model. 

Our simplified chemistry of graphite burning is exothermic, leading to a local temperature jump pre- and post-C depletion. In fiducial runs, this jump of $\sim$50 K does not significantly influence the overall thermomechanical evolution. However for C-rich CI bodies the reverse is true. To demonstrate this point, we run a $R_\mathrm{c}=$40 km body accreted at 2 Myrs after CAI with 3.5 wt\% initial C, and show its evolution in Figure \ref{fig:resolution}a-c. The planetesimal's almost complete C depletion is in line with the fiducial case (Figure \ref{fig:resolution}b), yet the temperature and porosity evolutions are distinct. C depletion leads to a rapid temperature rise of $\sim$600 K, which dominates the overall thermal evolution and drastically accelerates sintering in newly depleted layers (Figure \ref{fig:resolution}a and c). As a result, each layer of the planetesimal transitions sharply (within $\sim$5 thousand years) from a porous, C-rich state to a fully compacted, C-depleted one. 

Two difficulties arise from this scenario. One issue is the realism of this temperature jump due to C burning. Direct evidence of fast CC meteorite parent body metamorphism due to exothermic devolatilation is lacking (although there are meteoritic irons traced to differentiated CC parent bodies indicating high heat sources, consistent with this scenario, but can also be interpreted as early formation, e.g. \citealt{Neumann_2024NatSR..1414017N}). Whether the total thermal effect of C depletion is represented reasonably by graphite oxidation is also unclear, when there seems to be a lack of chondrite calorimetric data. Theoretically, the pyrolysis of IOM is likely endothermic as it involves breaking chemical bounds, but could also exhibit exothermicity if it behaves similarly to terrestrial biomass (e.g. \citealt{Di_blasi_biomass}). The subsequent equilibration between the product of pyrolysis and the matrix may also be less exothermic than graphite burning, since the former requires energy to liberate additional oxygen from the matrix, while the latter assumes a constant supply of oxygen. Taken together, our chemical treatment likely overestimates the exothermicity of C volatilization in chondrites. Thus, adopting a low C content for CI bodies mitigates the impact of this overestimation on the thermal evolution of these bodies. Further, as our results establish that greater heating leads to more severe C depletion, adopting a low C abundance for CI bodies thus decreases their total heat source and is conservative in C depletion. This is likely true for our entire range of radius, since the post-burning interior of a C-rich CI body can reach $\sim1400$K, which corresponds to an equilibrium gas pressure $P_{tot, eq}>$ 1 kbar, exceeding the highest confinement pressure in our simulations, $\sim 100\,$bars. 

A second, related issue is the effects of fast outgassing and high exothermality to the numerical convergence of CI outgassing outcomes. In Figure \ref{fig:resolution}d, we run the high C body with spatial resolutions of 200, 400 and 800 m, and show the progression of the burning front's radius as a function of time (dashed, dark green lines). In other words, we show the location of the same C concentration cliff as Figure \ref{fig:resolution}b, but for three spatial resolutions. Runs with higher spatial resolutions lead to sooner ignition and C depletion by up to 0.1 Myrs, which is a numerical artifact. This artifact is small in the fiducial, low C runs (brown lines in Figure \ref{fig:resolution}d). This artifact does not appreciably change the final C reservoir for each high C run, which are all close to 0, thanks to the high C burning heat effectively igniting the entire body. Yet, as discussed before, since our model likely overestimates the exothermality of C devolatilization, these numerical artifacts are likely not realistic either, again favoring our fiducial, low C treatment.

\subsection{Effects of aqueous processes on C loss}

The effects of accreting a significant amount of water are complex. For CCs that formed beyond the snow line, the latent heat of melting water ice likely buffers radiogenic heating and enables more C survival. However, as the body heats up further, a narrative of overpressure-driven venting, analogous to our model, might lead to rapid, global water loss \citep{Grimm_1989Icar_McSween...82..244G, Fu_2017pedc.book..115F}. This is because water vapor pressures can also easily reach $>$100 bars; if the body heats up beyond the critical point while preserving water, then supercritical water can generate even higher pressures \citep{Grimm_1989Icar_McSween...82..244G, Fu_2017pedc.book..115F}. Soluble organics may migrate with water \citep{Le_Guillou_2014GeCoA.131..344L}, leading to some C depletion. Yet distinct from C gas venting, water venting is endothermic and redistributes heat from the interior to the surface, and thus it may be self-limited. There is also competition between water migration and consumption through aqueous alteration reactions. The net effect of low-temperature aqueous processes on C depletion is therefore uncertain.

The impact of phyllosilicates' dehydration at higher temperatures is more straightforward to gauge. Again, one can envision a pressure-valve effect in which steam venting is efficient but perhaps energy-limited. Carbon-rich gases can be dragged away with steam in the open-system flow, resulting in more efficient carbon loss. Since our model shows that the enthalpy barrier of dehydration reactions is insufficient to prevent severe C depletion in CC planetesimals, the full effect of matrix dehydration likely favors more extensive C depletion.

\subsection{The NC-CC dichotomy and C destruction/recycling in the disk}

Not only are NCs and CCs different in redox and volatile, but they are also isotopically separate, which is widely accepted as indicative of their distinct formation locations in the protoplanetary disk: CCs in the outer solar system, while NCs in the inner Solar System (e.g., \citealt{Bermingham20}). The relationship between this dichotomy and the Earth's volatile accretion history is debated. Possible pathways of volatile delivery to the Earth include inheritance from local NCs (e.g \citealt{Piani_2020Sci...369.1110P, Thomassin_2023E&PSL.61618225T}), delivery of CCs either during or after the main accretion episode (e.g. \citealt{Braukmueller_2019NatGe..12..564B, Mahan_2018GeCoA.235...21M}), ingassing from the protoplanetary disk \citep{Ikoma_2006ApJ...648..696I, Genda_2008Icar..194...42G, Sharp_2022GeCoA.333..124S, Young_2023_Earth_Primordial_H2_atm}, or some combination of all three (e.g. \citealt{Rubie11, Rubie_15_redox_core_formation, Wu_2018JGRE..123.2691W}). 

In this theoretical landscape, planetesimal outgassing complicates the picture in two ways. Firstly, as also noted by \cite{Lichtenberg_2021Sci...371..365L}, early-accreted material is necessarily volatile-poor. In fact, some of the carbon locked by sintering in our simulations is likely still lost during core-mantle differentiation \citep{Wilson_1991E&PSL.104..505W, Muenow_1995Metic..30..639M, Fu_2014E&PSL.390..128F, Hirschmann_2021PNAS}. In this sense, our C depletion predictions are likely lower bounds.

Secondly, our work demonstrates that CC's oxidizing environment can promote significantly more efficient devolatilization. If the distribution of radii and times of accretion were the same across NC and CC, then our finding would favor an NC source of carbon for the terrestrial planets. Previous planetesimal thermal models fitted to meteorite samples indicate that NCs were formed 0 - 2.2 Myr after CAI, while CCs accreted at $<$0.6 - 4 Myr (\citealt{Lichtenberg_2023ASPC_review..534..907L, Neumann_2024NatSR..1414017N}). Comparing these time frames to the formation time lower bounds for 50\% C survival (Figure \ref{fig:outcome_prfl_t0_HCV_122024}, top), 1.9 Myr for OC and 2.6 Myr for CV, we see that the post-outgassing population of NC and CC planetesimals is likely a mix of C-depleted and C-rich bodies. 

Recent studies also probed C destruction in the protoplanetary disks to explain the Earth's C inventory. Yet robustly reproducing Earth-like C depletion remains difficult without invoking additional mechanisms, such as the chondrule formation events, possibly FU Ori-style outbursts (e.g. \citealt{GTDust, Klarmann_2018A&A...618L...1K_disk, Binkert_23_C_in_disc}), radial barriers of dust influx \citep{Klarmann_2018A&A...618L...1K_disk} and icy pebble sublimation across the snow line \citep{Okamoto_2024A&A...692A..11O}. 

Planetesimal outgassing, as a separate pathway of C depletion, relaxes the constrains on disk processes in explaining the C-poor Earth (also see \citealt{bergin15,Li_2021SciA_C_deficit_early_subli}). While in a traditional, hierarchical growth picture, where dust particles continuously grow into planets, planetesimal outgassing cannot contribute to the C depletion in chondrites relative to the ISM and the Sun. However the protracted timings of accretion inferred from chondritic samples \citep{Neumann_2024NatSR..1414017N} hint at the possibility of repeated planetesimal assembly and collisional disruption, where late-formed ``primitive" chondrites could be the product of multiple parent-body heating and outgassing events. 

\section{Summary and Conclusion}\label{sec:conclusion} 
We have developed a thermochemical model of C outgassing from porous, chondritic planetesimals before the onset of melting and differentiation. Our model posits the formation of global fracture networks when gas pressure exceeds the confinement level, resulting in open-system flows and efficient C depletion. We used the redox states, bulk densities, carbon contents, and radioactive isotope abundances of four typical chondrites: EH, OC, CV, and CI. We account for sintering, which can preserve some C by collapsing the pore spaces. We also account for the enthalpy of dehydration reactions from CC bodies of 15 wt\% water. Our conclusions are summarized as follows:

\begin{itemize}
    \item The fraction of carbon preserved is foremost predicated on composition. Generally speaking, regarding final C abundance, CI$\lesssim$CV$<$OC$<$EH. This is a result of the competition between C venting and sintering. While sintering occurs earlier at the center, C venting starts near the surface.  More oxidizing bodies start C venting at lower temperatures and earlier, therefore venting more C before the C depletion front is stopped by sintering. CI bodies experience significantly later sintering due to low bulk densities and radiogenic heat sources. 
    \item A body's radius also impacts its final C depletion. Larger bodies sinter earlier and reach ignition later, preserving more carbon at depth. Meanwhile, smaller bodies cool faster and maintain more carbon in the primordial surface layer. Conversely, larger bodies tend to have proportionally thinner primordial surface layers. For EH, OC, and CV bodies, a sweet spot size that maximizes C depletion exists at final radii of 30, 40, and 50 km, respectively. Yet, overall composition still has a greater impact than radius in informing C depletion. For CI bodies, we found $>90\%$ bulk C depletion for 10 - 100 km bodies. For planetesimals formed at 2 Myr after CAI formation, almost all CC bodies of 10 - 100 km radii deplete $>50\%$ of their initial reservoir, while nearly all NC bodies deplete $<50\%$. 
    \item Timing of formation also impacts final C survival. The earlier planetesimals accrete, the more internal heat sources are available, enabling both earlier C depletion and faster sintering; however, the net effect favors C depletion for both OC and CV bodies. Of the C preserved, the earlier a planetesimal accretes, the greater fraction of C is locked in by sintering; conversely, later-formed planetesimals preserve more C near the primordial surface. CV and OC bodies would preserve over half of their C inventory if formed later than 1.9 and 2.6 Myr after CAI. 
\end{itemize}

Since CCs are more prone to planetesimal outgassing, our results indicate that Earth may have accreted its C locally in the NC reservoir. We have demonstrated that planetesimal outgassing can play a significant role in explaining the volatile depletion in the Earth and terrestrial planets relative to primitive chondrites.   Given the protracted history of meteoritic parent body accretion, the planetesimal population was likely a combination of C-rich and C-depleted bodies in the NC and CC reservoirs. To improve planetesimal thermochemistry models like ours, we advocate for further experimental efforts in understanding the kinetics of chondritic organics' thermal processing and the interaction between the C-carrying phases and the matrix for diverse chondrite groups. 

\section{Acknowledgement}
  This work has been partially funded by the Natural Sciences and Engineering Research Council of Canada (grant RGPIN-2021-02706) and the Ontario Early Researcher Awards (grant ER18-14-131). We gratefully acknowledge Profs. Laura Schaefer and Edward Young for stimulating discussions. We acknowledge that our work was performed on land traditionally inhabited by the Wendat, the Anishinaabeg, Haudenosaunee, Metis, and the Mississaugas of the New Credit First Nation. 
\software{Numpy \citep{Numpy_harris2020array}, Scipy \citep{SciPy_2020-NMeth}, Matplotlib \citep{Matplotlib_Hunter:2007}}
\end{CJK*}
\bibliography{Pltm_C}

\begin{thebibliography}{}
\expandafter\ifx\csname natexlab\endcsname\relax\def\natexlab#1{#1}\fi
\providecommand{\url}[1]{\href{#1}{#1}}
\providecommand{\dodoi}[1]{doi:~\href{http://doi.org/#1}{\nolinkurl{#1}}}
\providecommand{\doeprint}[1]{\href{http://ascl.net/#1}{\nolinkurl{http://ascl.net/#1}}}
\providecommand{\doarXiv}[1]{\href{https://arxiv.org/abs/#1}{\nolinkurl{https://arxiv.org/abs/#1}}}

\bibitem[{C.~M.~O.~D. {Alexander} {et~al.}(2017){Alexander}, {Cody}, {De Gregorio}, {Nittler}, \& {Stroud}}]{Alexander_2017ChEG...77..227A}
{Alexander}, C.~M.~O.~D., {Cody}, G.~D., {De Gregorio}, B.~T., {Nittler}, L.~R., \& {Stroud}, R.~M. 2017, \bibinfo{title}{{The nature, origin and modification of insoluble organic matter in chondrites, the major source of Earth's C and N},} Chemie der Erde / Geochemistry, 77, 227, \dodoi{10.1016/j.chemer.2017.01.007}

\bibitem[{D.~E. {Anderson} {et~al.}(2017){Anderson}, {Bergin}, {Blake}, {Ciesla}, {Visser}, \& {Lee}}]{Anderson_2017ApJ_disk_C...845...13A}
{Anderson}, D.~E., {Bergin}, E.~A., {Blake}, G.~A., {et~al.} 2017, \bibinfo{title}{{Destruction of Refractory Carbon in Protoplanetary Disks},} \apj, 845, 13, \dodoi{10.3847/1538-4357/aa7da1}

\bibitem[{B.~A. {Anzures} {et~al.}(2024){Anzures}, {Telus}, {McCubbin}, {Thompson}, {Jakubek}, {Fries}, {Clark}, \& {Casbeer}}]{Anzures_2024LPICo3040.1250A}
{Anzures}, B.~A., {Telus}, M., {McCubbin}, F.~M., {et~al.} 2024, \bibinfo{title}{{Enstatite Chondrite Outgassing and Condensate Formation: Implications for Early Atmosphere Development},} in LPI Contributions, Vol. 3040, 55th Lunar and Planetary Science Conference, 1250

\bibitem[{A.~E. {Beck}(1976){Beck}}]{Beck_1976Geop...41..133B}
{Beck}, A.~E. 1976, \bibinfo{title}{{An Improved Method of Computing the Thermal Conductivity of Fluid-Filled Sedimentary Rocks},} Geophysics, 41, 133, \dodoi{10.1190/1.1440596}

\bibitem[{E.~A. {Bergin} {et~al.}(2015){Bergin}, {Blake}, {Ciesla}, {Hirschmann}, \& {Li}}]{bergin15}
{Bergin}, E.~A., {Blake}, G.~A., {Ciesla}, F., {Hirschmann}, M.~M., \& {Li}, J. 2015, \bibinfo{title}{{Tracing the ingredients for a habitable earth from interstellar space through planet formation},} Proceedings of the National Academy of Science, 112, 8965, \dodoi{10.1073/pnas.1500954112}

\bibitem[{K.~R. {Bermingham} {et~al.}(2020){Bermingham}, {F{\"u}ri}, {Lodders}, \& {Marty}}]{Bermingham20}
{Bermingham}, K.~R., {F{\"u}ri}, E., {Lodders}, K., \& {Marty}, B. 2020, \bibinfo{title}{{The NC-CC Isotope Dichotomy: Implications for the Chemical and Isotopic Evolution of the Early Solar System},} \ssr, 216, 133, \dodoi{10.1007/s11214-020-00748-w}

\bibitem[{F. {Binkert} \& T. {Birnstiel}(2023){Binkert} \& {Birnstiel}}]{Binkert_23_C_in_disc}
{Binkert}, F., \& {Birnstiel}, T. 2023, \bibinfo{title}{{Carbon depletion in the early Solar system},} \mnras, 520, 2055, \dodoi{10.1093/mnras/stad182}

\bibitem[{A. {Bischoff} {et~al.}(2006){Bischoff}, {Scott}, {Metzler}, \& {Goodrich}}]{Bischoff_2006_brecciation_mess.book..679B}
{Bischoff}, A., {Scott}, E.~R.~D., {Metzler}, K., \& {Goodrich}, C.~A. 2006, \bibinfo{title}{{Nature and Origins of Meteoritic Breccias},} in Meteorites and the Early Solar System II, ed. D.~S. {Lauretta} \& H.~Y. {McSween}, 679

\bibitem[{P.~A. {Bland} {et~al.}(2009){Bland}, {Jackson}, {Coker}, {Cohen}, {Webber}, {Lee}, {Duffy}, {Chater}, {Ardakani}, {McPhail}, {McComb}, \& {Benedix}}]{Bland}
{Bland}, P.~A., {Jackson}, M.~D., {Coker}, R.~F., {et~al.} 2009, \bibinfo{title}{{Why aqueous alteration in asteroids was isochemical: High porosity {\ensuremath{\neq}} high permeability},} Earth and Planetary Science Letters, 287, 559, \dodoi{10.1016/j.epsl.2009.09.004}

\bibitem[{N. {Braukm{\"u}ller} {et~al.}(2019){Braukm{\"u}ller}, {Wombacher}, {Funk}, \& {M{\"u}nker}}]{Braukmueller_2019NatGe..12..564B}
{Braukm{\"u}ller}, N., {Wombacher}, F., {Funk}, C., \& {M{\"u}nker}, C. 2019, \bibinfo{title}{{Earth's volatile element depletion pattern inherited from a carbonaceous chondrite-like source},} Nature Geoscience, 12, 564, \dodoi{10.1038/s41561-019-0375-x}

\bibitem[{A.~J. {Brearley} {et~al.}(1989){Brearley}, {Scott}, {Keil}, {Clayton}, {Mayeda}, {Boynton}, \& {Hill}}]{Brearley_1989GeCoA..53.2081B}
{Brearley}, A.~J., {Scott}, E.~R.~D., {Keil}, K., {et~al.} 1989, \bibinfo{title}{{Chemical, isotopic and mineralogical evidence for the origin of matrix in ordinary chondrites},} \gca, 53, 2081, \dodoi{10.1016/0016-7037(89)90326-8}

\bibitem[{D.~T. {Britt} \& G.~J. {Consolmagno}(2003){Britt} \& {Consolmagno}}]{Britt_2003M&PS...38.1161B}
{Britt}, D.~T., \& {Consolmagno}, G.~J. 2003, \bibinfo{title}{{Stony meteorite porosities and densities: A review of the data through 2001},} \maps, 38, 1161, \dodoi{10.1111/j.1945-5100.2003.tb00305.x}

\bibitem[{O. {Christ} {et~al.}(2024){Christ}, {Nestola}, \& {Alvaro}}]{Christ_2024CmChe...7..118C}
{Christ}, O., {Nestola}, F., \& {Alvaro}, M. 2024, \bibinfo{title}{{Open questions on carbonaceous matter in meteorites},} Communications Chemistry, 7, 118, \dodoi{10.1038/s42004-024-01200-8}

\bibitem[{G.~D. {Cody} {et~al.}(2008){Cody}, {Alexander}, {Yabuta}, {Kilcoyne}, {Araki}, {Ade}, {Dera}, {Fogel}, {Militzer}, \& {Mysen}}]{COdy_2008E&PSL.272..446C_Organic_reaction}
{Cody}, G.~D., {Alexander}, C.~M.~O., {Yabuta}, H., {et~al.} 2008, \bibinfo{title}{{Organic thermometry for chondritic parent bodies},} Earth and Planetary Science Letters, 272, 446, \dodoi{10.1016/j.epsl.2008.05.008}

\bibitem[{G. {Consolmagno} {et~al.}(2008){Consolmagno}, {Britt}, \& {Macke}}]{Consolmagno_2008ChEG...68....1C_porosity}
{Consolmagno}, G., {Britt}, D., \& {Macke}, R. 2008, \bibinfo{title}{{The significance of meteorite density and porosity},} Chemie der Erde / Geochemistry, 68, 1, \dodoi{10.1016/j.chemer.2008.01.003}

\bibitem[{C.~M. {Corrigan} {et~al.}(1997){Corrigan}, {Zolensky}, {Dahl}, {Long}, {Weir}, {Sapp}, \& {Burkett}}]{Corrigan_1997M&PS...32..509C_permeability}
{Corrigan}, C.~M., {Zolensky}, M.~E., {Dahl}, J., {et~al.} 1997, \bibinfo{title}{{The porosity and permeability of chondritic meteorites and interplanetary dust particles},} \maps, 32, 509, \dodoi{10.1111/j.1945-5100.1997.tb01296.x}

\bibitem[{C. Di~Blasi {et~al.}(2017)Di~Blasi, Branca, \& Galgano}]{Di_blasi_biomass}
Di~Blasi, C., Branca, C., \& Galgano, A. 2017, \bibinfo{title}{On the Experimental Evidence of Exothermicity in Wood and Biomass Pyrolysis,} Energy Technology, 5, 19, \dodoi{https://doi.org/10.1002/ente.201600091}

\bibitem[{P.~M. {Doyle} {et~al.}(2015){Doyle}, {Jogo}, {Nagashima}, {Krot}, {Wakita}, {Ciesla}, \& {Hutcheon}}]{Doyle_2015NatCo...6.7444D}
{Doyle}, P.~M., {Jogo}, K., {Nagashima}, K., {et~al.} 2015, \bibinfo{title}{{Early aqueous activity on the ordinary and carbonaceous chondrite parent bodies recorded by fayalite},} Nature Communications, 6, 7444, \dodoi{10.1038/ncomms8444}

\bibitem[{R.~A. Fischer {et~al.}(2020)Fischer, Cottrell, Hauri, Lee, \& Le~Voyer}]{Fischer20}
Fischer, R.~A., Cottrell, E., Hauri, E., Lee, K. K.~M., \& Le~Voyer, M. 2020, \bibinfo{title}{The carbon content of Earth and its core,} Proceedings of the National Academy of Sciences, 117, 8743, \dodoi{10.1073/pnas.1919930117}

\bibitem[{R.~R. {Fu} \& L.~T. {Elkins-Tanton}(2014){Fu} \& {Elkins-Tanton}}]{Fu_2014E&PSL.390..128F}
{Fu}, R.~R., \& {Elkins-Tanton}, L.~T. 2014, \bibinfo{title}{{The fate of magmas in planetesimals and the retention of primitive chondritic crusts},} Earth and Planetary Science Letters, 390, 128, \dodoi{10.1016/j.epsl.2013.12.047}

\bibitem[{R.~R. {Fu} {et~al.}(2017){Fu}, {Young}, {Greenwood}, \& {Elkins-Tanton}}]{Fu_2017pedc.book..115F}
{Fu}, R.~R., {Young}, E.~D., {Greenwood}, R.~C., \& {Elkins-Tanton}, L.~T. 2017, \bibinfo{title}{{Silicate Melting and Volatile Loss During Differentiation in Planetesimals},} in Planetesimals: Early Differentiation and Consequences for Planets, ed. L.~T. {Elkins-Tanton} \& B.~P. {Weiss}, 115--135, \dodoi{10.1017/9781316339794.006}

\bibitem[{R.~R. Fu {et~al.}(2017)Fu, Young, Greenwood, \& Elkins-Tanton}]{fu_2017}
Fu, R.~R., Young, E.~D., Greenwood, R.~C., \& Elkins-Tanton, L.~T. 2017, Silicate Melting and Volatile Loss During Differentiation in Planetesimals, ed. L.~T. Elkins-Tanton \& B.~P. Weiss, Cambridge Planetary Science (Cambridge University Press), 115--135, \dodoi{10.1017/9781316339794.006}

\bibitem[{H.-P. {Gail} \& M. {Trieloff}(2017){Gail} \& {Trieloff}}]{GTDust}
{Gail}, H.-P., \& {Trieloff}, M. 2017, \bibinfo{title}{{Spatial distribution of carbon dust in the early solar nebula and the carbon content of planetesimals},} \aap, 606, A16, \dodoi{10.1051/0004-6361/201730480}

\bibitem[{H. {Genda} \& M. {Ikoma}(2008){Genda} \& {Ikoma}}]{Genda_2008Icar..194...42G}
{Genda}, H., \& {Ikoma}, M. 2008, \bibinfo{title}{{Origin of the ocean on the Earth: Early evolution of water D/H in a hydrogen-rich atmosphere},} \icarus, 194, 42, \dodoi{10.1016/j.icarus.2007.09.007}

\bibitem[{A. {Giri} {et~al.}(2022){Giri}, {Dionne}, \& {Hopkins}}]{Giri_2022npjCM...8...55G}
{Giri}, A., {Dionne}, C.~J., \& {Hopkins}, P.~E. 2022, \bibinfo{title}{{Atomic coordination dictates vibrational characteristics and thermal conductivity in amorphous carbon},} npj Computational Mathematics, 8, 55, \dodoi{10.1038/s41524-022-00741-7}

\bibitem[{C.~A. {Goodrich} {et~al.}(2022){Goodrich}, {Collinet}, {Treiman}, {Prissel}, {Patzek}, {Ebert}, {Jercinovic}, {Bischoff}, {Pack}, {Barrat}, \& {Decker}}]{Goodrich_2022M&PS...57.1589G}
{Goodrich}, C.~A., {Collinet}, M., {Treiman}, A., {et~al.} 2022, \bibinfo{title}{{The first main group ureilite with primary plagioclase: A missing link in the differentiation of the ureilite parent body},} \maps, 57, 1589, \dodoi{10.1111/maps.13889}

\bibitem[{M.~M. {Grady} \& I.~P. {Wright}(2003){Grady} \& {Wright}}]{Grady_2003SSRv..106..231G}
{Grady}, M.~M., \& {Wright}, I.~P. 2003, \bibinfo{title}{{Elemental and Isotopic Abundances of Carbon and Nitrogen in Meteorites},} \ssr, 106, 231, \dodoi{10.1023/A:1024645906350}

\bibitem[{R.~E. {Grimm} \& H.~Y. {McSween}(1989){Grimm} \& {McSween}}]{Grimm_1989Icar_McSween...82..244G}
{Grimm}, R.~E., \& {McSween}, Jr., H.~Y. 1989, \bibinfo{title}{{Water and the thermal evolution of carbonaceous chondrite parent bodies},} \icarus, 82, 244, \dodoi{10.1016/0019-1035(89)90038-9}

\bibitem[{J.~T. {Gu} {et~al.}(2024){Gu}, {Peng}, {Ji}, {Zhang}, {Yang}, {Hoyos}, {Hirschmann}, {Kite}, \& {Fischer}}]{Gu_2024_Earth_Volatiles}
{Gu}, J.~T., {Peng}, B., {Ji}, X., {et~al.} 2024, \bibinfo{title}{{Composition of Earth's initial atmosphere and fate of accreted volatiles set by core formation and magma ocean redox evolution},} Earth and Planetary Science Letters, 629, 118618, \dodoi{10.1016/j.epsl.2024.118618}

\bibitem[{A.~F. {Gualtieri} {et~al.}(2012){Gualtieri}, {Giacobbe}, \& {Viti}}]{Gualtieri_2012AmMin..97..666G}
{Gualtieri}, A.~F., {Giacobbe}, C., \& {Viti}, C. 2012, \bibinfo{title}{{The dehydroxylation of serpentine group minerals},} American Mineralogist, 97, 666, \dodoi{10.2138/am.2012.3952}

\bibitem[{C.~R. Harris {et~al.}(2020)Harris, Millman, van~der Walt, Gommers, Virtanen, Cournapeau, Wieser, Taylor, Berg, Smith, Kern, Picus, Hoyer, van Kerkwijk, Brett, Haldane, del R{\'{i}}o, Wiebe, Peterson, G{\'{e}}rard-Marchant, Sheppard, Reddy, Weckesser, Abbasi, Gohlke, \& Oliphant}]{Numpy_harris2020array}
Harris, C.~R., Millman, K.~J., van~der Walt, S.~J., {et~al.} 2020, \bibinfo{title}{Array programming with {NumPy},} Nature, 585, 357, \dodoi{10.1038/s41586-020-2649-2}

\bibitem[{K. {Hashizume} \& N. {Sugiura}(1998){Hashizume} \& {Sugiura}}]{Hashizume_1998M&PS...33.1181H}
{Hashizume}, K., \& {Sugiura}, N. 1998, \bibinfo{title}{{Transportation of gaseous elements and isotopes in a thermally evolving chondritic planetesimal},} \maps, 33, 1181, \dodoi{10.1111/j.1945-5100.1998.tb01722.x}

\bibitem[{S. {Henke} {et~al.}(2012){Henke}, {Gail}, {Trieloff}, {Schwarz}, \& {Kleine}}]{henke12}
{Henke}, S., {Gail}, H.~P., {Trieloff}, M., {Schwarz}, W.~H., \& {Kleine}, T. 2012, \bibinfo{title}{{Thermal evolution and sintering of chondritic planetesimals},} \aap, 537, A45, \dodoi{10.1051/0004-6361/201117177}

\bibitem[{P.~J. {Hevey} \& I.~S. {Sanders}(2006){Hevey} \& {Sanders}}]{Hevey_2006M&PS_differentiation...41...95H}
{Hevey}, P.~J., \& {Sanders}, I.~S. 2006, \bibinfo{title}{{A model for planetesimal meltdown by $^{26}$Al and its implications for meteorite parent bodies},} \maps, 41, 95, \dodoi{10.1111/j.1945-5100.2006.tb00195.x}

\bibitem[{N. {Hirakawa} {et~al.}(2021){Hirakawa}, {Kebukawa}, {Furukawa}, {Kondo}, {Nakano}, \& {Kobayashi}}]{Hirakawa_2021EP&S...73...16H}
{Hirakawa}, N., {Kebukawa}, Y., {Furukawa}, Y., {et~al.} 2021, \bibinfo{title}{{Aqueous alteration without initial water: possibility of organic-induced hydration of anhydrous silicates in meteorite parent bodies},} Earth, Planets and Space, 73, 16, \dodoi{10.1186/s40623-020-01352-6}

\bibitem[{M.~M. Hirschmann(2012)Hirschmann}]{HIRSCHMANN2012}
Hirschmann, M.~M. 2012, \bibinfo{title}{Magma ocean influence on early atmosphere mass and composition,} Earth and Planetary Science Letters, 341-344, 48 , \dodoi{https://doi.org/10.1016/j.epsl.2012.06.015}

\bibitem[{M.~M. {Hirschmann}(2018){Hirschmann}}]{Hirschmann_2018_deep_Earth_HCN}
{Hirschmann}, M.~M. 2018, \bibinfo{title}{{Comparative deep Earth volatile cycles: The case for C recycling from exosphere/mantle fractionation of major (H$_{2}$O, C, N) volatiles and from H$_{2}$O/Ce, CO$_{2}$/Ba, and CO$_{2}$/Nb exosphere ratios},} Earth and Planetary Science Letters, 502, 262, \dodoi{10.1016/j.epsl.2018.08.023}

\bibitem[{M.~M. {Hirschmann} {et~al.}(2021){Hirschmann}, {Bergin}, {Blake}, {Ciesla}, \& {Li}}]{Hirschmann_2021PNAS}
{Hirschmann}, M.~M., {Bergin}, E.~A., {Blake}, G.~A., {Ciesla}, F.~J., \& {Li}, J. 2021, \bibinfo{title}{{Early volatile depletion on planetesimals inferred from C{\textendash}S systematics of iron meteorite parent bodies},} Proceedings of the National Academy of Science, 118, e2026779118, \dodoi{10.1073/pnas.2026779118}

\bibitem[{M.~L. {Huber} {et~al.}(2016){Huber}, {Sykioti}, {Assael}, \& {Perkins}}]{Huber_2016JPCRD..45a3102H}
{Huber}, M.~L., {Sykioti}, E.~A., {Assael}, M.~J., \& {Perkins}, R.~A. 2016, \bibinfo{title}{{Reference Correlation of the Thermal Conductivity of Carbon Dioxide from the Triple Point to 1100 K and up to 200 MPa},} Journal of Physical and Chemical Reference Data, 45, 013102, \dodoi{10.1063/1.4940892}

\bibitem[{J.~D. Hunter(2007)Hunter}]{Matplotlib_Hunter:2007}
Hunter, J.~D. 2007, \bibinfo{title}{Matplotlib: A 2D graphics environment,} Computing in Science \& Engineering, 9, 90, \dodoi{10.1109/MCSE.2007.55}

\bibitem[{M. {Ikoma} \& H. {Genda}(2006){Ikoma} \& {Genda}}]{Ikoma_2006ApJ...648..696I}
{Ikoma}, M., \& {Genda}, H. 2006, \bibinfo{title}{{Constraints on the Mass of a Habitable Planet with Water of Nebular Origin},} \apj, 648, 696, \dodoi{10.1086/505780}

\bibitem[{C. {Jain} \& J. {Korenaga}(2020){Jain} \& {Korenaga}}]{Jain_2020JGRB..12519896J}
{Jain}, C., \& {Korenaga}, J. 2020, \bibinfo{title}{{Synergy of Experimental Rock Mechanics, Seismology, and Geodynamics Reveals Still Elusive Upper Mantle Rheology},} Journal of Geophysical Research (Solid Earth), 125, e2020JB019896, \dodoi{10.1029/2020JB019896}

\bibitem[{C. {Jain} {et~al.}(2019){Jain}, {Korenaga}, \& {Karato}}]{Jain}
{Jain}, C., {Korenaga}, J., \& {Karato}, S.-i. 2019, \bibinfo{title}{{Global Analysis of Experimental Data on the Rheology of Olivine Aggregates},} Journal of Geophysical Research (Solid Earth), 124, 310, \dodoi{10.1029/2018JB016558}

\bibitem[{S. {Jakobsson} \& N. {Oskarsson}(1994){Jakobsson} \& {Oskarsson}}]{Jakobsson_1994GeCoA..58....9J}
{Jakobsson}, S., \& {Oskarsson}, N. 1994, \bibinfo{title}{{The system C-O in equilibrium with graphite at high pressure and temperature: An experimental study},} \gca, 58, 9, \dodoi{10.1016/0016-7037(94)90442-1}

\bibitem[{I.~C. {Jaramillo} {et~al.}(2014){Jaramillo}, {Gaddam}, {Vander Wal}, {Huang}, {Levinthal}, \& {Lighty}}]{Jaramillo_2014CoFl..161.2951J}
{Jaramillo}, I.~C., {Gaddam}, C.~K., {Vander Wal}, R.~L., {et~al.} 2014, \bibinfo{title}{{Soot oxidation kinetics under pressurized conditions},} Combustion and Flame, 161, 2951, \dodoi{10.1016/j.combustflame.2014.04.016}

\bibitem[{A. {Johansen} {et~al.}(2021){Johansen}, {Ronnet}, {Bizzarro}, {Schiller}, {Lambrechts}, {Nordlund}, \& {Lammer}}]{Johansen21}
{Johansen}, A., {Ronnet}, T., {Bizzarro}, M., {et~al.} 2021, \bibinfo{title}{{A pebble accretion model for the formation of the terrestrial planets in the Solar System},} Science Advances, 7, eabc0444, \dodoi{10.1126/sciadv.abc0444}

\bibitem[{S. Kanchika \& F. Wakai(2018)Kanchika \& Wakai}]{KANCHIKA_20184283_sintering_pressure}
Kanchika, S., \& Wakai, F. 2018, \bibinfo{title}{Surface tension-pressure superposition principle for anisotropic shrinkage of an ellipsoidal pore in viscous sintering,} Journal of the European Ceramic Society, 38, 4283, \dodoi{https://doi.org/10.1016/j.jeurceramsoc.2018.05.015}

\bibitem[{Y. {Kebukawa} {et~al.}(2010){Kebukawa}, {Nakashima}, \& {Zolensky}}]{Kebukawa_2010M&PS...45...99K}
{Kebukawa}, Y., {Nakashima}, S., \& {Zolensky}, M.~E. 2010, \bibinfo{title}{{Kinetics of organic matter degradation in the Murchison meteorite for the evaluation of parent-body temperature history},} \maps, 45, 99, \dodoi{10.1111/j.1945-5100.2009.01008.x}

\bibitem[{H. Keppler \& G. Golabek(2019)Keppler \& Golabek}]{keppler19}
Keppler, H., \& Golabek, G. 2019, \bibinfo{title}{Graphite floatation on a magma ocean and the fate of carbon during core formation,} Geochemical Perspectives Letters, 12--17, \dodoi{10.7185/geochemlet.1918}

\bibitem[{M. {Kimura} {et~al.}(2008){Kimura}, {Grossman}, \& {Weisberg}}]{kimura_2008M&PS...43.1161K_FeNi_in_chondrites}
{Kimura}, M., {Grossman}, J.~N., \& {Weisberg}, M.~K. 2008, \bibinfo{title}{{Fe-Ni metal in primitive chondrites: Indicators of classification and metamorphic conditions for ordinary and CO chondrites},} \maps, 43, 1161, \dodoi{10.1111/j.1945-5100.2008.tb01120.x}

\bibitem[{A.~J. {King} {et~al.}(2021){King}, {Schofield}, \& {Russell}}]{King_2021GeCoA.298..167K}
{King}, A.~J., {Schofield}, P.~F., \& {Russell}, S.~S. 2021, \bibinfo{title}{{Thermal alteration of CM carbonaceous chondrites: Mineralogical changes and metamorphic temperatures},} \gca, 298, 167, \dodoi{10.1016/j.gca.2021.02.011}

\bibitem[{K. {Kiryu} {et~al.}(2020){Kiryu}, {Kebukawa}, {Igisu}, {Shibuya}, {Zolensky}, \& {Kobayashi}}]{Kiryu_2020M&PS...55.1848K}
{Kiryu}, K., {Kebukawa}, Y., {Igisu}, M., {et~al.} 2020, \bibinfo{title}{{Kinetics in thermal evolution of Raman spectra of chondritic organic matter to evaluate thermal history of their parent bodies},} \maps, 55, 1848, \dodoi{10.1111/maps.13548}

\bibitem[{L. {Klarmann} {et~al.}(2018){Klarmann}, {Ormel}, \& {Dominik}}]{Klarmann_2018A&A...618L...1K_disk}
{Klarmann}, L., {Ormel}, C.~W., \& {Dominik}, C. 2018, \bibinfo{title}{{Radial and vertical dust transport inhibit refractory carbon depletion in protoplanetary disks},} \aap, 618, L1, \dodoi{10.1051/0004-6361/201833719}

\bibitem[{T.~S. {Kruijer} {et~al.}(2017){Kruijer}, {Burkhardt}, {Budde}, \& {Kleine}}]{Kruijer_2017PNAS..114.6712K}
{Kruijer}, T.~S., {Burkhardt}, C., {Budde}, G., \& {Kleine}, T. 2017, \bibinfo{title}{{Age of Jupiter inferred from the distinct genetics and formation times of meteorites},} Proceedings of the National Academy of Science, 114, 6712, \dodoi{10.1073/pnas.1704461114}

\bibitem[{C. {Le Guillou} \& A. {Brearley}(2014){Le Guillou} \& {Brearley}}]{Le_Guillou_2014GeCoA.131..344L}
{Le Guillou}, C., \& {Brearley}, A. 2014, \bibinfo{title}{{Relationships between organics, water and early stages of aqueous alteration in the pristine CR3.0 chondrite MET 00426},} \gca, 131, 344, \dodoi{10.1016/j.gca.2013.10.024}

\bibitem[{J.-E. {Lee} {et~al.}(2010){Lee}, {Bergin}, \& {Nomura}}]{LeeDisk}
{Lee}, J.-E., {Bergin}, E.~A., \& {Nomura}, H. 2010, \bibinfo{title}{{The Solar Nebula on Fire: A Solution to the Carbon Deficit in the Inner Solar System},} \apjl, 710, L21, \dodoi{10.1088/2041-8205/710/1/L21}

\bibitem[{C. {Li} \& T.~C. {Brown}(2001){Li} \& {Brown}}]{Li_2001Carbo_kinetics..39..725L}
{Li}, C., \& {Brown}, T.~C. 2001, \bibinfo{title}{{Carbon oxidation kinetics from evolved carbon oxide analysis during temperature-programmed oxidation},} Carbon, 39, 725, \dodoi{10.1016/S0008-6223(00)00189-5}

\bibitem[{J. {Li} {et~al.}(2021){Li}, {Bergin}, {Blake}, {Ciesla}, \& {Hirschmann}}]{Li_2021SciA_C_deficit_early_subli}
{Li}, J., {Bergin}, E.~A., {Blake}, G.~A., {Ciesla}, F.~J., \& {Hirschmann}, M.~M. 2021, \bibinfo{title}{{Earth's carbon deficit caused by early loss through irreversible sublimation},} Science Advances, 7, eabd3632, \dodoi{10.1126/sciadv.abd3632}

\bibitem[{T. {Lichtenberg} {et~al.}(2021){Lichtenberg}, {Dr{\k{a}}{\.z}kowska}, {Sch{\"o}nb{\"a}chler}, {Golabek}, \& {Hands}}]{Lichtenberg_2021Sci...371..365L}
{Lichtenberg}, T., {Dr{\k{a}}{\.z}kowska}, J., {Sch{\"o}nb{\"a}chler}, M., {Golabek}, G.~J., \& {Hands}, T.~O. 2021, \bibinfo{title}{{Bifurcation of planetary building blocks during Solar System formation},} Science, 371, 365, \dodoi{10.1126/science.abb3091}

\bibitem[{T. {Lichtenberg} {et~al.}(2023){Lichtenberg}, {Schaefer}, {Nakajima}, \& {Fischer}}]{Lichtenberg_2023ASPC_review..534..907L}
{Lichtenberg}, T., {Schaefer}, L.~K., {Nakajima}, M., \& {Fischer}, R.~A. 2023, \bibinfo{title}{{Geophysical Evolution During Rocky Planet Formation},} in Astronomical Society of the Pacific Conference Series, Vol. 534, Protostars and Planets VII, ed. S.~{Inutsuka}, Y.~{Aikawa}, T.~{Muto}, K.~{Tomida}, \& M.~{Tamura}, 907, \dodoi{10.48550/arXiv.2203.10023}

\bibitem[{P. {Lindgren} {et~al.}(2020){Lindgren}, {Lee}, {Sparkes}, {Greenwood}, {Hanna}, {Franchi}, {King}, {Floyd}, {Martin}, {Hamilton}, \& {Haberle}}]{Lindgren_2020GeCoA.289...69L}
{Lindgren}, P., {Lee}, M.~R., {Sparkes}, R., {et~al.} 2020, \bibinfo{title}{{Signatures of the post-hydration heating of highly aqueously altered CM carbonaceous chondrites and implications for interpreting asteroid sample returns},} \gca, 289, 69, \dodoi{10.1016/j.gca.2020.08.021}

\bibitem[{K. {Lodders}(2003){Lodders}}]{Lodders_03_SolarAbun}
{Lodders}, K. 2003, \bibinfo{title}{{Solar System Abundances and Condensation Temperatures of the Elements},} \apj, 591, 1220, \dodoi{10.1086/375492}

\bibitem[{R.~J. {Macke} {et~al.}(2011{\natexlab{a}}){Macke}, {Consolmagno}, \& {Britt}}]{Macke_2011M&PS...46.1842M}
{Macke}, R.~J., {Consolmagno}, G.~J., \& {Britt}, D.~T. 2011{\natexlab{a}}, \bibinfo{title}{{Density, porosity, and magnetic susceptibility of carbonaceous chondrites},} \maps, 46, 1842, \dodoi{10.1111/j.1945-5100.2011.01298.x}

\bibitem[{R.~J. {Macke} {et~al.}(2011{\natexlab{b}}){Macke}, {Consolmagno}, \& {Britt}}]{Mache_2011M&PS...46.1842M_porosity_CC}
{Macke}, R.~J., {Consolmagno}, G.~J., \& {Britt}, D.~T. 2011{\natexlab{b}}, \bibinfo{title}{{Density, porosity, and magnetic susceptibility of carbonaceous chondrites},} \maps, 46, 1842, \dodoi{10.1111/j.1945-5100.2011.01298.x}

\bibitem[{J.~K. {Mackenzie} \& R. {Shuttleworth}(1949){Mackenzie} \& {Shuttleworth}}]{Mackenzie_1949PPSB_sintering_theory}
{Mackenzie}, J.~K., \& {Shuttleworth}, R. 1949, \bibinfo{title}{{A Phenomenological Theory of Sintering},} Proceedings of the Physical Society B, 62, 833, \dodoi{10.1088/0370-1301/62/12/310}

\bibitem[{G.~J. {MacPherson} \& A.~N. {Krot}(2014){MacPherson} \& {Krot}}]{MacPherson_2014M&PS...49.1250M_permeability}
{MacPherson}, G.~J., \& {Krot}, A.~N. 2014, \bibinfo{title}{{The formation of Ca-, Fe-rich silicates in reduced and oxidized CV chondrites: The roles of impact-modified porosity and permeability, and heterogeneous distribution of water ices},} \maps, 49, 1250, \dodoi{10.1111/maps.12316}

\bibitem[{B. {Mahan} {et~al.}(2018){Mahan}, {Siebert}, {Blanchard}, {Borensztajn}, {Badro}, \& {Moynier}}]{Mahan_2018GeCoA.235...21M}
{Mahan}, B., {Siebert}, J., {Blanchard}, I., {et~al.} 2018, \bibinfo{title}{{Constraining compositional proxies for Earth's accretion and core formation through high pressure and high temperature Zn and S metal-silicate partitioning},} \gca, 235, 21, \dodoi{10.1016/j.gca.2018.04.032}

\bibitem[{U. {Marboeuf} {et~al.}(2012){Marboeuf}, {Schmitt}, {Petit}, {Mousis}, \& {Fray}}]{Marboeuf_2012A&A...542A..82M_comet_flow}
{Marboeuf}, U., {Schmitt}, B., {Petit}, J.~M., {Mousis}, O., \& {Fray}, N. 2012, \bibinfo{title}{{A cometary nucleus model taking into account all phase changes of water ice: amorphous, crystalline, and clathrate},} \aap, 542, A82, \dodoi{10.1051/0004-6361/201118176}

\bibitem[{S. {Mathur} {et~al.}(1967){Mathur}, {Tondon}, \& {Saxena}}]{Mathur_1967MolPh..12..569M}
{Mathur}, S., {Tondon}, P.~K., \& {Saxena}, S.~C. 1967, \bibinfo{title}{{Thermal conductivity of binary, ternary and quaternary mixtures of rare gases},} Molecular Physics, 12, 569, \dodoi{10.1080/00268976700100731}

\bibitem[{T. {Matsui} {et~al.}(1986){Matsui}, {Sugiura}, \& {Brar}}]{Matsui_1986Metic..21..109M_permeability}
{Matsui}, T., {Sugiura}, N., \& {Brar}, N.~S. 1986, \bibinfo{title}{{Gas Permeability of Shocked Chondrites},} Meteoritics, 21, 109, \dodoi{10.1111/j.1945-5100.1986.tb01229.x}

\bibitem[{M. {Matsuoka} {et~al.}(2022){Matsuoka}, {Nakamura}, {Miyajima}, {Hiroi}, {Imae}, \& {Yamaguchi}}]{Matsuoka_2022GeCoA.316..150M}
{Matsuoka}, M., {Nakamura}, T., {Miyajima}, N., {et~al.} 2022, \bibinfo{title}{{Spectral and mineralogical alteration process of naturally-heated CM and CY chondrites},} \gca, 316, 150, \dodoi{10.1016/j.gca.2021.08.042}

\bibitem[{M. {Mentser} \& S. {Ergun}(1967){Mentser} \& {Ergun}}]{Mentser_1967Carbo...5..331M}
{Mentser}, M., \& {Ergun}, S. 1967, \bibinfo{title}{{Kinetics of oxygen exchange between CO $_{2}$ and CO on carbon},} Carbon, 5, 331, \dodoi{10.1016/0008-6223(67)90049-8}

\bibitem[{J. {Millat} \& W.~A. {Wakeham}(1989){Millat} \& {Wakeham}}]{Millat_1989JPCRD..18..565M}
{Millat}, J., \& {Wakeham}, W.~A. 1989, \bibinfo{title}{{The Thermal Conductivity of Nitrogen and Carbon Monoxide in the Limit of Zero Density},} Journal of Physical and Chemical Reference Data, 18, 565, \dodoi{10.1063/1.555827}

\bibitem[{C.~B. {Moore} \& C.~F. {Lewis}(1967){Moore} \& {Lewis}}]{Moore_1967JGR....72.6289M_C_in_OC}
{Moore}, C.~B., \& {Lewis}, C.~F. 1967, \bibinfo{title}{{Total carbon content of ordinary chondrites},} \jgr, 72, 6289, \dodoi{10.1029/JZ072i024p06289}

\bibitem[{N. {Moskovitz} \& E. {Gaidos}(2011){Moskovitz} \& {Gaidos}}]{Moskovitz_2011M&PS...46..903M}
{Moskovitz}, N., \& {Gaidos}, E. 2011, \bibinfo{title}{{Differentiation of planetesimals and the thermal consequences of melt migration},} \maps, 46, 903, \dodoi{10.1111/j.1945-5100.2011.01201.x}

\bibitem[{D.~W. {Muenow} {et~al.}(1995){Muenow}, {Keil}, \& {McCoy}}]{Muenow_1995Metic..30..639M}
{Muenow}, D.~W., {Keil}, K., \& {McCoy}, T.~J. 1995, \bibinfo{title}{{Volatiles in unequilibrated ordinary chondrites: Abundances, sources and implications for explosive volcanism on differentiated asteroids},} Meteoritics, 30, 639, \dodoi{10.1111/j.1945-5100.1995.tb01161.x}

\bibitem[{T. {Nakamura}(2005){Nakamura}}]{Nakamura_2005JMPeS.100..260N}
{Nakamura}, T. 2005, \bibinfo{title}{{Post-hydration thermal metamorphism of carbonaceous chondrites},} Journal of Mineralogical and Petrological Sciences, 100, 260, \dodoi{10.2465/jmps.100.260}

\bibitem[{W. {Neumann} {et~al.}(2012){Neumann}, {Breuer}, \& {Spohn}}]{Neumann_2012A&A...543A.141N}
{Neumann}, W., {Breuer}, D., \& {Spohn}, T. 2012, \bibinfo{title}{{Differentiation and core formation in accreting planetesimals},} \aap, 543, A141, \dodoi{10.1051/0004-6361/201219157}

\bibitem[{W. {Neumann} {et~al.}(2014){Neumann}, {Breuer}, \& {Spohn}}]{neumann14}
{Neumann}, W., {Breuer}, D., \& {Spohn}, T. 2014, \bibinfo{title}{{Modelling of compaction in planetesimals},} \aap, 567, A120, \dodoi{10.1051/0004-6361/201423648}

\bibitem[{W. {Neumann} {et~al.}(2018){Neumann}, {Kruijer}, {Breuer}, \& {Kleine}}]{neumann17}
{Neumann}, W., {Kruijer}, T.~S., {Breuer}, D., \& {Kleine}, T. 2018, \bibinfo{title}{{Multistage Core Formation in Planetesimals Revealed by Numerical Modeling and Hf-W Chronometry of Iron Meteorites},} Journal of Geophysical Research (Planets), 123, 421, \dodoi{10.1002/2017JE005411}

\bibitem[{W. {Neumann} {et~al.}(2024){Neumann}, {Ma}, {Bouvier}, \& {Trieloff}}]{Neumann_2024NatSR..1414017N}
{Neumann}, W., {Ma}, N., {Bouvier}, A., \& {Trieloff}, M. 2024, \bibinfo{title}{{Recurrent planetesimal formation in an outer part of the early solar system},} Scientific Reports, 14, 14017, \dodoi{10.1038/s41598-024-63768-4}

\bibitem[{H. {Okamoto}(1992){Okamoto}}]{Okamoto_1992_CFe}
{Okamoto}, H. 1992, \bibinfo{title}{{The C-Fe (carbon-iron) system},} Jounral of Phase Equilibria, 13, 543, \dodoi{https://doi.org/10.1007/BF02665767}

\bibitem[{T. {Okamoto} \& S. {Ida}(2024){Okamoto} \& {Ida}}]{Okamoto_2024A&A...692A..11O}
{Okamoto}, T., \& {Ida}, S. 2024, \bibinfo{title}{{Effects from different grades of stickiness between icy and silicate particles on carbon depletion in protoplanetary disks},} \aap, 692, A11, \dodoi{10.1051/0004-6361/202451908}

\bibitem[{R. {Okazaki} {et~al.}(2023){Okazaki}, {Marty}, {Busemann}, {Hashizume}, {Gilmour}, {Meshik}, {Yada}, {Kitajima}, {Broadley}, {Byrne}, {F{\"u}ri}, {Riebe}, {Krietsch}, {Maden}, {Ishida}, {Clay}, {Crowther}, {Fawcett}, {Lawton}, {Pravdivtseva}, {Miura}, {Park}, {Bajo}, {Takano}, {Yamada}, {Kawagucci}, {Matsui}, {Yamamoto}, {Righter}, {Sakai}, {Iwata}, {Shirai}, {Sekimoto}, {Inagaki}, {Ebihara}, {Yokochi}, {Nishiizumi}, {Nagao}, {Lee}, {Kano}, {Caffee}, {Uemura}, {Nakamura}, {Naraoka}, {Noguchi}, {Yabuta}, {Yurimoto}, {Tachibana}, {Sawada}, {Sakamoto}, {Abe}, {Arakawa}, {Fujii}, {Hayakawa}, {Hirata}, {Hirata}, {Honda}, {Honda}, {Hosoda}, {Iijima}, {Ikeda}, {Ishiguro}, {Ishihara}, {Iwata}, {Kawahara}, {Kikuchi}, {Kitazato}, {Matsumoto}, {Matsuoka}, {Michikami}, {Mimasu}, {Miura}, {Morota}, {Nakazawa}, {Namiki}, {Noda}, {Noguchi}, {Ogawa}, {Ogawa}, {Okada}, {Okamoto}, {Ono}, {Ozaki}, {Saiki}, {Sakatani}, {Senshu}, {Shimaki}, {Shirai}, {Sugita}, {Takei}, {Takeuchi}, {Tanaka}, {Tatsumi}, {Terui},
  {Tsukizaki}, {Wada}, {Yamada}, {Yamada}, {Yamamoto}, {Yano}, {Yokota}, {Yoshihara}, {Yoshikawa}, {Yoshikawa}, {Furuya}, {Hatakeda}, {Hayashi}, {Hitomi}, {Kumagai}, {Miyazaki}, {Nakato}, {Nishimura}, {Soejima}, {Iwamae}, {Yamamoto}, {Yogata}, {Yoshitake}, {Fukai}, {Usui}, {Connolly}, {Lauretta}, {Watanabe}, \& {Tsuda}}]{Okazaki_2023Sci...379.0431O_Ryugu}
{Okazaki}, R., {Marty}, B., {Busemann}, H., {et~al.} 2023, \bibinfo{title}{{Noble gases and nitrogen in samples of asteroid Ryugu record its volatile sources and recent surface evolution},} Science, 379, abo0431, \dodoi{10.1126/science.abo0431}

\bibitem[{J. {Palguta} {et~al.}(2010){Palguta}, {Schubert}, \& {Travis}}]{Palguta_2010E&PSL.296..235P}
{Palguta}, J., {Schubert}, G., \& {Travis}, B.~J. 2010, \bibinfo{title}{{Fluid flow and chemical alteration in carbonaceous chondrite parent bodies},} Earth and Planetary Science Letters, 296, 235, \dodoi{10.1016/j.epsl.2010.05.003}

\bibitem[{L. {Piani} {et~al.}(2020){Piani}, {Marrocchi}, {Rigaudier}, {Vacher}, {Thomassin}, \& {Marty}}]{Piani_2020Sci...369.1110P}
{Piani}, L., {Marrocchi}, Y., {Rigaudier}, T., {et~al.} 2020, \bibinfo{title}{{Earth{\textquoteright}s water may have been inherited from material similar to enstatite chondrite meteorites},} Science, 369, 1110, \dodoi{10.1126/science.aba1948}

\bibitem[{L. {Piani} {et~al.}(2012){Piani}, {Robert}, {Beyssac}, {Binet}, {Bourot-Denise}, {Derenne}, {Le Guillou}, {Marrocchi}, {Mostefaoui}, {Rouzaud}, \& {Thomen}}]{Piani_2012M&PS...47....8P}
{Piani}, L., {Robert}, F., {Beyssac}, O., {et~al.} 2012, \bibinfo{title}{{Structure, composition, and location of organic matter in the enstatite chondrite Sahara 97096 (EH3)},} \maps, 47, 8, \dodoi{10.1111/j.1945-5100.2011.01306.x}

\bibitem[{L. {Remusat} {et~al.}(2012){Remusat}, {Rouzaud}, {Charon}, {Le Guillou}, {Guan}, \& {Eiler}}]{Remusat_2012GeCoA..96..319R}
{Remusat}, L., {Rouzaud}, J.~N., {Charon}, E., {et~al.} 2012, \bibinfo{title}{{D-depleted organic matter and graphite in the Abee enstatite chondrite},} \gca, 96, 319, \dodoi{10.1016/j.gca.2012.07.031}

\bibitem[{R.~A. {Robie} \& B.~S. {Hemingway}(1995){Robie} \& {Hemingway}}]{Robie_1995_book_therm}
{Robie}, R.~A., \& {Hemingway}, B.~S. 1995, \bibinfo{title}{{Thermodynamic properties of minerals and related substances at 298.15 K and 1 bar (10\^5 Pascals) pressure and at higher temperatures},} U.S. Geol. Survey Bull, 2131, 461

\bibitem[{D.~C. Rubie {et~al.}(2011)Rubie, Frost, Mann, Asahara, Nimmo, Tsuno, Kegler, Holzheid, \& Palme}]{Rubie11}
Rubie, D.~C., Frost, D.~J., Mann, U., {et~al.} 2011, \bibinfo{title}{Heterogeneous accretion, composition and core--mantle differentiation of the Earth,} Earth and Planetary Science Letters, 301, 31 , \dodoi{https://doi.org/10.1016/j.epsl.2010.11.030}

\bibitem[{D.~C. {Rubie} {et~al.}(2015){Rubie}, {Jacobson}, {Morbidelli}, {O'Brien}, {Young}, {de Vries}, {Nimmo}, {Palme}, \& {Frost}}]{Rubie_15_redox_core_formation}
{Rubie}, D.~C., {Jacobson}, S.~A., {Morbidelli}, A., {et~al.} 2015, \bibinfo{title}{{Accretion and differentiation of the terrestrial planets with implications for the compositions of early-formed Solar System bodies and accretion of water},} \icarus, 248, 89, \dodoi{10.1016/j.icarus.2014.10.015}

\bibitem[{L. {Schaefer} \& J. {Fegley}(2017){Schaefer} \& {Fegley}}]{Schaefer17}
{Schaefer}, L., \& {Fegley}, Bruce, J. 2017, \bibinfo{title}{{Redox States of Initial Atmospheres Outgassed on Rocky Planets and Planetesimals},} \apj, 843, 120, \dodoi{10.3847/1538-4357/aa784f}

\bibitem[{M.~B. {Schwenn} \& C. {Goetze}(1978){Schwenn} \& {Goetze}}]{Schwenn_1978Tectp..48...41S}
{Schwenn}, M.~B., \& {Goetze}, C. 1978, \bibinfo{title}{{Creep of olivine during hot-pressing},} Tectonophysics, 48, 41, \dodoi{10.1016/0040-1951(78)90085-9}

\bibitem[{Z.~D. {Sharp} \& P.~L. {Olson}(2022){Sharp} \& {Olson}}]{Sharp_2022GeCoA.333..124S}
{Sharp}, Z.~D., \& {Olson}, P.~L. 2022, \bibinfo{title}{{Multi-element constraints on the sources of volatiles to Earth},} \gca, 333, 124, \dodoi{10.1016/j.gca.2022.07.007}

\bibitem[{A.~J. {Soini} {et~al.}(2020){Soini}, {Kukkonen}, {Kohout}, \& {Luttinen}}]{Soini_2020M&PS...55..402S}
{Soini}, A.~J., {Kukkonen}, I.~T., {Kohout}, T., \& {Luttinen}, A. 2020, \bibinfo{title}{{Thermal and porosity properties of meteorites: A compilation of published data and new measurements},} \maps, 55, 402, \dodoi{10.1111/maps.13441}

\bibitem[{J. {Storz} {et~al.}(2021){Storz}, {Ludwig}, {Bischoff}, {Schwarz}, \& {Trieloff}}]{Storz_2021GeCoA.307...86S}
{Storz}, J., {Ludwig}, T., {Bischoff}, A., {Schwarz}, W.~H., \& {Trieloff}, M. 2021, \bibinfo{title}{{Graphite in ureilites, enstatite chondrites, and unique clasts in ordinary chondrites - Insights from the carbon-isotope composition},} \gca, 307, 86, \dodoi{10.1016/j.gca.2021.05.028}

\bibitem[{N. {Sugiura} {et~al.}(1986{\natexlab{a}}){Sugiura}, {Arkani-Hamed}, \& {Strangway}}]{Sugiura_1986E&PSL..78..148S}
{Sugiura}, N., {Arkani-Hamed}, J., \& {Strangway}, D.~W. 1986{\natexlab{a}}, \bibinfo{title}{{Possible transport of carbon in meteorite parent bodies},} Earth and Planetary Science Letters, 78, 148, \dodoi{10.1016/0012-821X(86)90056-7}

\bibitem[{N. {Sugiura} {et~al.}(1986{\natexlab{b}}){Sugiura}, {Arkani-Hamed}, \& {Strangway}}]{Sugiura86}
{Sugiura}, N., {Arkani-Hamed}, J., \& {Strangway}, D.~W. 1986{\natexlab{b}}, \bibinfo{title}{{Possible transport of carbon in meteorite parent bodies},} Earth and Planetary Science Letters, 78, 148, \dodoi{10.1016/0012-821X(86)90056-7}

\bibitem[{N. {Sugiura} {et~al.}(1984){Sugiura}, {Brar}, {Strangway}, \& {Matsui}}]{Sugiura_1984LPSC...14..641S}
{Sugiura}, N., {Brar}, N.~S., {Strangway}, D.~W., \& {Matsui}, T. 1984, \bibinfo{title}{{Degassing of meteorite parent bodies},} Lunar and Planetary Science Conference Proceedings, 89, B641, \dodoi{10.1029/JB089iS02p0B641}

\bibitem[{N. {Sugiura} {et~al.}(1986{\natexlab{c}}){Sugiura}, {Matsui}, \& {Strangway}}]{Sugiura_1986PolRe..41..358S_permeability}
{Sugiura}, N., {Matsui}, T., \& {Strangway}, D.~W. 1986{\natexlab{c}}, \bibinfo{title}{{Gas permeability of some Antarctic chondrites},} National Institute Polar Research Memoirs, 41, 358

\bibitem[{M.~V. {Swain} \& B.~K. {Atkinson}(1978){Swain} \& {Atkinson}}]{Swain_Atkinson_1978PApGe_Olivine_SE}
{Swain}, M.~V., \& {Atkinson}, B.~K. 1978, \bibinfo{title}{{Fracture surface energy of olivine},} Pure and Applied Geophysics, 116, 866, \dodoi{10.1007/BF00876542}

\bibitem[{G.~J. {Taylor}(1992){Taylor}}]{Taylor_1992JGR....9714717T}
{Taylor}, G.~J. 1992, \bibinfo{title}{{Core formation in asteroids},} \jgr, 97, 14717, \dodoi{10.1029/92JE01501}

\bibitem[{N.~B.~L. Thermodynamics Research~Center(2021)Thermodynamics Research~Center}]{NIST}
Thermodynamics Research~Center, N. B.~L. 2021, {Thermodynamics Source Database}, \dodoi{10.18434/T4D303}

\bibitem[{D. {Thomassin} {et~al.}(2023){Thomassin}, {Piani}, {Villeneuve}, {Caumon}, {Bouden}, \& {Marrocchi}}]{Thomassin_2023E&PSL.61618225T}
{Thomassin}, D., {Piani}, L., {Villeneuve}, J., {et~al.} 2023, \bibinfo{title}{{The high-temperature origin of hydrogen in enstatite chondrite chondrules and implications for the origin of terrestrial water},} Earth and Planetary Science Letters, 616, 118225, \dodoi{10.1016/j.epsl.2023.118225}

\bibitem[{M.~A. {Thompson} {et~al.}(2021){Thompson}, {Telus}, {Schaefer}, {Fortney}, {Joshi}, \& {Lederman}}]{Thompson_2021NatAs...5..575T}
{Thompson}, M.~A., {Telus}, M., {Schaefer}, L., {et~al.} 2021, \bibinfo{title}{{Composition of terrestrial exoplanet atmospheres from meteorite outgassing experiments},} Nature Astronomy, 5, 575, \dodoi{10.1038/s41550-021-01338-8}

\bibitem[{J.~M. {Tucker} \& S. {Mukhopadhyay}(2014){Tucker} \& {Mukhopadhyay}}]{Tucker_14_multipleOutgassing}
{Tucker}, J.~M., \& {Mukhopadhyay}, S. 2014, \bibinfo{title}{{Evidence for multiple magma ocean outgassing and atmospheric loss episodes from mantle noble gases},} Earth and Planetary Science Letters, 393, 254, \dodoi{10.1016/j.epsl.2014.02.050}

\bibitem[{E. {Vaccaro} {et~al.}(2023){Vaccaro}, {Wozniakiewicz}, {Franchi}, {Starkey}, \& {Russell}}]{Vaccaro_2023M&PS...58..688V}
{Vaccaro}, E., {Wozniakiewicz}, P., {Franchi}, I.~A., {Starkey}, N., \& {Russell}, S.~S. 2023, \bibinfo{title}{{The grain size distribution of matrix in primitive chondrites},} \maps, 58, 688, \dodoi{10.1111/maps.13979}

\bibitem[{L. Vacher {et~al.}(2024)Vacher, Eschrig, Bonal, Fujiya, Flandinet, \& Beck}]{VACHER_2024_Geochimica_et_Cosmochimica_Acta}
Vacher, L., Eschrig, J., Bonal, L., {et~al.} 2024, \bibinfo{title}{Thermal metamorphism and volatile evolution in unequilibrated ordinary chondrites: Implications for the delivery of hydrogen to terrestrial planets,} Geochimica et Cosmochimica Acta, \dodoi{https://doi.org/10.1016/j.gca.2024.12.016}

\bibitem[{P. Virtanen {et~al.}(2020)Virtanen, Gommers, Oliphant, Haberland, Reddy, Cournapeau, Burovski, Peterson, Weckesser, Bright, {van der Walt}, Brett, Wilson, Millman, Mayorov, Nelson, Jones, Kern, Larson, Carey, Polat, Feng, Moore, {VanderPlas}, Laxalde, Perktold, Cimrman, Henriksen, Quintero, Harris, Archibald, Ribeiro, Pedregosa, {van Mulbregt}, \& {SciPy 1.0 Contributors}}]{SciPy_2020-NMeth}
Virtanen, P., Gommers, R., Oliphant, T.~E., {et~al.} 2020, \bibinfo{title}{{{SciPy} 1.0: Fundamental Algorithms for Scientific Computing in Python},} Nature Methods, 17, 261, \dodoi{10.1038/s41592-019-0686-2}

\bibitem[{C. {Viti}(2010){Viti}}]{Viti_2010AmMin..95..631V}
{Viti}, C. 2010, \bibinfo{title}{{Serpentine minerals discrimination by thermal analysis},} American Mineralogist, 95, 631, \dodoi{10.2138/am.2010.3366}

\bibitem[{T. {Wada} {et~al.}(1971){Wada}, {Wada}, {Elliott}, \& {Chipman}}]{Wada_1971_FeNiC}
{Wada}, T., {Wada}, H., {Elliott}, J.~F., \& {Chipman}, J. 1971, \bibinfo{title}{{Thermodynamics of the Fcc Fe−Ni−C and Ni−C alloys},} Metallurgical Transactions, 2, 2199, \dodoi{https://doi.org/10.1007/BF02917551}

\bibitem[{J.~T. {Wasson}(1974){Wasson}}]{Wasson_1974_Meteorites_book}
{Wasson}, J.~T. 1974, {Meteorites : classification and properties}

\bibitem[{M.~K. {Weisberg} {et~al.}(2006){Weisberg}, {McCoy}, \& {Krot}}]{Weisberg_2006mess.book...19W}
{Weisberg}, M.~K., {McCoy}, T.~J., \& {Krot}, A.~N. 2006, \bibinfo{title}{{Systematics and Evaluation of Meteorite Classification},} in Meteorites and the Early Solar System II, ed. D.~S. {Lauretta} \& H.~Y. {McSween} (University of Arizona Press), 19

\bibitem[{L. {Wilson} \& K. {Keil}(1991){Wilson} \& {Keil}}]{Wilson_1991E&PSL.104..505W}
{Wilson}, L., \& {Keil}, K. 1991, \bibinfo{title}{{Consequences of explosive eruptions on small Solar System bodies: the case of the missing basalts on the aubrite parent body},} Earth and Planetary Science Letters, 104, 505, \dodoi{10.1016/0012-821X(91)90225-7}

\bibitem[{J. {Wu} {et~al.}(2018){Wu}, {Desch}, {Schaefer}, {Elkins-Tanton}, {Pahlevan}, \& {Buseck}}]{Wu_2018JGRE..123.2691W}
{Wu}, J., {Desch}, S.~J., {Schaefer}, L., {et~al.} 2018, \bibinfo{title}{{Origin of Earth's Water: Chondritic Inheritance Plus Nebular Ingassing and Storage of Hydrogen in the Core},} Journal of Geophysical Research (Planets), 123, 2691, \dodoi{10.1029/2018JE005698}

\bibitem[{E.~D. {Young}(2001){Young}}]{Young_2001RSPTA.359.2095Y}
{Young}, E.~D. 2001, \bibinfo{title}{{The hydrology of carbonaceous chondrite parent bodies and the evolution of planet progenitors},} Philosophical Transactions of the Royal Society of London Series A, 359, 2095, \dodoi{10.1098/rsta.2001.0900}

\bibitem[{E.~D. {Young} {et~al.}(1999){Young}, {Ash}, {England}, \& {Rumble}}]{Young_1999Sci...286.1331Y}
{Young}, E.~D., {Ash}, R.~D., {England}, P., \& {Rumble}, III, D. 1999, \bibinfo{title}{{Fluid flow in chondritic parent bodies: deciphering the compositions of planetesimals.},} Science, 286, 1331, \dodoi{10.1126/science.286.5443.1331}

\bibitem[{E.~D. {Young} {et~al.}(2023){Young}, {Shahar}, \& {Schlichting}}]{Young_2023_Earth_Primordial_H2_atm}
{Young}, E.~D., {Shahar}, A., \& {Schlichting}, H.~E. 2023, \bibinfo{title}{{Earth shaped by primordial H$_{2}$ atmospheres},} \nat, 616, 306, \dodoi{10.1038/s41586-023-05823-0}

\bibitem[{E.~D. {Young} {et~al.}(2003){Young}, {Zhang}, \& {Schubert}}]{Young_2003E&PSL.213..249Y}
{Young}, E.~D., {Zhang}, K.~K., \& {Schubert}, G. 2003, \bibinfo{title}{{Conditions for pore water convection within carbonaceous chondrite parent bodies - implications for planetesimal size and heat production},} Earth and Planetary Science Letters, 213, 249, \dodoi{10.1016/S0012-821X(03)00345-5}

\end{thebibliography}
\bibliographystyle{aasjournalv7}

\end{document}